\documentclass[a4paper,fleqn,usenatbib]{mnras}

\usepackage{newtxtext,newtxmath}

\usepackage[T1]{fontenc}
\usepackage{ae,aecompl}

\usepackage{graphicx}	
\usepackage{amsmath}	

\title[HR 8799 variability]{A High-Contrast Search for Variability in HR 8799bc with VLT-SPHERE}

\author[B.A. Biller et al.]{B.A. Biller,$^{1,2}$\thanks{E-mail: bb@roe.ac.uk}
D. Apai$^{3}$, M. Bonnefoy$^{4}$, S. Desidera$^{5}$, R. Gratton$^{5}$, M. Kasper$^{6}$,\newauthor M. Kenworthy$^{7}$, A.M. Lagrange$^{4}$, C. Lazzoni$^{5,8}$, D. Mesa$^{5}$, A. Vigan$^{9}$, \newauthor K. Wagner$^{3,10}$, J.M. Vos$^{11}$, A. Zurlo$^{9,12}$ \\
$^{1}$ SUPA, Institute for Astronomy, University of Edinburgh, Blackford Hill, Edinburgh EH9 3HJ, UK \\
$^{2}$ Centre for Exoplanet Science, University of Edinburgh, Edinburgh, UK \\
$^{3}$ Steward Observatory, The University of Arizona, Tucson, AZ 85721, USA \\
$^{4}$ Univ. Grenoble Alpes, CNRS, IPAG, F-38000 Grenoble, France \\
$^{5}$ INAF - Osservatorio Astronomico di Padova, Vicolo della Osservatorio 5, 35122, Padova, Italy \\
$^{6}$ European Southern Observatory (ESO), Karl-Schwarzschild-Str.
2,85748 Garching, Germany \\
$^{7}$ Leiden Observatory, Leiden University, PO Box 9513, 2300 RA, Leiden, The Netherlands \\
$^{8}$ Dipartimento di Fisica a Astronomia "G. Galilei", Universita' di Padova, Via Marzolo, 8, 35121 Padova, Italy \\
$^{9}$ Aix Marseille Univ, CNRS, CNES, LAM, Marseille, France \\
$^{10}$ NASA Hubble/Sagan Fellow \\
$^{11}$ Department of Astrophysics, American Museum of Natural History, Central Park West at 79th Street, New York, NY 10024, USA \\
$^{12}$ Escuela de Ingenier\'ia Industrial, Facultad de Ingenier\'ia y Ciencias, Universidad Diego Portales, Av. Ejercito 441, Santiago, Chile \\
}

\begin{document}
\label{firstpage}
\pagerange{\pageref{firstpage}--\pageref{lastpage}}
\maketitle

\begin{abstract}
The planets HR8799bc display nearly identical colours and spectra as variable young exoplanet analogues such as VHS 1256-1257ABb and PSO J318.5-22, and are likely to be similarly variable. Here we present results from a 5-epoch SPHERE IRDIS broadband-$H$ search for variability in these two planets. HR 8799b aperture photometry and HR 8799bc negative simulated planet photometry share similar trends within uncertainties.  Satellite spot lightcurves share the same trends as the planet lightcurves in the August 2018 epochs, but diverge in the October 2017 epochs. We consider $\Delta(mag)_{b} - \Delta(mag)_{c}$ to trace non-shared variations between the two planets, and rule out non-shared variability in $\Delta(mag)_{b} - \Delta(mag)_{c}$ to the 10-20$\%$ level over 4-5 hours.  To quantify our sensitivity to variability, we simulate variable lightcurves by inserting and retrieving a suite of simulated planets at similar radii from the star as HR 8799bc, but offset in position angle.  For HR 8799b, for periods $<$10 hours, we are sensitive to variability with amplitude $>5\%$.  For HR 8799c, our sensitivity is limited to variability $>25\%$ for similar periods. 
\end{abstract}



\section{Introduction}

Over the last 15 years, a small but growing cohort of young, widely-separated giant planets, such as HR 8799bcde \citep{Mar08, Maro10} and $\beta$ Pic b \citep{Lag10}, have been directly imaged via their own thermal emission in the infrared \citep[see~e.g.~][]{Bowl16}.  Instruments such as SPHERE at the VLT \citep{Beuzit2019}, GPI at Gemini \citep{Mac14}, SCExAO-Charis at Subaru \citep{Jovanovic2015, Currie2019}, and Gravity at the VLT \citep{Gravity2019} have enabled spectroscopy and deep characterization of the atmospheres of these planets \citep{Konopacky2013, Sne14,Barman2015, Bon14, Bon16, Zur16, DeR16, Ing14,Raj15, Lavie17, Wang2018a, Greenbaum2018, Ruffio2019AJ, Gravity2019}.  However, most of these studies have captured only a snapshot of the planet's atmospheric state, taken over a 1-3 hour long observation, or have combined multiple observations taken at different points in the planet's rotation, creating essentially an "integrated" spectrum.  Only one directly imaged exoplanet has a measured rotation period, $\beta$ Pic b, which is a fast rotator, with a 7-9 hour period \citep{Sne14}, assuming a similar inclination as the disc.  Top-of-atmosphere asymmetric structure (e.g. from spots, patchy clouds and various other mechanisms) could cause significant changes in flux as a function of time and wavelength, as different structures rotate in and out of view. These structures may also evolve dynamically over multiple rotation periods, producing longer-term changes in planet brightness.  To determine if either of these phenomena are present requires time-resolved observations of exoplanets, preferably covering multiple rotation periods.  Detecting and characterizing such features, especially as a function of wavelength, will provide strong tests for atmospheric models of these planets.


Studies of field brown dwarfs, the more massive cousins of these young planets, suggest that we would expect to find significant changes in the flux of young planets with time.  Quasiperiodic variability is commonly found in field L and T type brown dwarfs, with periods from 1.5-30 hours, and amplitudes ranging from 0.01-27$\%$ \citep{Rad12, Rad14a, Rad14b, Wil14, Met15,Eriksson2019}.  These objects are fast rotators \citep[3-20~hours,~see~e.g.][]{Zap06} with silicate clouds expected in their atmospheres \citep[e.g.~horizontal~banded~clouds~recently~detected~in~Luhman~16A~via~linear~polarization~by][]{Millar-Blanchaer2020}. These potentially inhomogeneous cloud features likely drive the observed variability, modulated by the fast rotation of these objects.  Spectroscopic variability monitoring has shown, for instance, that the variability in high amplitude L/T transition brown dwarfs such as 2MASS J21392676+0220226 (henceforth 2M2139) and SIMP J013656.5+093347 (henceforth SIMP 0136) is best reproduced by inhomogeneous coverage of both thin and thicker cloud components \citep{Apa13}. 

Directly imaged young, giant exoplanets share similar effective temperatures and compositions as these L and T dwarfs, and likely possess comparable or even higher amplitude variability.  
One key physical difference between brown dwarfs and young giant exoplanets is the much lower surface gravity of the young planets compared to more massive, older brown dwarfs.  
Surface gravity strongly affects the placement of clouds in these atmospheres, with ramifications for the variability properties of younger, lower mass objects \citep{Mar12}. The variability amplitudes of young, free-floating planetary mass  objects appear to be significantly boosted relative to older, higher mass brown dwarf counterparts with similar spectral types.  \citet{Vos2019} find that 30$\%$ of low surface-gravity L dwarfs are variable, compared to 11$\%$ of field L dwarfs.  The near-IR variability detections in the L6-L7 exoplanet analogues VHS J125601.92-125723.9 \citep[henceforth~VHS~1256-1257ABb,][]{Gauza2015, Bowler2020, Zhou2020}, PSO J318.5338-22.8603
\citep[henceforth~PSO~J318.5-22~,$\sim8~M_{Jup}~\beta$~Pic~moving~group~member][]{Liu13, All16}, 
WISEPJ004701.06+680352.1 \citep[henceforth~W0047,~AB~Dor~moving~group,~][]{Giz15}, and 2MASS J2244316+204343 \citep[henceforth 2M2244,~AB~Dor~moving~group,~][]{Vos2018} have peak-to-trough  amplitudes $>$5$\%$ 
\citep{Biller2015, Biller2018, Vos2018, Lew16}.  These are the {\bf highest-amplitude} variability detections for {\bf any} L spectral type object, with VHS 1256-1257ABb potentially the most variable planetary mass or substellar object yet observed 
\citep[$\sim$25$\%$~variability~observed~over~an~8-hour~HST~observation,][]{Bowler2020}. 
These objects have spectra that are nearly identical to those of the
outer two HR 8799 planets \citep{Zur16, Bon16} and HIP 65426b \citep{Chauvin2017}.
Given their close spectral match, the HR 8799 planets are potentially equally intrinsically variable \citep[although~the~observable~variability~amplitude~ is~likely~smaller,~as~the~HR~8799~planets~are~probably~viewed~close~to~pole-on,][]{Ruffio2019AJ, Wang2018b}. 

The extreme contrast difference between star and planet has hindered studies of variability in exoplanet companions such as HR 8799bcde.  Planet imagers such as SPHERE at the VLT \citep{Beuzit2019} and GPI at Gemini \citep{Mac14} now enable short cadence observations of bright giant planets -- previous generations of imagers required $\geq$1 hour for a high S/N ratio image of such a planet.  In contrast, HR 8799bcd are visible even in $<$1 min SPHERE $H$-band single exposures.  \citet{Apa16}
conducted a pilot variability study ($<$2 hours on sky) of the HR 8799 planets as part of SPHERE science verification and conclusively show that variability monitoring is possible using high-contrast coronagraphic imaging on SPHERE.  Here we present the first in-depth, long ($>$4 hour) search for variability in the HR 8799 planets. 

\section{Prospects for Variability monitoring of the HR 8799 system}
\label{sec:system}

The star HR 8799 is a F0V member of the 42$^{+6}_{-4}$ Myr Columba moving group \citep{Torres2008, Zuckerman2011, Bell2015}, with an estimated mass of $\sim$1.5 solar masses.  HR 8799 is a $\gamma$ Doradus pulsating variable. As in all $\gamma$ Doradus stars, many pulsation modes are excited; the mode with highest amplitude has a period of $\sim$0.5 day \citep{Sodor2014}, unfortunately close to the potential rotation period of the planets.  The optical lightcurve from the Microvariability and Oscillations in STars (MOST) space telescope from Figure 1 of \citet{Sodor2014} displays clear peaks at the $\sim$0.5 day period; these peaks, however, have variable heights due to constructive and destructive interference from other weaker pulsation modes. At its most variable, HR 8799 displays a peak-to-valley amplitude of $\sim$0.1 mag in the visible \citep{Sodor2014}.  While we expect a smaller amplitude at longer wavelengths \citep[and~indeed,~observations~of~other~$\gamma$~Dor~variables~with~TESS~suggest~this~is~the~case,][]{Antoci2019}, the red optical or near-IR variability of HR 8799 has not been directly measured.  Thus, the fact that SPHERE enables monitoring of the multiple HR 8799 planets simultaneously as well as the ability to produce artificial "satellite spots" as photometric references \citep{Lang13} crucially allow us to disentangle any intrinsic variability of the star from variability of the planets.  As we do not expect the planets to have the same rotational period, phase or variability amplitude, any variability with the same period and phase shared by multiple planets would then probably either be intrinsic to the star or caused by changing observing conditions.  Similarly, the satellite spot light curves should follow the trend produced by the changing conditions as well as any astrophysical variability from the star, providing a reference lightcurve that can be used to detrend single planet lightcurves.  Detrending using a lightcurve from a photometric reference has been successfully implemented in numerous ground-based variability monitoring campaigns for isolated objects \citep{Rad14a, Vos2019}, building a "calibration curve" from other stars in the field to remove variations due to changing conditions; here we will attempt a similar detrending using both the satellite spot lightcurves and the other planets as the photometric references.    

The HR 8799 system also hosts a 3-component debris disk (a warm inner belt separated by the orbits of the planets from a cold planetesimal belt and an outer halo of small grains) imaged at mid-IR, submillimeter, and mm wavelengths \citep{Su2009, Matthews2014, Booth2016, Wilner2018}.  The disk is viewed nearly face-on \citep{Matthews2014}.
Current orbital fits suggest that the planet orbits are close to coplanar with the disk, with a small inclination of $\sim$27$^{\circ}$ \citep{Ruffio2019AJ, Wang2018b, Konopacky2016, Pueyo2015, Maire2015}.  If the 4 planets indeed share a similar inclination with the disk, they are likely viewed close to pole-on.  This is a non-ideal viewing angle for variability detection. 
The brown dwarfs with the highest observed variability amplitudes are generally those which are observed at or close to equator-on \citep[90$^{\circ}$~inclination,][]{Vos2017}, as inhomogeneous features on objects observed at lower inclinations will appear smaller due to projection effects.  Thus, given the probable pole-on viewing angle of the HR 8799 planets, any variability amplitude we potentially measure is likely only a fraction of the intrinsic variability amplitude.  

With the current generation of near-IR extreme adaptive optics (AO) planet-finding cameras, high-contrast ground-based variability monitoring is feasible at wavelengths of 1-4 $\mu$m.  After correcting for inclination effects, field brown dwarfs \citep{Rad14a,Met15} display a monotonic decrease in variability amplitude across this wavelength range, with the highest amplitude variability in $J$ band ($\sim$1.1-1.3 $\mu$m), decreasing somewhat through the rest of the near-IR, and with considerably lower mid-IR (3-5 $\mu$m) amplitudes; thus, SPHERE's near-IR capabilities are ideal for a first search for variability in the HR 8799 planets, which share similar atmospheric properties to these objects.  In our initial July 2015 observation, we chose to observe in the SPHERE $J23$ filter, as close as possible to the expected $J$-band peak of variability for similar objects.  However, as HR 8799bcde are intrinsically fainter at $J$ than in $H$ or $K$, we found that the planets were not detected with sufficient S/N in the July 2015 data to enable variability studies, hence for the rest of the epochs, the broadband $H$ filter was chosen as the best compromise.

\section{Observations and Data Reduction}

From July 2015 to August 2018, we obtained 7 epochs of SPHERE-IRDIS variability monitoring for the HR 8799 planets (programs 095.C-0689(A), 099.C-0588(A), and 0101.C-0315(A)). A summary of the observations is provided in Table~\ref{tab:observations}. SPHERE-IRDIS is a dual-band imaging (DBI) camera, enabling coronagraphic observations in two wavelength bands simultaneously \citep{Dohlen2008,Vigan2010, Beuzit2019}.  In July 2015, we collected one epoch of simultaneous imaging in the SPHERE dual-band $J23$ filters ($J2$: $\lambda_c$=1189.5 nm, $\Delta\lambda$=47.6 nm, $J3$: $\lambda_c$=1269.8 nm, $\Delta\lambda$=50.8 nm) and a second epoch of simultaneous imaging in the $K12$ filters ($K1$: $\lambda_c$=2102.5 nm, $\Delta\lambda$=102 nm, $K2$: $\lambda_c$=2255 nm, $\Delta\lambda$=109 nm).  All other epochs were obtained using the SPHERE broadband $H$ filter ($\lambda_c$=1625.5 nm, $\Delta\lambda$=291 nm) in both of IRDIS' wavelength channels\footnote{We report results from IRDIS channel 1 here, as the data obtained from channel 2 is functionally identical to the channel 1 data.}.  HR 8799 is relatively far north for the VLT, thus we were only able to observe for 4-6 hours per epoch with airmass $<$2.5.  Observations were taken in pupil-tracking mode, allowing the field to rotate with parallactic angle on the sky, in order to build an accurate point spread function (PSF) for quasistatic speckle suppression \citep[see~e.g.][]{Mar06, Laf07, Sou12, Ama12}.  We used the $N\_ALC\_YJH\_S$ coronagraphic configuration (inner working angle of 0.0925") of SPHERE's apodized pupil Lyot coronagraph \citep[][]{Carbillet2011, Guerri2011} in all epochs except for the $K12$ observations taken on 31 July 2015, which utilized the $N\_ALC\_Ks$ coronagraphic configuration instead (inner working angle of 0.12").  The 2015 epochs were taken in designated visitor modes for full half nights, enabling longer overall observations (albeit sometimes at higher airmasses).  All other epochs were taken in service mode.  Conditions on 20 August 2018 were excellent, leading the service observer to adjust the base exposure time (DIT) on the fly to reduce saturation of HR 8799.  Unfortunately, the number of exposures (NEXP) taken was not also adjusted to compensate for the decreased DIT, leading to a significant loss in time coverage at this epoch.  On some nights, initial exposures were taken with different DIT values or satellite spot illumination; these frames have been removed from the subsequent analysis.

To ensure a simultaneous photometric reference at all epochs, observations were conducted using the \textit{star center} template that provides satellite spots.  Adding an additional periodic modulation to the deformable mirror (above and beyond that used to correct the incoming wavefront) produces 4 satellite spots in a cross pattern on the detector \citep{Lang13}.  These satellite spots appear relatively circular when using a narrow-band filter and elongated when using a wide-band filter.  They share the spectra of the star and can potentially be used as both photometric and astrometric references \citep{Lang13, Wang2014}.  In the July 2015 observations, off-axis point spread function (PSF) images were also interspersed every 20-30 minutes throughout the observation to use as a photometric reference.  We found that this compromised the stability of the observations, so in later epochs we only obtained off-axis PSF images at the start and finish of each 4-5 hour observing sequence.

\begin{figure*}
\includegraphics[scale=0.3]{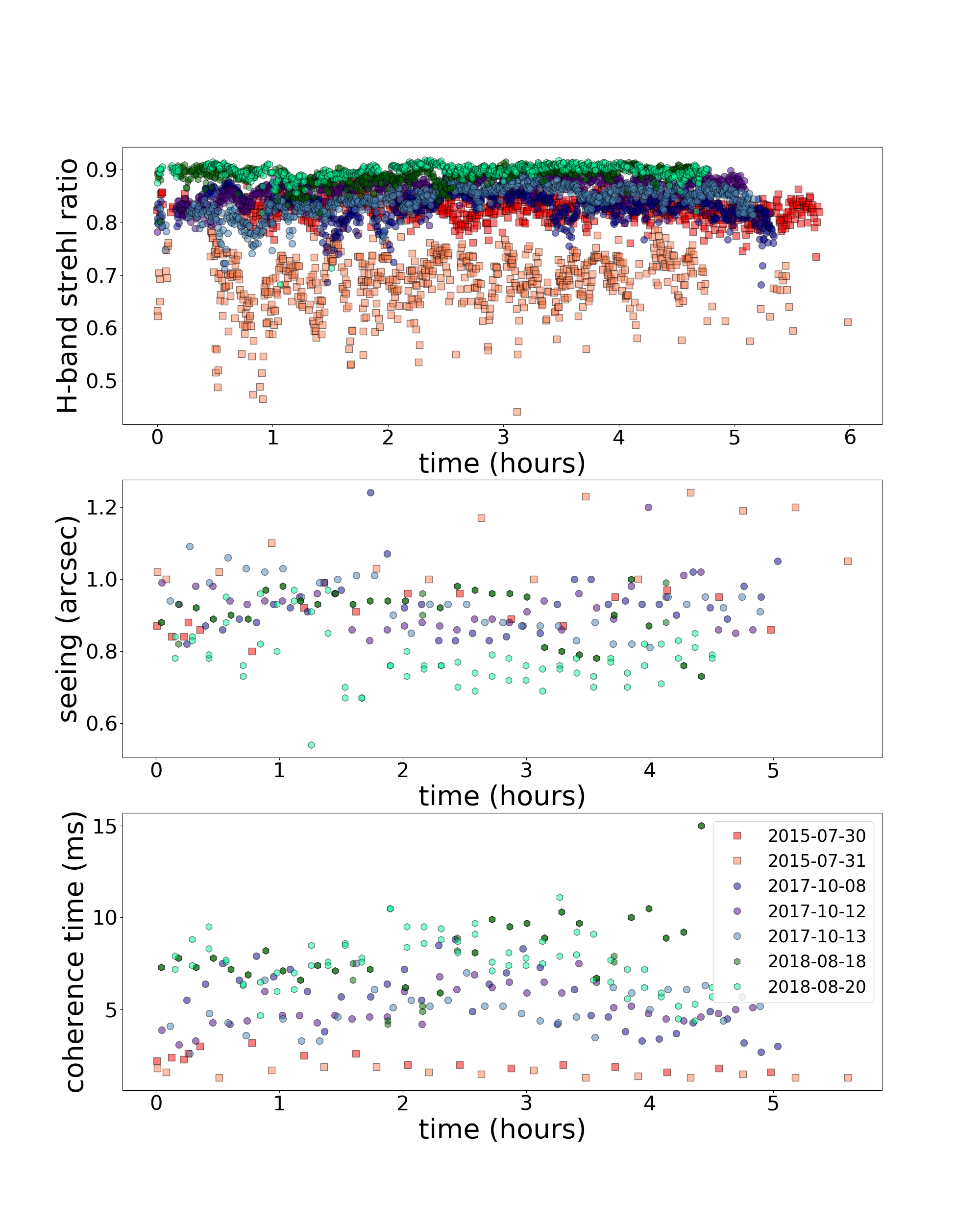}
\caption{$H$-band Strehl ratio, seeing, and coherence time as a function of time during each of our
7 variability monitoring observations.  Ranking observations in order of observing conditions: 
the 2015 observations were taken in the worst conditions among those probed by our 
variability search and are excluded from our variability analysis.  The October 2017 observations were taken in very good conditions and the August 2018 observations were taken in excellent conditions.
}
\label{fig:SR_seeing_tau0}       
\end{figure*}

\begin{figure*}
\includegraphics[scale=0.3]{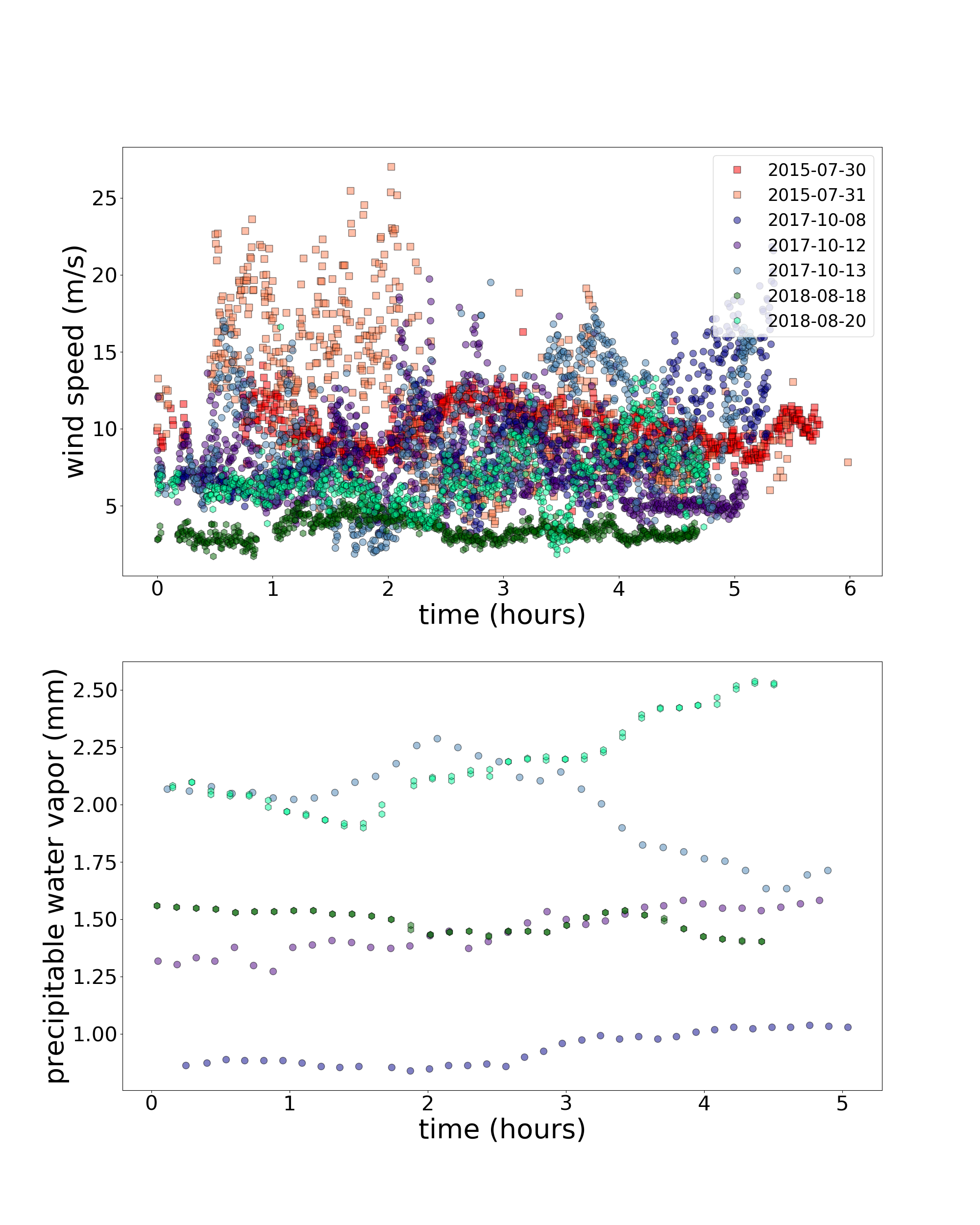}
\caption{Windspeed and precipitable water vapor as a function of time during each of our 
7 variability monitoring observations.  Precipitable water vapor measurements were not available
for the 2015 epochs, and thus are shown here only for the October 2017 and August 2018 observations.
The windspeed measurement presented here is estimated from the SPARTA AO real-time computer.  In the 2015-07-31 epoch, this windspeed value is likely an overestimate of the actual windspeed, as telescopes on Paranal cease operations at windspeeds above 18 m/s.
}
\label{fig:windspeed_iwv}       
\end{figure*}

Conditions varied considerably between epochs.  Obtained \textit{H}-band Strehl ratio, seeing, and coherence time $\tau_0$ over each observation are plotted in Fig.~\ref{fig:SR_seeing_tau0} and wind speed and precipitable water vapor are plotted in Fig.~\ref{fig:windspeed_iwv}.  The \textit{H}-band Strehl ratio and wind speeds are estimates from the SPARTA AO real-time computer \citep{Beuzit2019}; seeing (from the observatory Differential Image Motion Monitor measurements), coherence time $\tau_0$ and precipitable water vapor are drawn from the raw data file headers. Most epochs had quite good conditions and AO performance, with seeing $<$1" and Strehl ratios between 0.8 and 0.9.  The \textit{K} band epoch taken on 31 July 2015, however, was obtained in considerably worse conditions (seeing$>$1") and with notably lower Strehl ratios from 0.5 to 0.7.  The \textit{J} band epoch taken on 2015-07-30 was obtained in reasonably good conditions, but, unlike other epochs, HR 8799b is not apparent in single frame images as the planet is fainter in $J$ than in $H$ or $K$ band.  Thus, the 2015 epochs either do not have sufficient photometric stability or sufficiently high signal-to-noise detection per planet to enable variability monitoring and are excluded from further variability analysis. 

When the wind speed at 10 m or 30 m drops below 3 m/s, SPHERE observations can be affected by the "Low Wind Effect" (LWE) \citep{Sauvage2016, Milli2018}, causing a highly distorted PSF shape.  For most of our observations, measured wind speed remained above 3 m/s, dropping below this speed for most of the observation on 2018-08-18 and a portion of the observation on 2017-10-13.  We visually inspected PSF images for these epochs obtained via the SPARTA AO data handling system to determine if these observations were affected by LWE and found that this effect was mild or non-existent in our observations.  

\begin{figure*}
\includegraphics[scale=0.3]{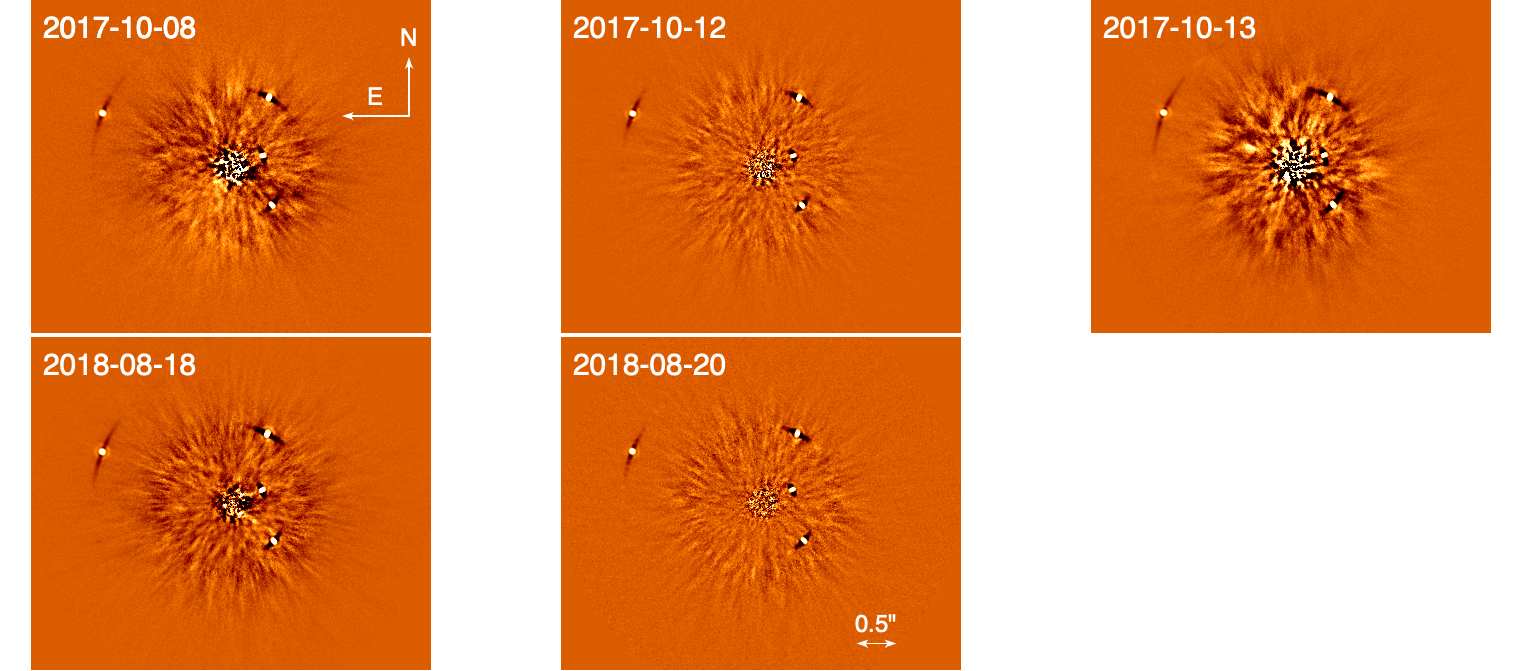}
\caption{Full sequence reduced images for data taken in October 2017 and August 2018.  Data were processed using the VIP PCA pipeline \citep{Gomez2017}, removing 25 principal component modes.  All four planets are easily detected; the image scale runs from -10 to 10 ADU and has been adjusted to highlight the residual speckle noise remaining after subtraction of the principal modes.
}
\label{fig:fullsequence}       
\end{figure*}

Data were reduced and aligned by the SPHERE Data Center using the standard SPHERE pipeline.  For details on the SPHERE Data Center and the pipeline, please see \citet{Delo17} and \citet{Pavlov2008}.  We adopted standard values for pixel scale of 12.25 mas / pixel and true north orientation of -1.75$^{\circ}$ in these reductions \citep{Maire2016SPIE}.  
Fully reduced images for the full observing sequences for the October 2017 and August 2018 data are presented in Fig.~\ref{fig:fullsequence}; these were processed using the Vortex Image Processing Principle Component Analysis (VIP-PCA) pipeline \citep{Gomez2017}.  This PCA analysis utilizes the library of frames composing the full observing sequence to build a series of principal component images (henceforth PCA modes), with the majority of the speckle pattern well-modeled with the first few components.  Subtracting out an image built from these PCA modes will then remove the speckle pattern, with an increasing number of modes further improving the modeling of the speckle pattern, at the expense of subtracting out some planet light as well.  For the full observing sequence, we found that removing 25 principal components provided the best balance between fully suppressing the speckle pattern without causing significant planet self-subtraction.

\section{Satellite Spot Time-Series Photometry}

Our observations were acquired with satellite spots present in all frames.  The standard SPHERE pipeline uses images taken with satellite spots before and after the main observing sequence (in most cases, without satellite spots) as astrometric references to measure the stellar position behind the coronagraph and finely centre images, while photometric calibration is performed relative to unsaturated images of the star taken before and after the main observing sequence.  Since variations in seeing, Strehl, and airmass over an observation will cause brightness fluctuations in the stellar PSF, the planets, and the speckle pattern, establishing a photometric reference free of variability from the planets is vital for detrending and interpreting our exoplanet observations.  As the star is behind the coronagraph during our observations, it is unfeasible to obtain unsaturated images of the star to use as a photometric reference.  Hence, here we establish whether or not the satellite spots serve as appropriate photometric references and a proxy for measurements of unsaturated stellar images. 

\begin{figure*}
\includegraphics[scale=0.4]{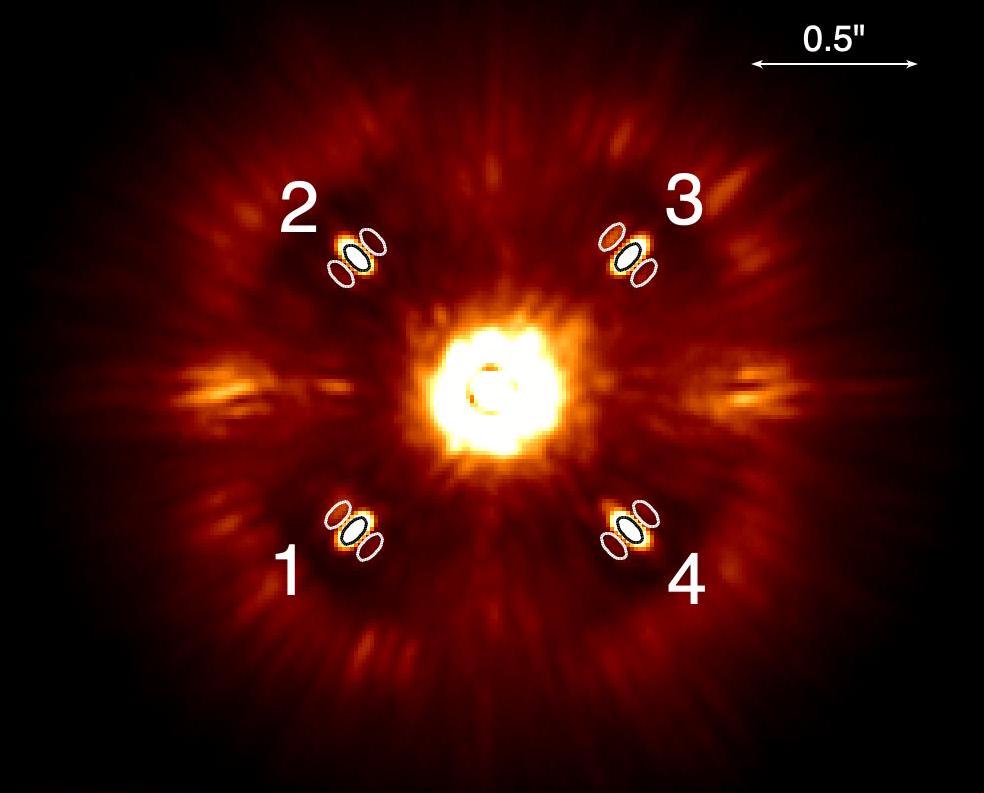}
\caption{Satellite spots aperture photometry extraction regions and naming convention, overplotted on a single frame of data from 2018-08-18.
}
\label{fig:SS_regions}       
\end{figure*}

\begin{figure*}
\includegraphics[scale=0.4]{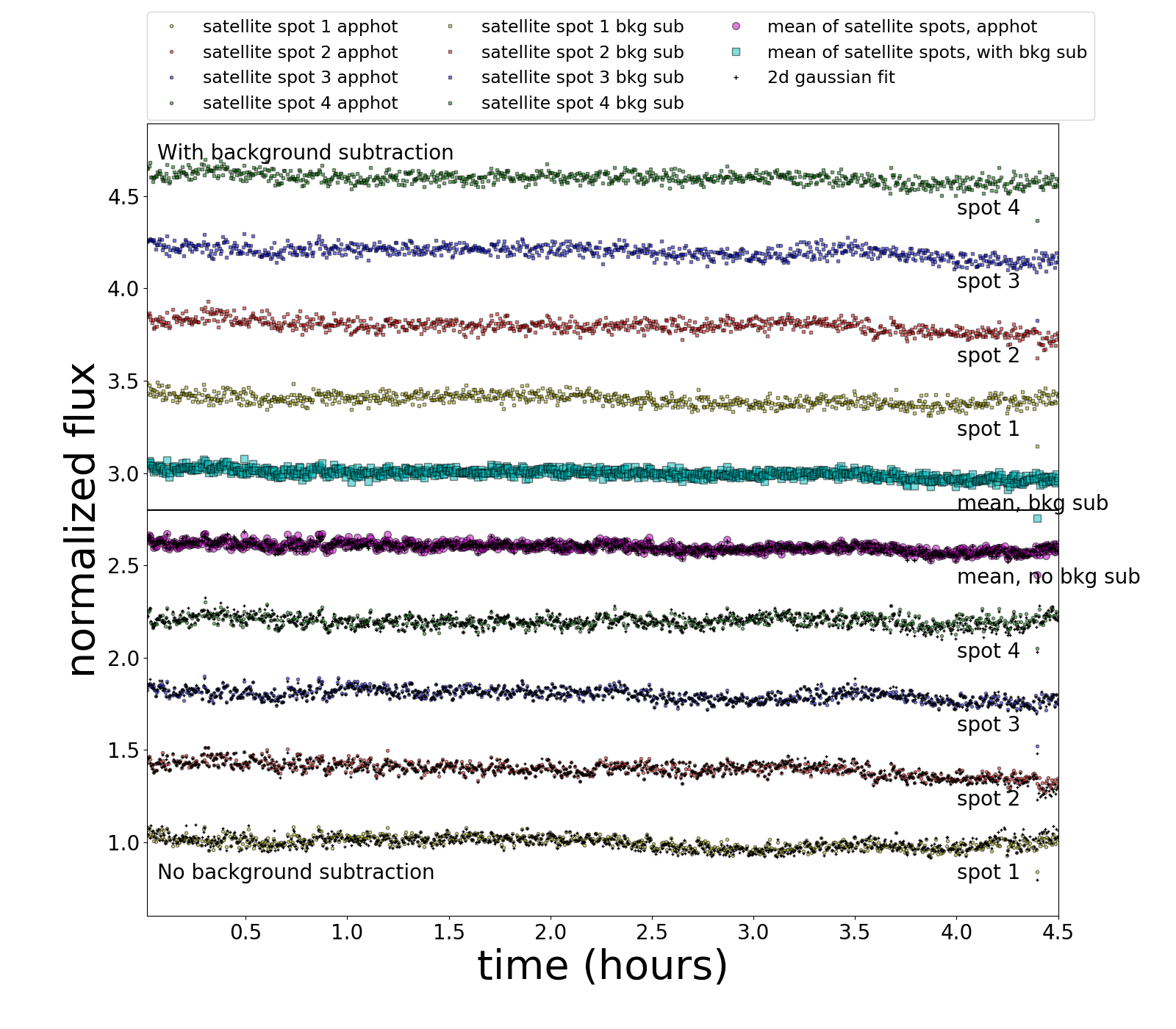}
\caption{Satellite spot photometry for 2018-18-18 (normalized
by the median value across the whole observation for each
satellite spot) as a function of time for each satellite spot, with
an offset in flux included between each satellite spot lightcurve for clarity.
Results with background subtraction are shown on the top half of the plot, while results without 
background subtraction are plotted on the bottom half of the plot.
The amplitudes from a 2-d gaussian fit to each satellite spot are plotted as 
black crosses, elliptical aperture photometry is plotted
as filled circles, and elliptical aperture photometry with an 
additional background subtraction is plotted as filled squares.  
The mean of the elliptical aperture photometry for 
all 4 satellite spots are plotted in the middle of the 
figure, with magenta circles for the elliptical aperture photometry 
without background subtraction, and cyan square for elliptical 
aperture photometry with background subtraction.}
\label{fig:SS_Aug18}       
\end{figure*}

While satellite spots appear relatively circular when using a narrow-band filter, with the broadband $H$ filter they appear as roughly elliptical spots.  Thus, to determine satellite spot position in each frame in our time series data, we fit a 2-dimensional elliptical Gaussian to each of the four satellite spots using the astropy photutils modeling package  \citep{Bradley2019}.  We then extracted aperture photometry using the astropy photutils aperphot package in an elliptical aperture with semi-major and semi-minor axes of 4 and 2 pixels, rotated to the correct angle to recover the flux from each satellite spot, and centered on the satellite spot positions from the 2-d elliptical Gaussian fits.  We also extracted aperture photometry from "background" elliptical apertures placed 6 pixels away on either side of each satellite spot.  The background as a function of time was then calculated as the average of these two background elliptical apertures.  The geometry of the extraction apertures and the numbering convention for satellite spots is shown in Fig.~\ref{fig:SS_regions}.  In Fig.~\ref{fig:SS_Aug18}, we plot photometry (normalized by the median value across the whole observation for each satellite spot) as a function of time for each satellite spot for the epoch of 2018-08-18 (taken in excellent conditions), for the various cases of: 1) the amplitude of the 2d-gaussian fit (shown as crosses), 2) elliptical aperture photometry (shown as filled circles), and 3) elliptical aperture photometry, with the background subtracted from the background elliptical apertures (shown as filled squares).  We plot as well the median of the normalized satellite spot curves for each case. Photometry for the other epochs of observations can be found in Appendix~\ref{app:ssphot}.  In all cases, we find that the 2-d Gaussian fit and the elliptical aperture photometry yield similar results, as expected.  We found amplitudes for the 2-d Gaussian fits of 200-300 counts for the 2017-10-08 epoch and 2000-3000 counts for the other epochs.  Likely owing to better observing conditions, the satellite spot photometry for the August 2018 epochs is notably more stable and less variable than that for the October 2017 epochs.   

In most cases, subtracting the background photometry did not significantly change the shape of the satellite spot light curves. Notably, however, for the 2017-10-08 epoch, where the satellite spots were under-illuminated, the satellite spot photometry seemed to follow that of the background regions and was significantly flattened and "detrended" via the background subtraction.  Also, for satellite spot 4 in the 2017-10-13 epoch, background subtraction appears to have corrected a linear trend seen in this satellite spot but not in the other 3 satellite spots during this epoch.  We extracted photometry from a set of 5 circular "background regions" for comparison as well, with one region centered on the core of the star PSF, and the other four spaced evenly around the image.  Backgrounds extracted from each of these regions were similar as a function of time and resembled the background photometry obtained in the elliptical background regions selected close to the satellite spots. In Appendix~\ref{app:ssphotconditions}, we overplot the median flux in the satellite spots, measured background flux, obtained Strehl ratio and seeing, to search for correlations in the satellite spot lightcurves as a function of ambient conditions.  Other than the 2017-10-08 epoch (when, as noted previously, the satellite spots were under-illuminated), the satellite spot light curves do not appear to be strongly correlated to the Strehl ratio or anti-correlated with the background level. 

\subsection{Correlations between satellite spots}
\label{sec:correlations}

\begin{figure*}
\includegraphics[scale=0.24]{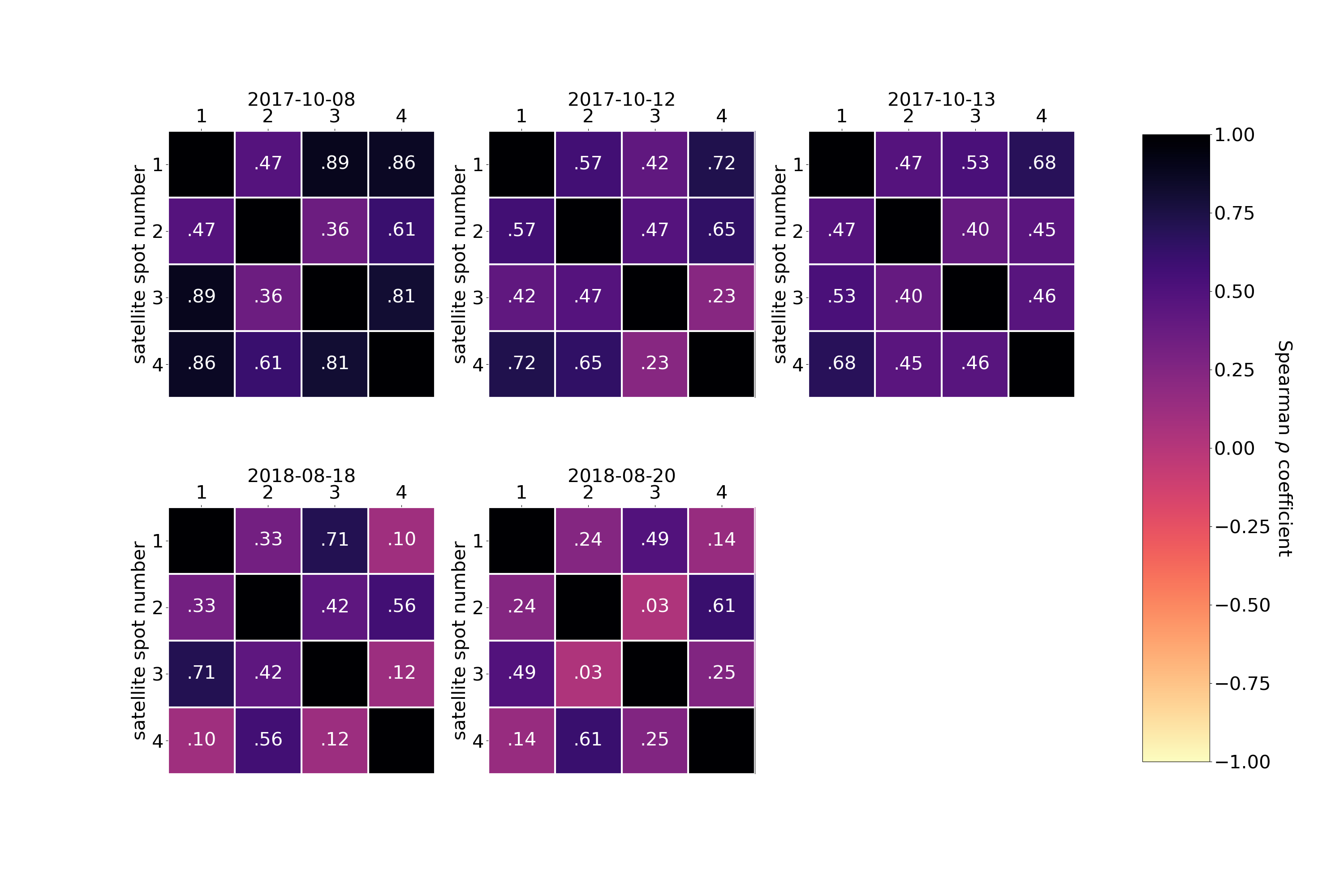}
\caption{Spearman $\rho$ coefficients between satellite spots, with no division by the mean lightcurve.  A $\rho$ value of 1 implies perfect correlation; a $\rho$ value of -1 implies perfect anti-correlation.  Conditions were particularly excellent for our August 2018 datasets; our satellite spot photometry thus covered a limited range of values, meaning that correlation between satellite spots is not a particularly good proxy for the quality of the observation.
}
\label{fig:spearmanr_nomean}       
\end{figure*}

Following \citet{Apa16}, we calculate the degree of correlation between individual satellite spots using the Spearman's correlation test.  The Spearman's correlation test ranks two sets of data in order of ascending data values, and then assesses how correlated the ranked sets of data are -- this is the Spearman $\rho$ value.  If the datasets are perfectly correlated (a higher data value in one dataset always corresponds to a higher value in the other), this will produce a $\rho$ value of 1.  Perfectly anti-correlated datasets will produce a $\rho$ value of -1.
The Spearman's correlation test requires that the relationship between two variables follows a monotonic function, unlike a Pearson correlation test which specifically requires a linear relationship.  We used spearmanr from from scipy.stats in python to calculate the Spearman $\rho$ values between different normalized satellite spot lightcurves.
The Spearman $\rho$ values between different satellite spots are shown in Fig.~\ref{fig:spearmanr_nomean}.  

We find that the satellite spots are mildly correlated in the October 2017 data, but at a low level and with a wide spread, with correlations ranging from 0.23 to 0.90, but generally around 0.5.  This is significantly lower than the typical correlations of 0.7-0.9 found by \citet{Apa16}.  
In our August 2018 data, we find much smaller correlations between satellite spots, even though these data were taken in considerably better conditions, with correlations ranging from 0.03 to 0.71, with values of around 0.3-0.4 for most of the pairs of satellite spots.  
These low Spearman $\rho$ values likely stem from the very small spread in satellite spot photometric values for these data, due to the excellent conditions, causing the Spearman's correlation test to be an imperfect proxy for observation quality in this case.  This can be seen more clearly if we examine the correlation plots between satellite spots directly -- an example of which is given in Fig.~\ref{fig:corr_sides_nomed2}, and the rest of which appear in Appendix~\ref{app:sscor}.  Due to larger changes in observing conditions and poorer Strehl values overall, values for satellite spot photometry varied more over the October 2017 observations compared to the August 2018, leading to higher values of the Spearman $\rho$ value (in other words, the greater range of values sampled allows a seemingly larger correlation between values).

\begin{figure*}
\includegraphics[scale=0.25]{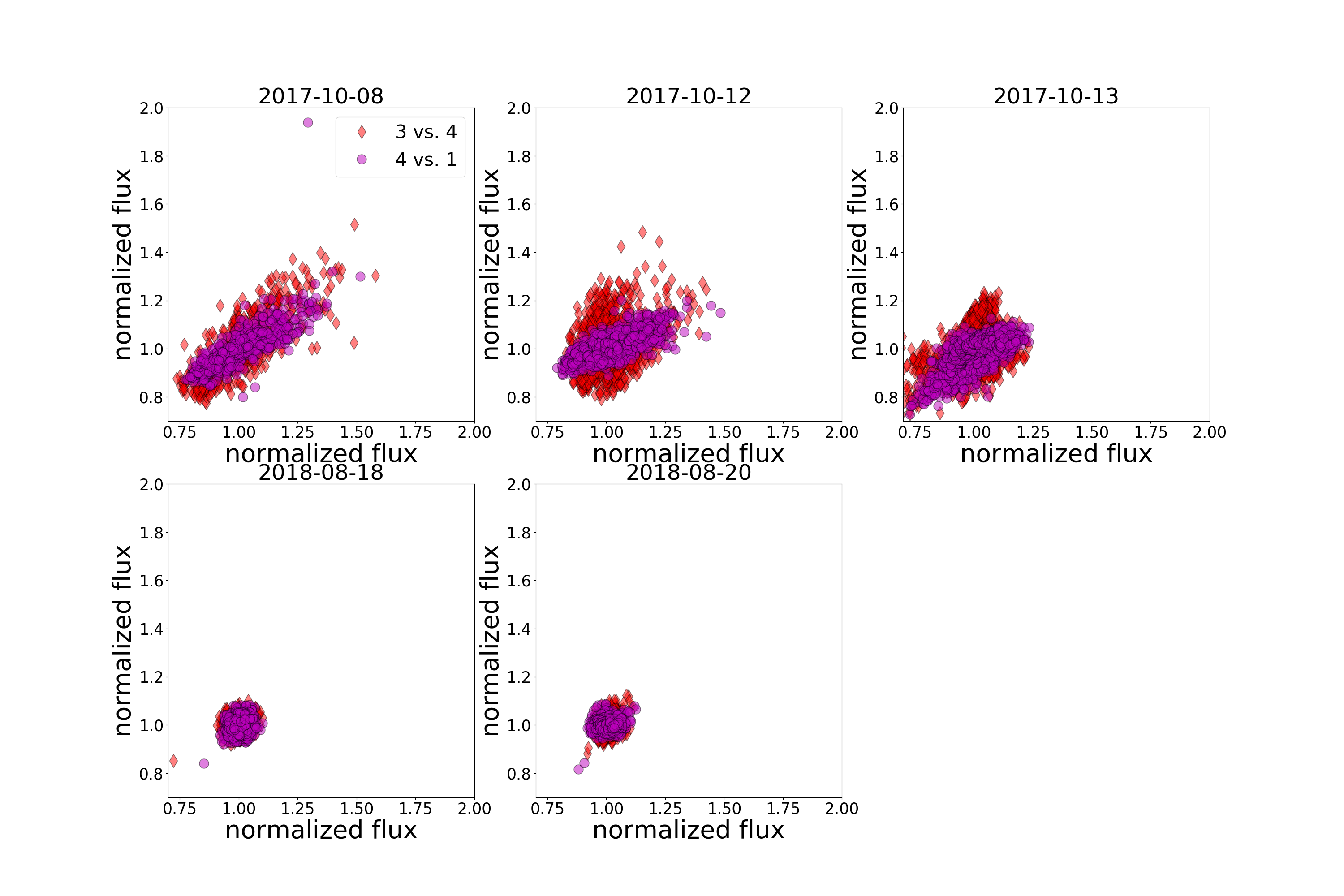}
\caption{Correlation plots for satellite spots 3 vs. 4 and 4 vs. 1.  The satellite spot photometry covers a much more limited range of values on nights with excellent conditions (August 2018) compared to nights with more moderate conditions (October 2017), leading to lower Spearman $\rho$ values for August 2018 vs. August 2017, despite the higher quality of the August 2018 data.
}
\label{fig:corr_sides_nomed2}       
\end{figure*}

Following \citet{Apa16}, we also calculate Spearman $\rho$ values after dividing through by the mean satellite spot lightcurve.  This removes shared variations between satellite spots, thus highlighting the difference in behavior between the individual satellite spots. The Spearman $\rho$ values between different satellite spots after division by the mean are shown in Fig.~\ref{fig:spearmanr_mean}, with the corresponding correlation plots shown in Appendix~\ref{app:sscor}.  \citet{Apa16} found a fairly constant pattern of no or weak anti-correlation between most of the satellite spots, with relatively strong anti-correlations between one set of directly opposite spots, and two sets of adjacent spots.  We find more night-by-night diversity in our results, which are discussed in detail in Appendix~\ref{app:sscor} -- however, given the small range of flux values covered in our observations, we do not consider these Spearman $\rho$ values to be robust. 

\begin{figure*}
\hspace{-1cm}
\includegraphics[scale=0.24]{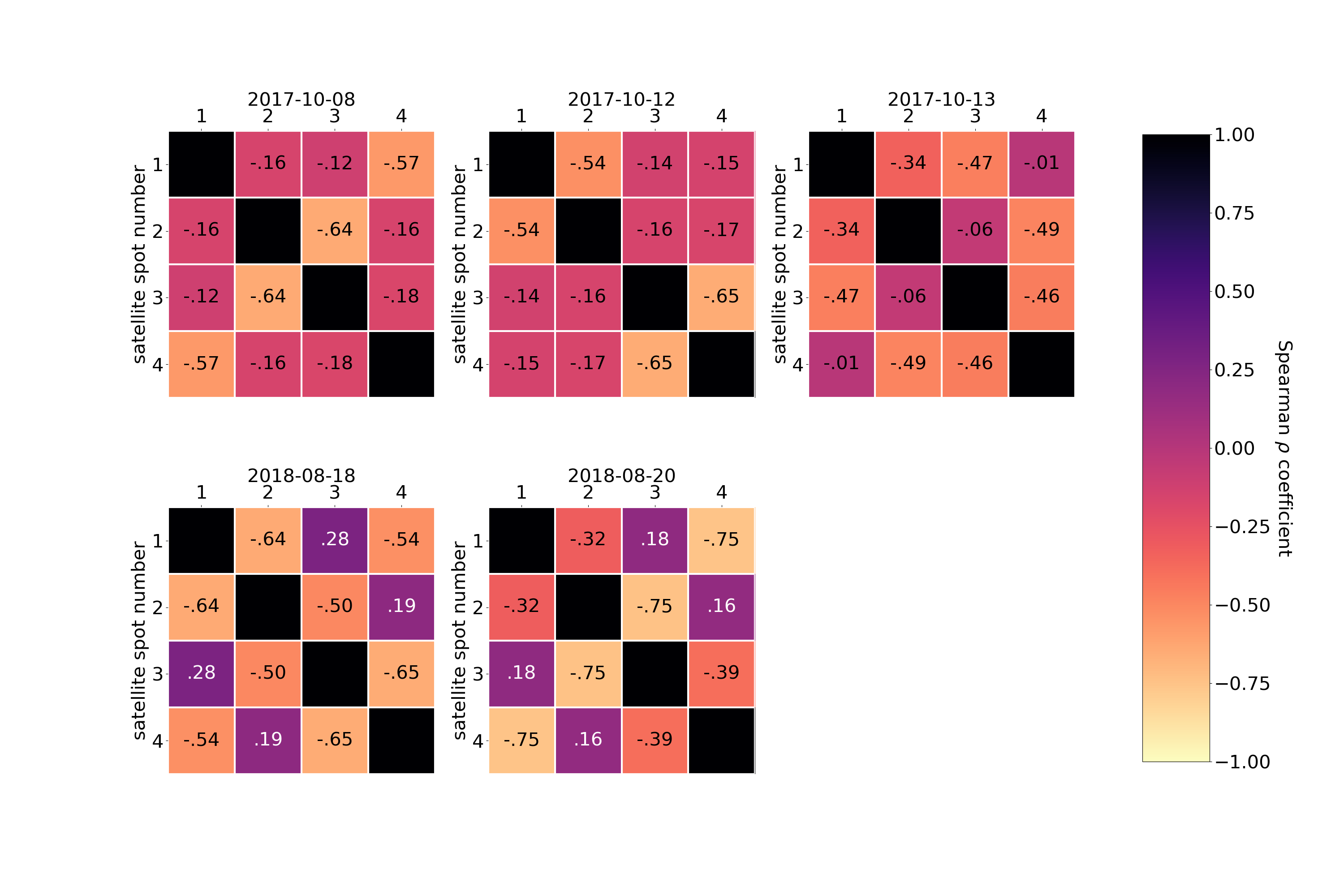}
\caption{Spearman $\rho$ coefficients between satellite spots, after division by the mean lightcurve (thus removing common variations).  A $\rho$ value of 1 implies perfect correlation; a $\rho$ value of -1 implies perfect anti-correlation.   
}
\label{fig:spearmanr_mean}       
\end{figure*}

\section{Exoplanet Time Series Photometry}

\subsection{Achievable time resolution \label{sec:timeresolution}}

To minimize self-subtraction and ensure sufficient speckle suppression, coverage of a sufficient range in parallactic angle is necessary within each temporal bin within the full observation to move each planet off of itself over the course of that segment of the observation.  We fit a 2-dimensional Gaussian to the PSF images taken before and after each temporal sequence.  On all 5 nights, the FWHM of the PSF was measured to be roughly 4 pixels.  Thus, ideally, we would cover at least 20 pixels of on-sky rotation ($\geq$5 PSF FWHMs) in each temporal bin to minimize planet self-subtraction.  This is a quite conservative choice; the parallactic angle range we adopt for each temporal bin is essentially that of the shortest useful angular differential imaging \citep[ADI~see~e.g.][]{Mar06, Laf07} observation for that planet-star separation.  We found that speckle evolution in our datasets was sufficiently rapid that frames taken too far before or after a given frame were not useful to include in building the PSF using PCA, thus the frames selected for each temporal bin were run through the PCA analysis separately.  The parallactic angle range we adopt for each temporal bin  ensures that for each frame in the temporal bin, a sufficient number of frames taken far enough away in parallactic angle are available to build an accurate PSF model without significant self-subtraction.  Each of our observations covered $>$70$^{\circ}$ of sky rotation for HR 8799, over 4-5 hours; the parallactic angle coverage necessary for 20 pixels of on-sky rotation will vary according to the planet-star separation.

For HR 8799b, we adopt 10$^{\circ}$ of rotation on sky per temporal bin as this minimum coverage; at a planet-star separation of $\sim$1.7", 10$^{\circ}$ of sky rotation corresponds to a shift of 24 pixels on sky, sufficient to shift the planet $>$5 PSF-widths away from its initial position.  Thus, we binned the observing sequences into 6 equal parallactic angle bins, ensuring a minimum variation in parallactic angle of 10$^{\circ}$ in each bin.  For HR 8799c at a planet-star separation of $\sim$1.0", 10$^{\circ}$ of sky rotation corresponds to a shift of 14 pixels on sky ($\sim$3.5 PSF-widths), which may not be sufficient to completely prevent self-subtraction.  Thus, for HR 8799c, we adopt 14$^{\circ}$ of rotation per sky per temporal bin, corresponding to 5 equal parallactic angle bins and a shift of $\sim$20 pixels ($\sim$5 PSF-widths) on sky at $\sim$1.0".  

For the inner two planets, HR 8799d and HR8799e, at separations of $\sim$0.6" and $\sim$0.4", 10$^{\circ}$ of sky rotation corresponds to shifts of $\sim$9 pixels and $\sim$6 pixels on sky, insufficient to avoid significant self-subtraction and preventing an appropriate cadence for variability monitoring.
In Section~\ref{sec:innerplanets}, we consider whether useful variability measurements can be obtained for the inner two planets if we adopt 20$^{\circ}$ of rotation on sky per temporal bin as the minimum coverage, and hence only consider a total of 3 equal parallactic angle bins per observation. 

\subsection{HR 8799b direct aperture photometry time-series\label{directphot}}

At $\sim$1.7" from the star, as the widest-separated of the 4 planets, HR 8799b lies outside of the AO-corrected region, where uncorrected atmospheric halo dominates over the static instrumental speckles.  Unlike the closer-in planets, which are swimming in a sea of speckles and require a full suite of post-processing techniques for accurate photometry, deriving photometry for HR 8799b is amenable to simpler approaches.  Therefore, we extracted aperture photometry directly for HR 8799b. The observing sequences were binned in 6 equal parallactic angle bins, and then we simply derotated and stacked data within each parallactic angle bin.  We first fit a 2-d gaussian to the planet position in each binned data slice using photutils.modeling to determine the best planet center position and then extracted photometry in a 3-pixel radius circular aperture centered on the 2-d gaussian fit position.  Since we have not removed the stellar PSF, the background varies considerably as a function of radius at the position of the planet, hence we chose to use a quite small aperture here -- the 3-pixel aperture is sufficiently small to minimize variation in the background at the planet position.  To correct for this radially-dependent background, we subtracted both the median of an annulus (10 pixel width) at the planet radial separation to estimate the background at this position and also the median in a circular annulus aperture from 3-5 pixels around the 2-d Gaussian best fit position to remove any remaining trend in the background. Aperture photometry for HR 8799b as well as negative simulated planet photometry for HR 8799bc are presented in Figure~\ref{fig:bclightcurve_allnights}.

\subsection{Negative simulated planet subtraction}

We must robustly remove speckle noise while preserving a photometric reference in order to determine well-calibrated time-series photometry for the HR 8799 planets.  To do so, we adopt the standard technique of negative simulated planet subtraction.  We first divide the observation into a number of temporal bins, which vary according to the planet considered (see Section~\ref{sec:timeresolution}). For each temporal bin, we then inserted negative simulated planets at the approximate positions of each of the four planets in a grid of radius, planet position angle (henceforth $\theta$), and $\Delta$(magnitude) into each frame.  Three unsaturated point-spread function (PSF) images were taken before and after the time-series data.  Planets were simulated by median combining these unsaturated PSF images and scaling them according to the $\Delta(mag)$ of the planet.      

We performed the insertion in two steps:

{\it 1) Coarse Grid} -- in order to precisely determine the position of each planet, we ran an initial grid in radius and $\theta$, with planet $\Delta$(mag) kept fixed and set to literature values.  After inserting the negative simulated planets, the VIP PCA pipeline \citep{Gomez2017} was used to remove 10 principal modes, de-rotate, and stack data from each bin.  From the de-rotated and stacked data, we then calculated the $\chi^{2}$ residual for each planet at each grid point.  Over the full grid, we then find the minimum $\chi^{2}$ and hence the best solution for planet position.  The $\chi^{2}$ value for each grid point was calculated as the square of the sum of the counts in a circular region with radius of 5 pixels centered on the planet divided by the standard deviation in a circular noise annulus around the planet (5-10 pixels).  For the negative simulated planet subtraction, the inhomogeneous background due to the PSF of the star is removed via the PCA analysis, so we chose a larger aperture size in comparison to the direct aperture photometry of HR 8799b described in Section~\ref{directphot}, in order to obtain better statistics for evaluating the quality of our PSF subtraction.  In radius, we ran the wide grid over a radial offset of -4 to +4 pixels from the initial radius guess, in steps of 0.2 pixels.  In angle, we ran the wide grid over an angular offset of -2 to +2 degrees from the initial angle guess, in steps of 0.1 degrees. 

{\it 2) Fine Grid} -- after estimating planet position with the coarse grid, we ran a finer grid in position as well as $\Delta$(mag), centered on the best position found during the coarse grid step. From the fine-grid $\chi^{2}$ grid, we then converted to likelihood (e.g. $exp(-\chi^{2})$) and renormalized the posterior probability distribution function (henceforth PDF) to 1, assuming that we had estimated planet position and $\Delta(mag)$ with sufficient accuracy to ensure that each planet was within our fine grid. We then marginalized over radius and $\theta$, and fit a 1-d Gaussian to the 1-d posterior.  The peak and width of the best-fit Gaussian was adopted as the best $\Delta(mag)$ and error on $\Delta(mag)$ respectively; in fact, this ability to robustly estimate errors is why we chose to calculate the posterior across the full fine grid, instead of using e.g. an amoeba downhill optimization procedure to select the highest probability set of parameters.  In radius, we ran the fine grid over a radial offset from the best radius found by the wide grid, running from -1 to +1 pixel, in steps of 0.1 pixels.  We ran the fine grid over an angular offset running from -1 to +1 degrees from the best angular position found by the wide grid, in steps of 0.1 degrees.  In $\Delta(mag)$, we ran over a grid of offsets from -0.5 to 0.5 mag, in steps of 0.05 mag for the October 2017 epochs and from -0.333 to 0.333, in steps of 0.033 mag for the August 2018 epochs.       

We selected 10 PCA modes as the ideal choice to best subtract speckles while preserving planet flux -- we motivate this choice further in Sections~\ref{sec:nmodes} and~\ref{sec:simulations}. For HR 8799b and HR8799c, we then calculated best $\Delta(mag)$ values for 6 and 5 temporal bins respectively, evenly spaced in parallactic angle across the observation.  For the night of 18 August 2018, we show a zoomed-in view of HR 8799b or c respectively in Figures~\ref{fig:b6slicesubtraction} and~\ref{fig:c5slicesubtraction}, before and after removing the best-fit negative simulated planet.  The image scale in these figures runs from -10 to 10 ADU; while at the position of HR 8799b, speckle noise has been fully removed and noise statistics can be assumed to be Gaussian, a considerable amount of speckle noise remains at the separation of HR 8799c even after subtraction of 10 PCA modes.  

\begin{figure*}
\includegraphics[scale=0.4]{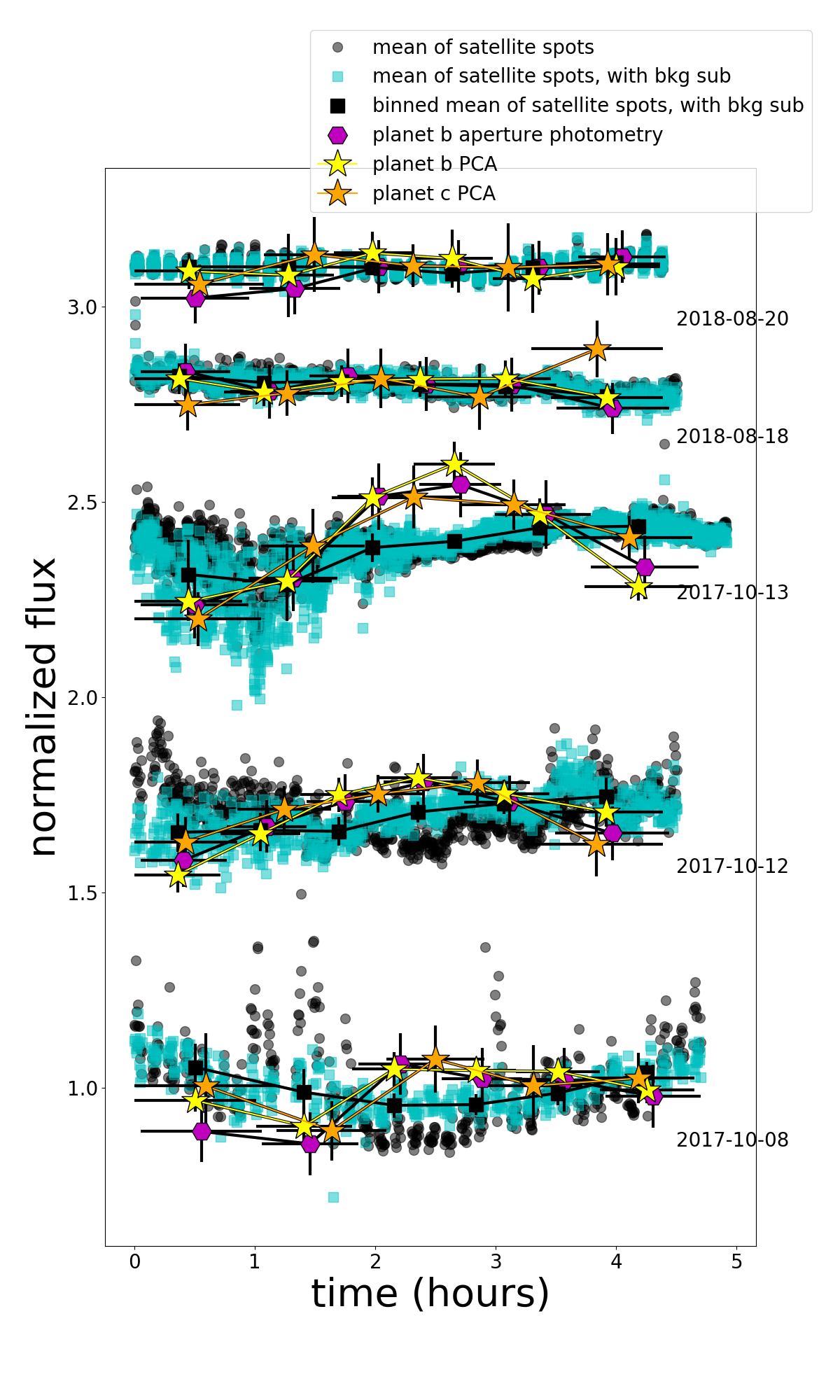}
\caption{Satellite spot aperture photometry (with and without background subtraction), planet b aperture photometry, and planet b and c psf-fitting photometry as a function of time during the observation for October 2017 and August 2018 broadband-$H$ datasets.  Each lightcurve has been normalized using its mean flux value. Planet b aperture and psf-fitting photometry have been slightly offset in time to improve error bar legibility.  Planets b and c almost always display the same trends as a function of time; in the August 2018 data these trends are also well-correlated with the trends seen in the satellite spot photometry.
}
\label{fig:bclightcurve_allnights}       
\end{figure*}

\begin{figure*}
\includegraphics[scale=0.2]{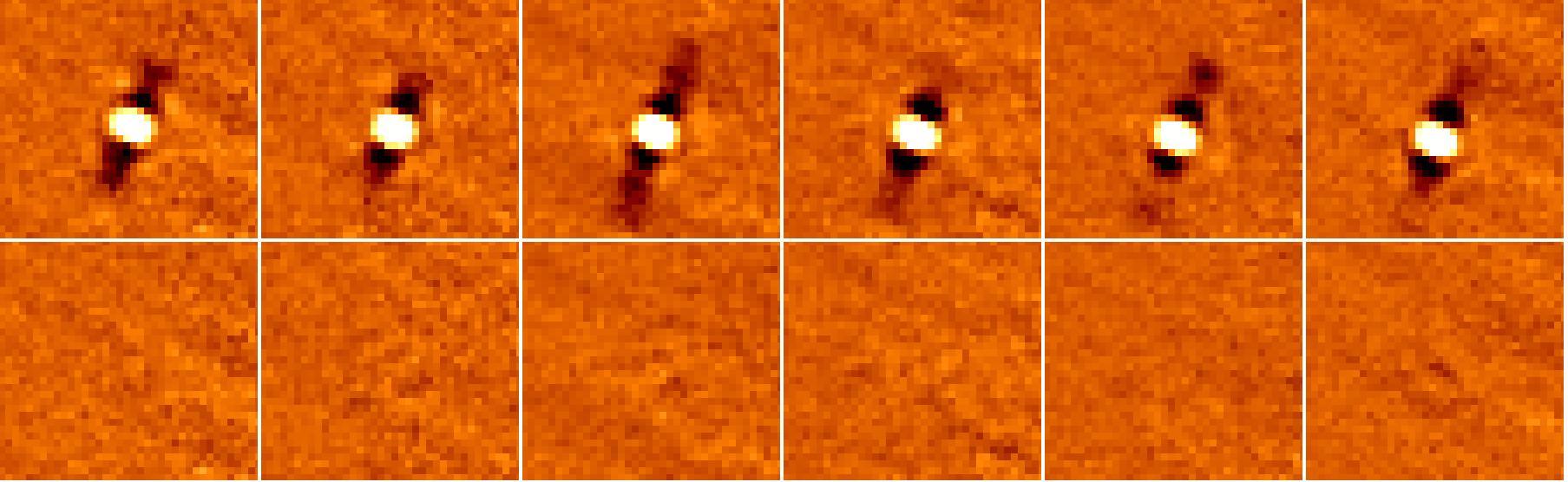}
\caption{Best negative simulated planet subtractions for HR 8799b on 18 August 2018.  The full $\sim$4.5 hour observation has been divided into 6 equal parallactic angle bins, shown in temporal order from left to right.  The image stretch runs from -10 to 10 ADU in a linear scale.
}
\label{fig:b6slicesubtraction}       
\end{figure*}

\begin{figure*}
\includegraphics[scale=0.2]{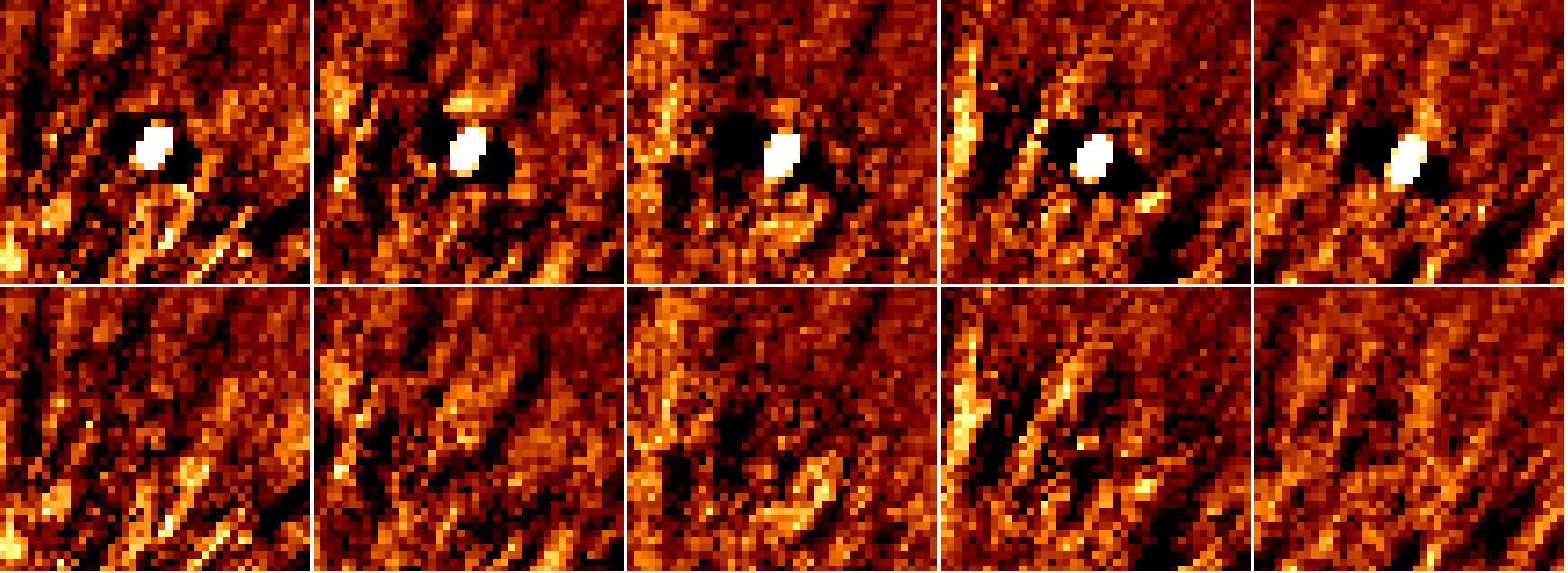}
\caption{Best negative simulated planet subtractions for HR 8799c on 18 August 2018.  The full $\sim$4.5 hour observation has been divided into 5 equal parallactic angle bins, shown in temporal order from left to right.  The image stretch runs from -10 to 10 ADU in a linear scale, with contrast adjusted to highlight the residual speckle noise after subtraction of 10 PCA modes.
}
\label{fig:c5slicesubtraction}       
\end{figure*}

Negative simulated planet photometry as a function of time (and converted to flux units) for HR 8799bc are presented in Figure~\ref{fig:bclightcurve_allnights}.  For each photometric measurement, to convert errors calculated in $\Delta(mag)$ to flux units, we drew 100000 samples from a Gaussian centered on the best $\Delta(mag)$ value and with $\sigma$ given by the error on $\Delta(mag)$, then converted to flux units and adopted the standard deviation of the 100000 samples as our error in flux.

From Figure~\ref{fig:bclightcurve_allnights}, it is clear that some nights yielded considerably better data than others -- specifically, compared to the October 2017 epochs, the August 2018 epochs have much less spread in satellite spot measurements and correspondingly smaller spread in planet photometry measurements.  Unsurprisingly, the two August 2018 epochs had the longest coherence times and highest Strehl ratios across all epochs (see Fig.~\ref{fig:SR_seeing_tau0}). For the August 2018 epochs, the planet and satellite spots measurements generally agree.  This implies that in these epochs, all or at least the bulk of the variation in planet flux we measure here is simply systematic trends as seeing, Strehl, etc., vary across the observation.  In October 2017, however, we find significant divergence between planet and satellite spot lightcurves.  Measurements for HR 8799b and HR 8799c also appear to be in agreement, except for epoch 2018-08-18, where the first and last measurements for HR 8799c are divergent from the HR 8799b and satellite spot lightcurves. However, we do not believe this is a bonafide detection of variability, as this result appears to be PCA-mode dependent and these two time bins were taken right at the beginning and end of the observation, where the airmass was at its highest and sky rotation at its lowest.  We discuss this further in Sections~\ref{sec:nmodes} and~\ref{sec:simulations}.

\subsection{Effect of number of PCA modes removed \label{sec:nmodes}}

\begin{figure*}
\includegraphics[scale=0.35]{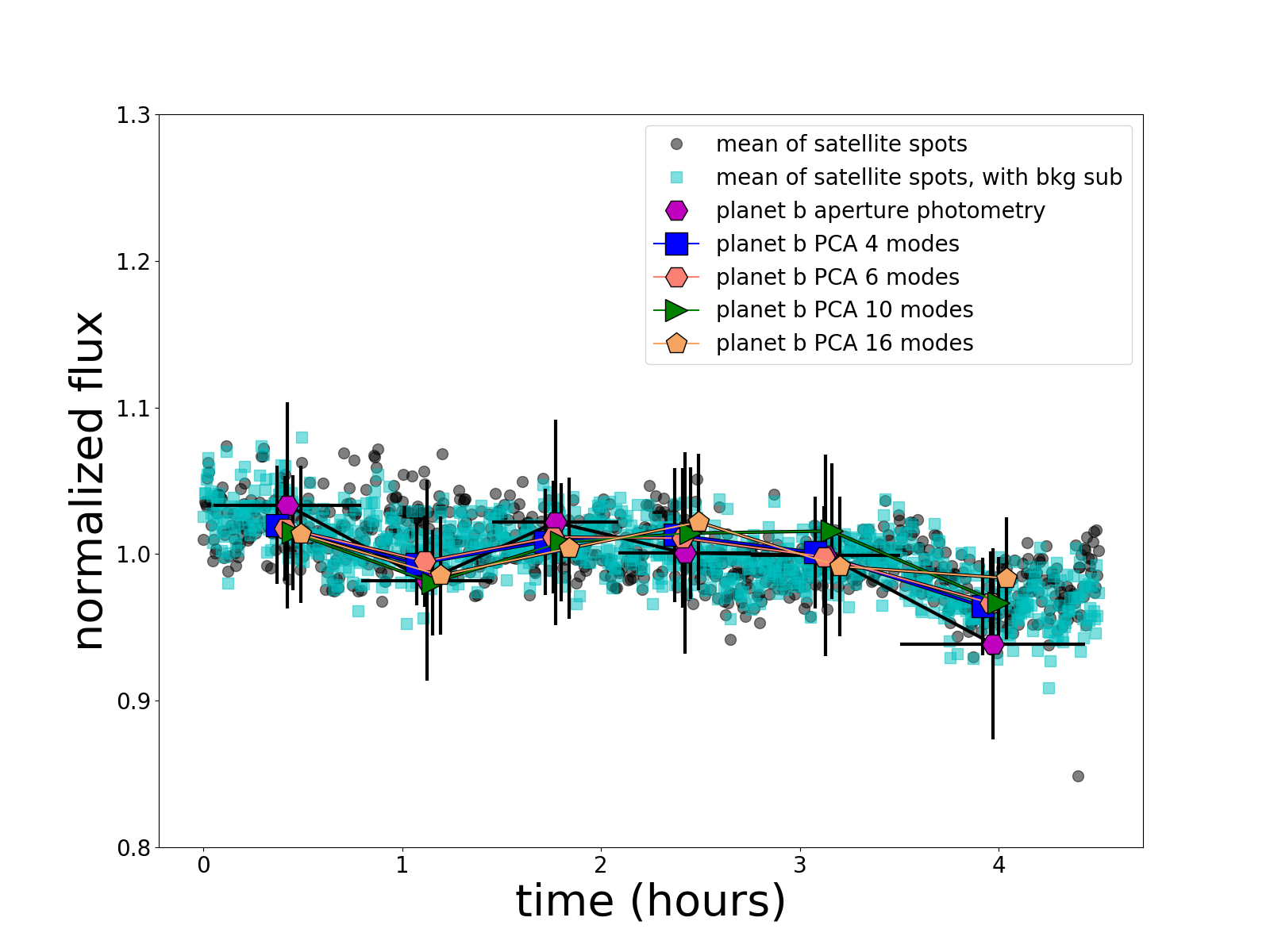}
\includegraphics[scale=0.35]{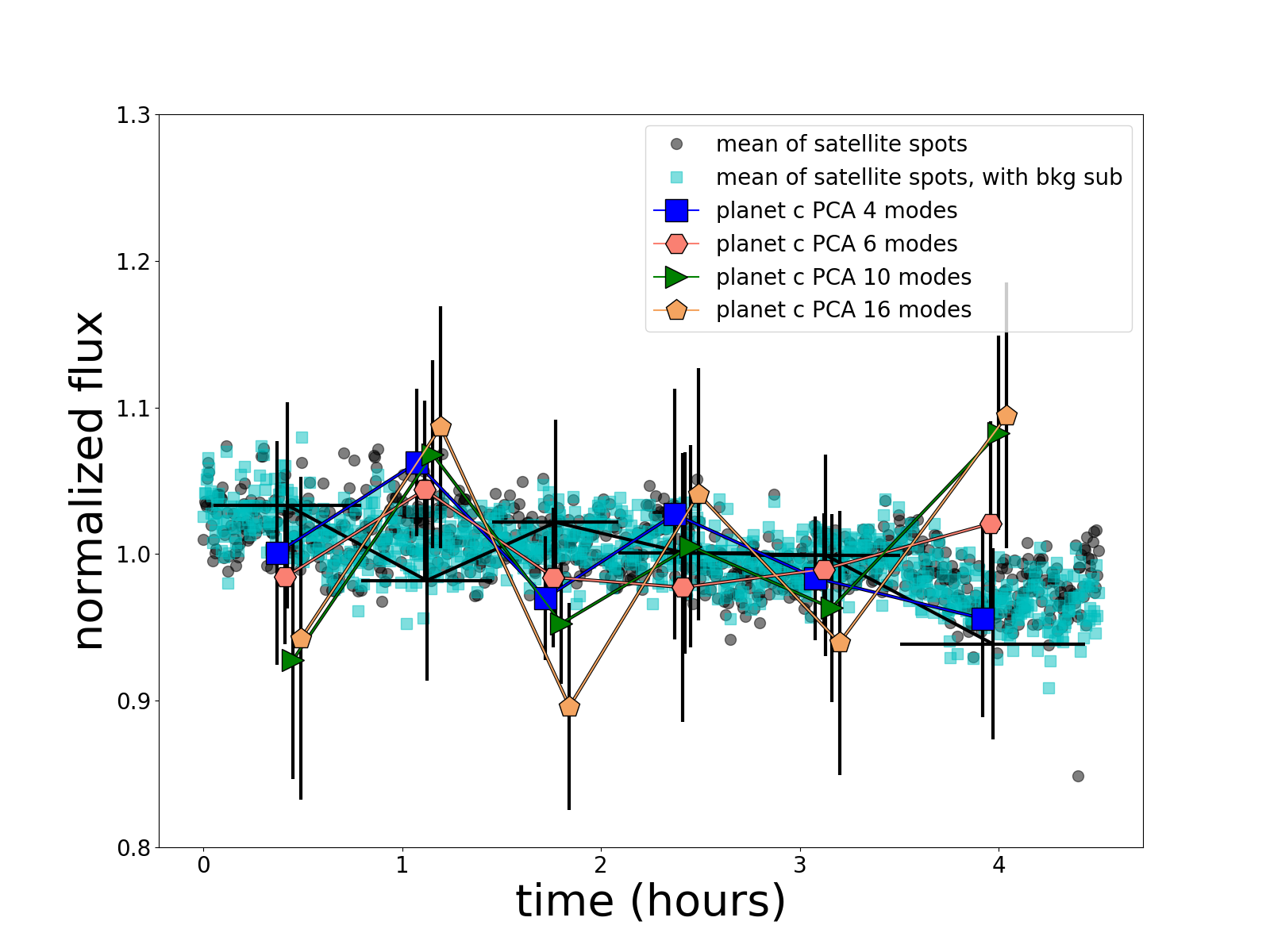}
\caption{The effect of utilizing different numbers of PCA modes on HR8799b and c lightcurves for 18 August 2018.  The lightcurve for planet b is consistent independent of the number of modes used.  The retrieved photometry for HR 8799c is largely self-consistent using a wide range of number of PCA modes, with the exception that the first and last parallactic angle bins diverge by $\sim$1$\sigma$ from the trend given by the satellite spots if the number of PCA modes used is $\geq$10.  However, these two time bins were taken at the extrema of the observation, where the airmass was at its highest. 
}
\label{fig:modecomparisonAug17}       
\end{figure*}

Obtaining accurate photometry for planets embedded in speckle noise requires balancing the number of PCA modes utilized with the acceptable level of planet self-subtraction.  Utilizing a greater number of PCA modes will more fully remove the speckle noise, but at the expense of 
removing flux from the planet itself.  We tested the effect of using different numbers of PCA modes with the 18 August 2018 dataset (which appears by eye to be the highest-quality and most stable of the datasets) by re-running the fine-grid planet photometry pipeline with 4, 6, 10, and 16 modes.  We used 6 equally spaced parallactic angle temporal bins for both b and c (as opposed to 5 for c) so we could directly compare results for the same sets of input data frames.  Lightcurves are plotted in Fig.~\ref{fig:modecomparisonAug17}.  Negative simulated planet photometry gives similar results for HR 8799b, independent of the number of PCA modes used.  However, we found that the negative simulated planet photometry for c varies more strongly with number of PCA modes chosen.  For the most part, the retrieved photometry for HR 8799c is still consistent using a wide range of number of PCA modes, with the exception that the first and last parallactic angle bins diverge by $\sim$1$\sigma$ from the trend given by the satellite spots if the number of PCA modes used is $\geq$10.  However, these two points do agree with the satellite spot trend within 2$\sigma$.  Given that they were taken at the extrema of the observation, at high airmasses and with a slow rate of sky rotation, they simply may not be reliable or may have true uncertainties higher than the uncertainty estimated here.  In Section~\ref{sec:simulations}, we further consider the ideal number of PCA modes to adopt in this case, using simulated planet tests. 

%

\subsection{Inner Planets \label{sec:innerplanets}}

\begin{figure*}
\includegraphics[scale=0.3]{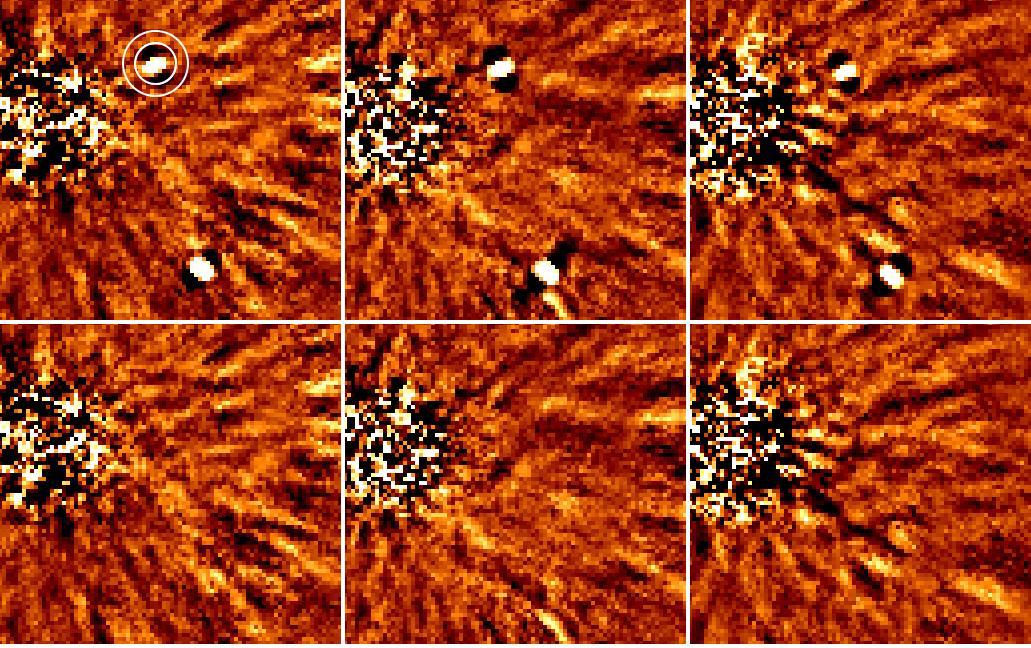}
\caption{Best negative simulated planet subtractions for HR 8799d and HR 8799e on 18 August 2018.  The full $\sim$4.5 hour observation has been divided into 3 equal parallactic angle bins, shown in temporal order from left to right.  The image stretch runs from -10 to 10 ADU in a linear scale, with contrast adjusted to highlight the residual spectral noise after subtraction of 10 PCA modes.  Apertures used for calculating planet flux and background flux in the negative simulated planet algorithm are shown as a white circle and annulus respectively in the image for the first parallactic angle bin.
}
\label{fig:Aug17deresidualimage}       
\end{figure*}

Even with 4-5 hour long observations spanning $>$70$^{\circ}$ degrees in parallactic angle, we do not achieve sufficient sky-rotation to obtain high-sensitivity variability measurements for the innermost HR 8799 planets.  Binning the observation into 3 equal parallactic angle bins will ensure $\geq$23$^{\circ}$ degrees of sky rotation.  For the inner two planets, HR 8799d and HR8799e, at separations of $\sim$0.6" and $\sim$0.4", 23$^{\circ}$ of sky rotation corresponds to shifts of $\sim$20 pixels and $\sim$14 pixels on sky.  Examples of final images after negative simulated planet removal for 2018-08-18 are presented in Figures~\ref{fig:Aug17deresidualimage}.  Like with HR 8799c, significant speckle noise was still present after removing 10 PCA modes. 

\begin{figure*}
\includegraphics[scale=0.3]{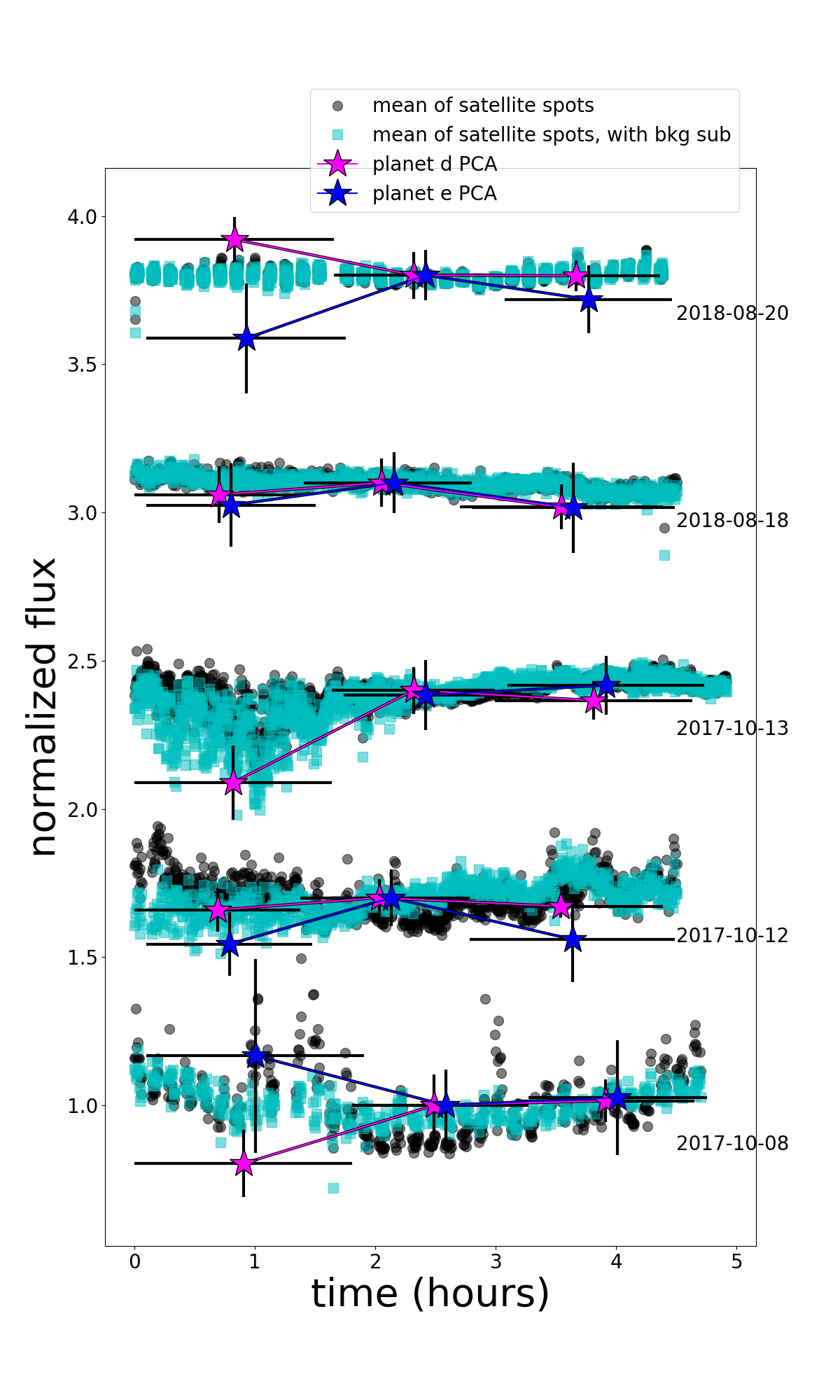}
\caption{Satellite spot aperture photometry (with and without background subtraction), planet b aperture photometry, and planet d and e psf-fitting photometry as a function of time during the observation for October 2017 and August 2018 broadband-$H$ datasets. Planets d and e 
are consistent within errors with the satellite spot photometry, however, our uncertainties are much higher for HR 8799de relative to HR 8799b, as this region of our images still show considerable speckle residuals even after removing 10 PCA modes.}
\label{fig:delightcurve}       
\end{figure*}

Lightcurves for HR 8799de are presented in Fig.~\ref{fig:delightcurve}, along with satellite spot photometry.
Lightcurves for HR 8799de have been normalized to the middle temporal bin out of the three bins and are consistent
within errors with satellite spot trends.  However, the errors on the planet photometry are considerably larger 
than those on the satellite spot photometry, preventing any fine determination of variability relative to the 
satellite spots.  The photometry measurements for HR 8799de are roughly consistent with each other to within 0.25 mag, thus we can 
rule out variability at the $>$30$\%$ level on 4-5 hour-long timescales.

\subsection{Constraints on b-c colors -- detrending b with c}

\begin{figure*}
\includegraphics[scale=0.3]{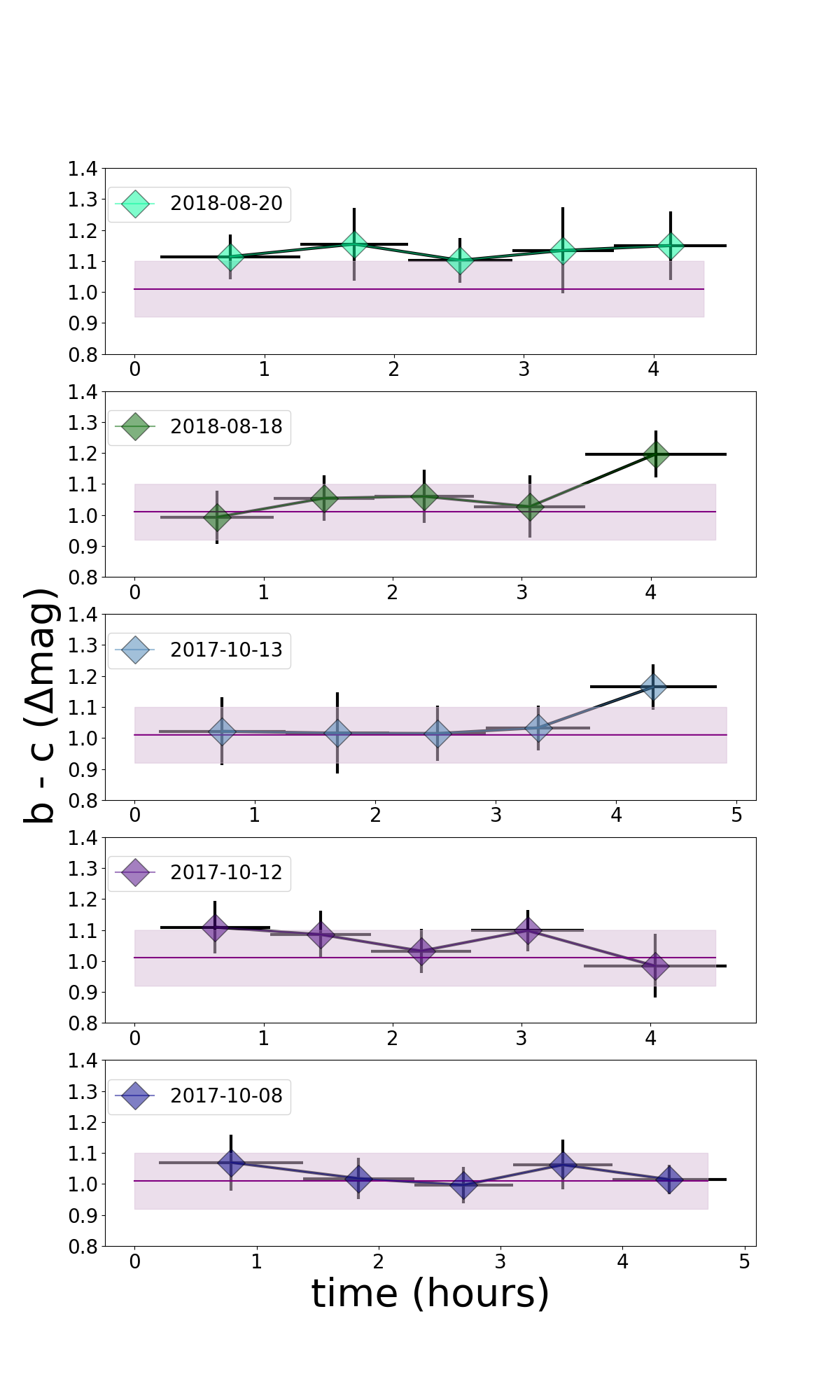}
\caption{$\Delta$(mag) of HR 8799b - $\Delta$(mag) of HR 8799c as a function of time during each broadband-$H$ observation, compared to the results from \citet{Apa16} (purple line and shaded rectangle).  $\Delta$(mag) of HR 8799b - $\Delta$(mag) is consistent with being constant within errors.
}
\label{fig:bcdeltadeltamag}       
\end{figure*}

We note that the lightcurves of HR 8799bc are almost always similar to each other, but in the October 2017 epochs, diverge significantly from the satellite spot lightcurves.  Thus, while we initially considered detrending the planet photometry using the satellite spot photometry, it is clear that the satellite spot lightcurves do not provide an appropriate photometric reference for the planet photometry at all epochs.  Instead, we follow the approach of \citet{Apa16} and calculated $\Delta(mag)_{b} - \Delta(mag)_{c}$ -- essentially detrending the b lightcurve using that of c, or vice versa.  This traces non-shared variations between the two planets.  We considered five temporal bins here, to assure 20 pixels of movement on the sky for HR 8799b and c, with difference lightcurves presented in Fig.~\ref{fig:bcdeltadeltamag}.  $\Delta(mag)_{b} - \Delta(mag)_{c}$ measured by \citet{Apa16} is overplotted as a purple horizontal line.  Within errors, our results are in agreement with those from \citet{Apa16}.  The uncertainty on the measurement is dominated by the uncertainty in our measurement of $\Delta(mag)_{c}$. 
Depending on epoch, we rule out non-shared variability in $\Delta(mag)_{b} - \Delta(mag)_{c}$ to the 10-20$\%$ level over 4-5 hours.  

\subsection{Simulated variability results\label{sec:simulations}}

To quantify our sensitivity to variability in the 2018-08-18 epoch (the best quality data of our search), we simulate variable and non-variable lightcurves by inserting and retrieving a suite of simulated planets at a similar radius from the star as HR 8799bc, but offset in position angle from the true planet position angle.  Thus, we sample a similar region of the speckle pattern, but place each simulated planet sufficiently far from its real counterpart that they do not interfere with each other during the PCA analysis. In the simulated lightcurves, we estimate systematic noise sources to first order (e.g. varying Strehl ratio, seeing, etc) expected in the data by scaling the simulated planet flux by the normalized satellite spot aperture photometry (averaged across all 4 satellite spots) after background subtraction.  While this is not a rigorous attempt to model all potential noise sources, it is sufficient to set constraints on the amplitudes and periods of variable signals that we can potentially detect.  Note the satellite spot aperture photometry is only an appropriate noise model for the August 2018 epochs (and not for the October 2017 epochs), as in August 2018, planet photometry and satellite spot photometry largely followed similar trends. We then simulate either a constant planet flux or a sinusoidal variability signal.  Since the positions of the simulated planets are known a priori, we do not run the full 3-d planet retrieval grid, but only consider a 1-d grid of $\Delta$mag at the a priori planet position. 

\begin{figure*}
\includegraphics[scale=0.3]{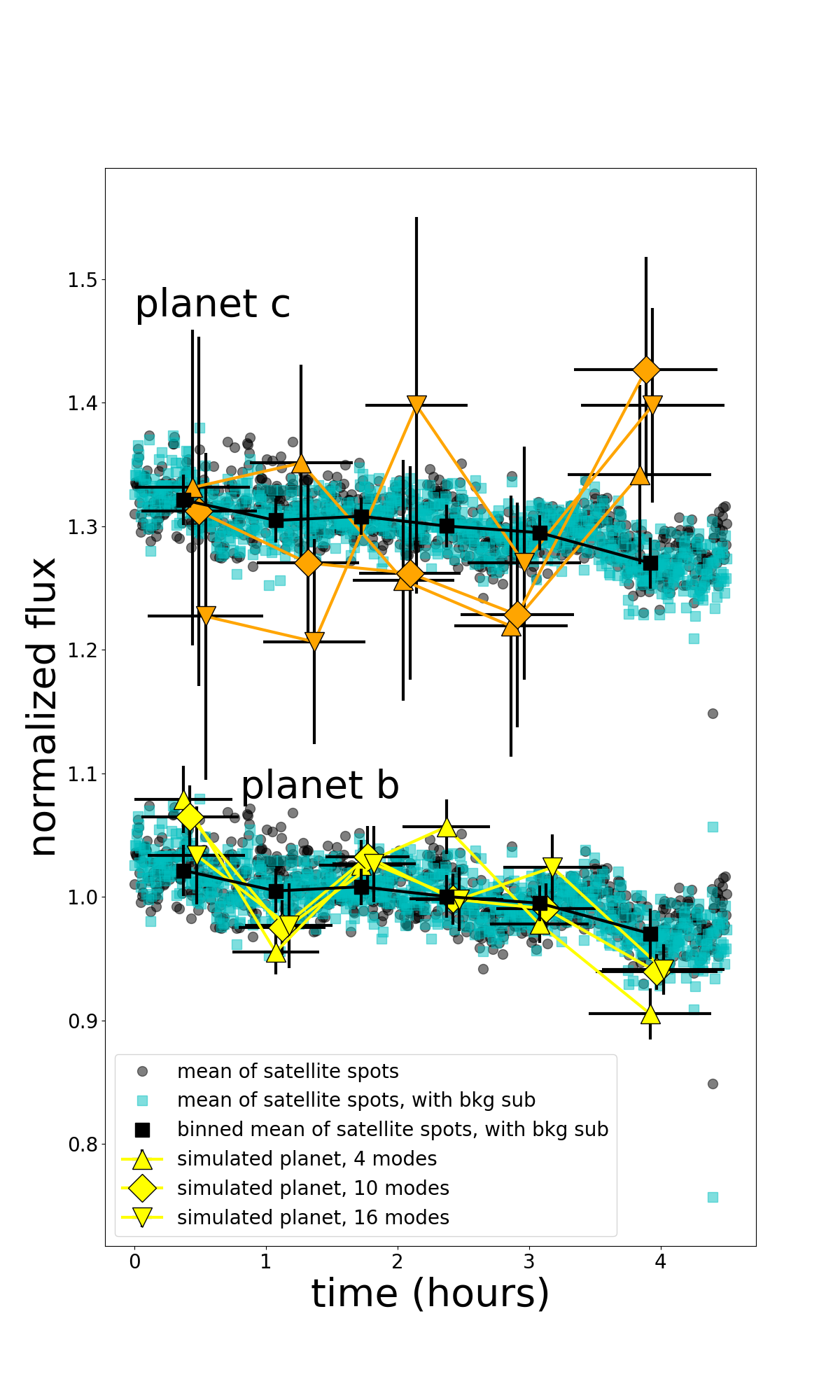}
\caption{The effect of utilizing different numbers of PCA modes on simulated constant-flux HR8799b and c lightcurves for 18 August 2018 (noise simulated using the satellite spot lightcurve, inserted at the same radius but differing PA compared to the actual planets).  The planet b simulated lightcurve is consistent independent of the number of modes used.  The retrieved photometry for HR 8799c is much noisier, with the 10 mode lightcurve providing the best trade-off between preserving planet flux and suppressing speckles.  The final photometric point in HR 8799c simulated lightcurve is considerably brighter than the satellite spot lightcurve at the same bin using 4, 10, or 16; this mirrors what is seen in the actual HR 8799c lightcurve.  At this radius from the star and this temporal point in the observations, speckle residuals remain that affect our negative simulated planet subtraction, but are not accounted for in the satellite spot lightcurves. 
}
\label{fig:simulatedmodecomparison}       
\end{figure*}

We first consider the effect of using different numbers of PCA modes when retrieving simulated planets.  Retrieved lightcurves (using 4, 10, or 16 PCA modes) for simulated constant-flux versions of HR 8799bc are presented in Fig.~\ref{fig:simulatedmodecomparison}.  The simulated HR 8799b lightcurve is robustly retrieved using 4, 10, or 16 modes, but the 4 and 16 mode retrievals look considerably noisier than the 10 mode retrieval, further justifying our adoption of 10 modes as standard.  The retrieved lightcurves for HR 8799c are noisier than those for HR 8799b.  Using 4 or 10 modes, the HR 8799c simulated lightcurves match the satellite spot lightcurve within errors.  The 16 mode lightcurve is extremely noisy and is clearly oversubtracting the planet flux, resulting in significantly higher uncertainties.  In all cases, the final photometric point in the HR 8799c simulated lightcurve is brighter than the satellite spot lightcurve at the same bin; this mirrors what is found in the actual HR 8799c lightcurve.  At this radius from the star and this temporal point in the observations, we suspect that speckle residuals remain that are not accounted for in the satellite spot lightcurves.  In other words, the satellite spot lightcurve does not serve as an appropriate photometric reference for this particular data point.

\begin{figure*}
\includegraphics[scale=0.25]{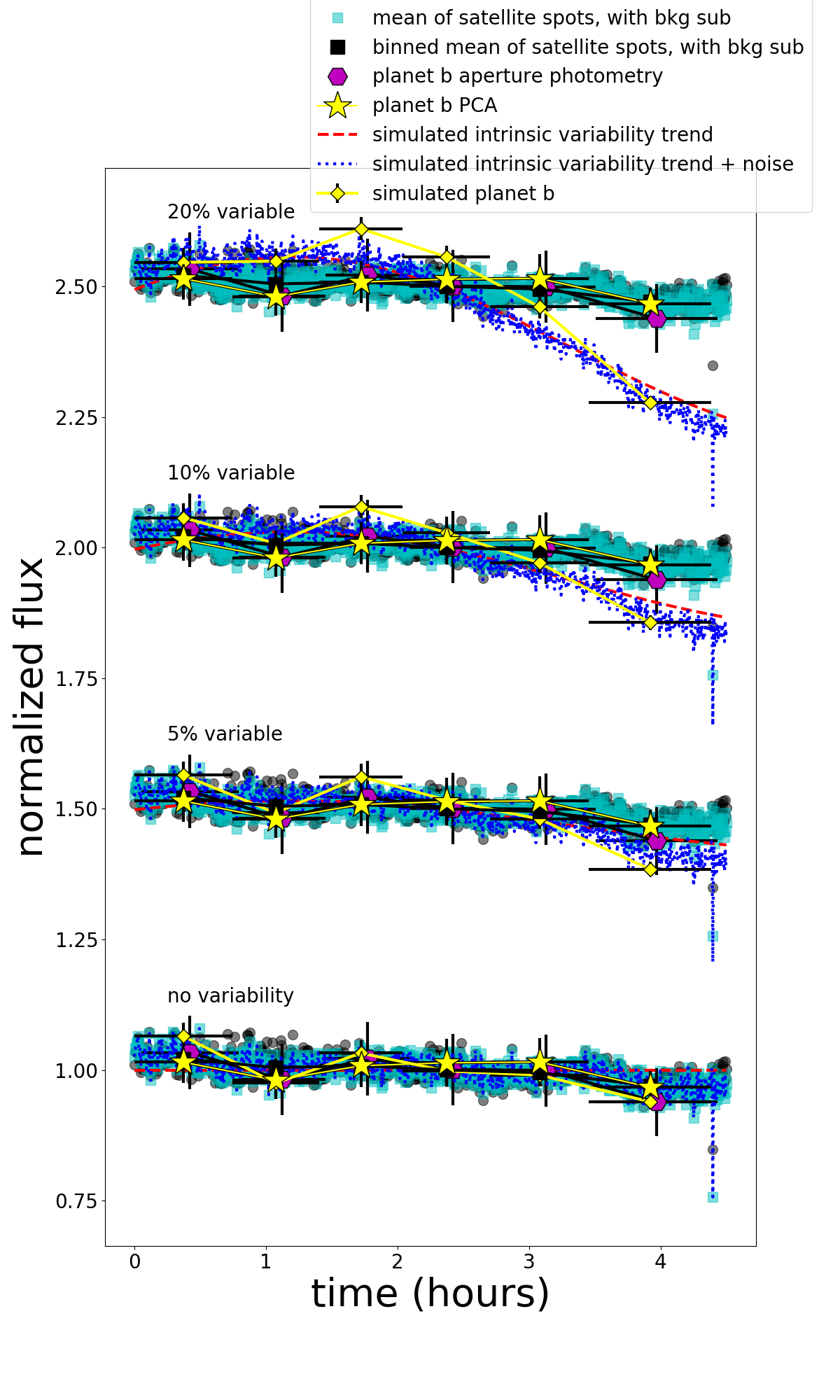}
\includegraphics[scale=0.25]{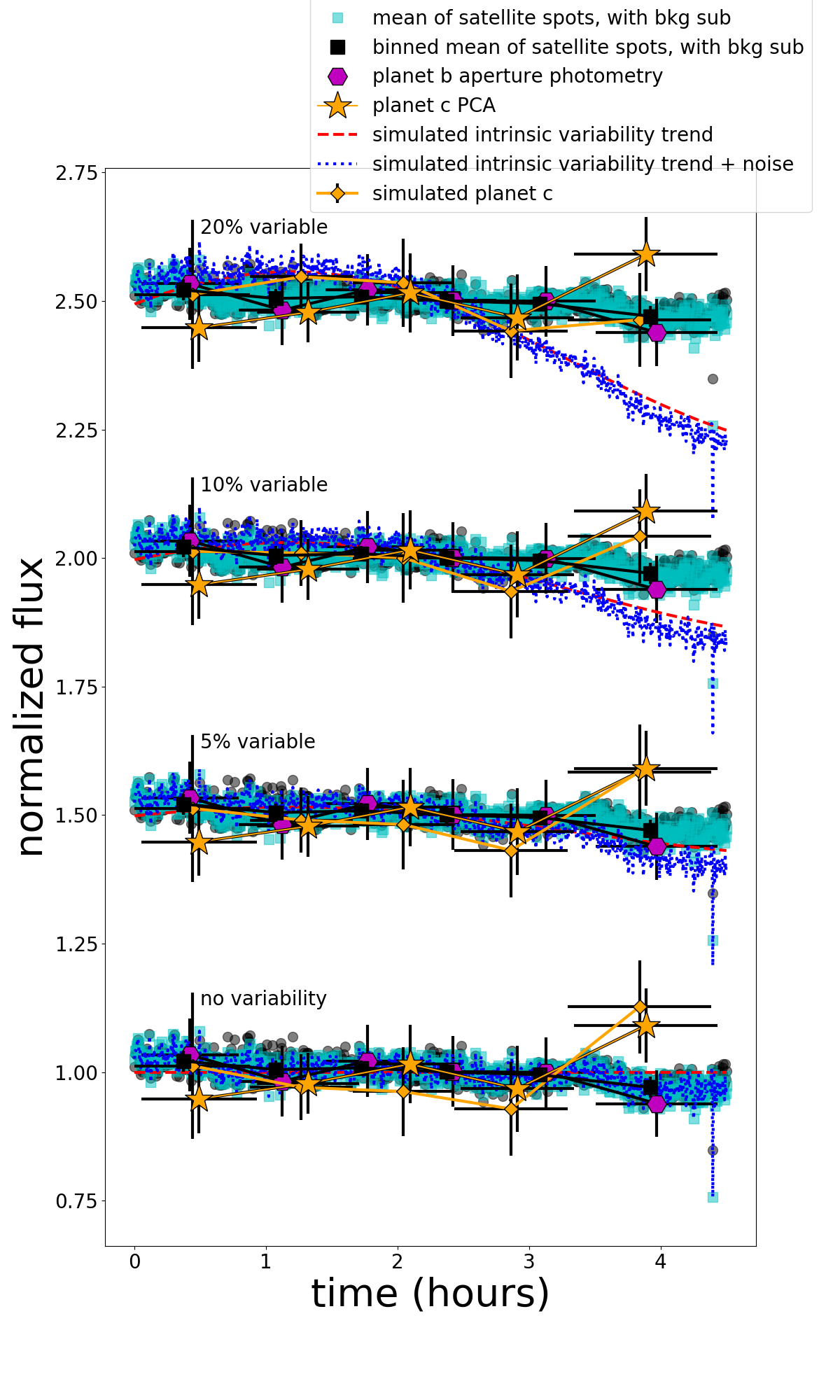}
\caption{Simulated lightcurves for HR 8799bc with sinusoidal variability with a period of 8.6 hours, 
and amplitudes of 5$\%$, 10$\%$, and 20$\%$.  Variability is apparent by eye in the simulated HR 8799b lightcurves with amplitudes $\geq$10$\%$, but there is no clear detection of variability for the simulated HR 8799c lightcurves even for the 20$\%$ amplitude curve.
}
\label{fig:simulatedvariability}       
\end{figure*}

In Fig.~\ref{fig:simulatedvariability}, we present simulated lightcurves for HR 8799bc with a period of 8.6 hours, similar to the period of the variable free-floating planetary mass object PSO J318.5-22 \citep{Biller2018}, a phase of 40$^{\circ}$, and with variability amplitudes of 0$\%$, 5$\%$, 10$\%$, and 20$\%$.  As before, we use 6 and 5 equal parallactic angle bins for HR 8799b and c respectively, to ensure at least 5 FWHMs of motion on the sky in each parallactic angle bin.  Variability is clearly apparent in the simulated HR 8799b lightcurves with amplitudes $\geq$10$\%$.  Variability detection is much more difficult for the noisier HR 8799c lightcurves, notably stymied by the high final photometric point.

\begin{figure*}
\includegraphics[scale=0.3]{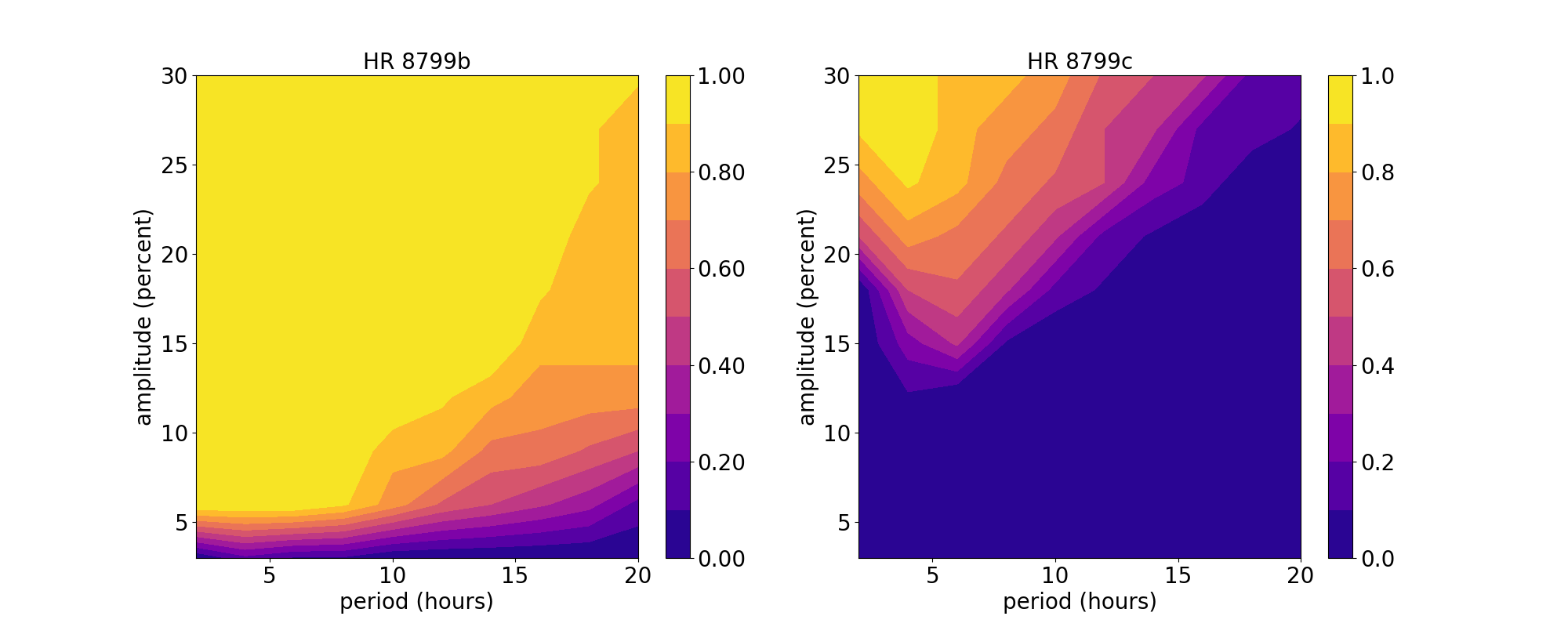}
\caption{Simulated planet fractional detection maps for HR 8799bc in our 2018-08-18 data.  The color scale is set by the fraction of simulated lightcurves detected at each combination of period and amplitude.  For HR 8799b, for periods $<$10 hours, we are generally quite sensitive to variability with amplitude $>5\%$.  For HR 8799c, our sensitivity is much lower, limited to variability $>25\%$ for similar periods.  
}
\label{fig:detectionmap}       
\end{figure*}

To estimate our sensitivity to variability across a wide range of different amplitudes, periods and phases, we ran a complete grid of simulated planet lightcurves using 5 equal parallactic angle temporal bins, with periods from 2 to 20 hours (in steps of 2 hours), amplitudes from 3$\%$ to 30$\%$ (in steps of 3$\%$), and phases from 0$^{\circ}$ to 330$^{\circ}$ (in steps of 30$^{\circ}$).  We then detrend each simulated HR8799bc lightcurve by dividing through by the mean satellite spot lightcurve, with errors on the detrended lightcurve given as the sum in quadrature of the errors on the simulated planet and satellite spot lightcurves.  For each detrended simulated lightcurve at a specific value of period, amplitude, and phase, we set up a metric for the detection of variability in the detrended lightcurve as follows:
If (maximum(flux) - minimum(flux)) / maximum(error) $>$2, we consider variability to be detected (e.g. a 2-$\sigma$ variability detection).  This metric was tested by eye for a significant number of simulated lightcurves, and generally agrees with visual assessment. To build the detection map, we assign each combination of period, amplitude, and phase the value 1 if variability is detected, and 0 if variability is not detected.  We then sum along the phase axis and divide by the number of phases simulated to produce a fractional detection map as a function of period and amplitude.  Detection maps for HR 8799bc in the 2018-08-18 epoch are presented in Fig.~\ref{fig:detectionmap}.  For HR 8799b, for periods $<$10 hours, we are generally quite sensitive to variability with amplitude $>5\%$.  For HR 8799c, our sensitivity is much lower, limited to variability $>25\%$ for similar periods.

\section{Epoch-by-Epoch Results}

In this section, we present epoch-by-epoch astrometry and photometry, and compare to literature values.
For this purpose, we used only the data frames from the hour approaching meridian crossing and the hour after meridian crossing, thus covering the time period during each observation with the most field rotation and best airmass.  We again ran our full negative simulated planet code to retrieve the best $\Delta$(mag), radius, and position angle value for each planet.  

\begin{figure*}
\includegraphics[scale=0.24]{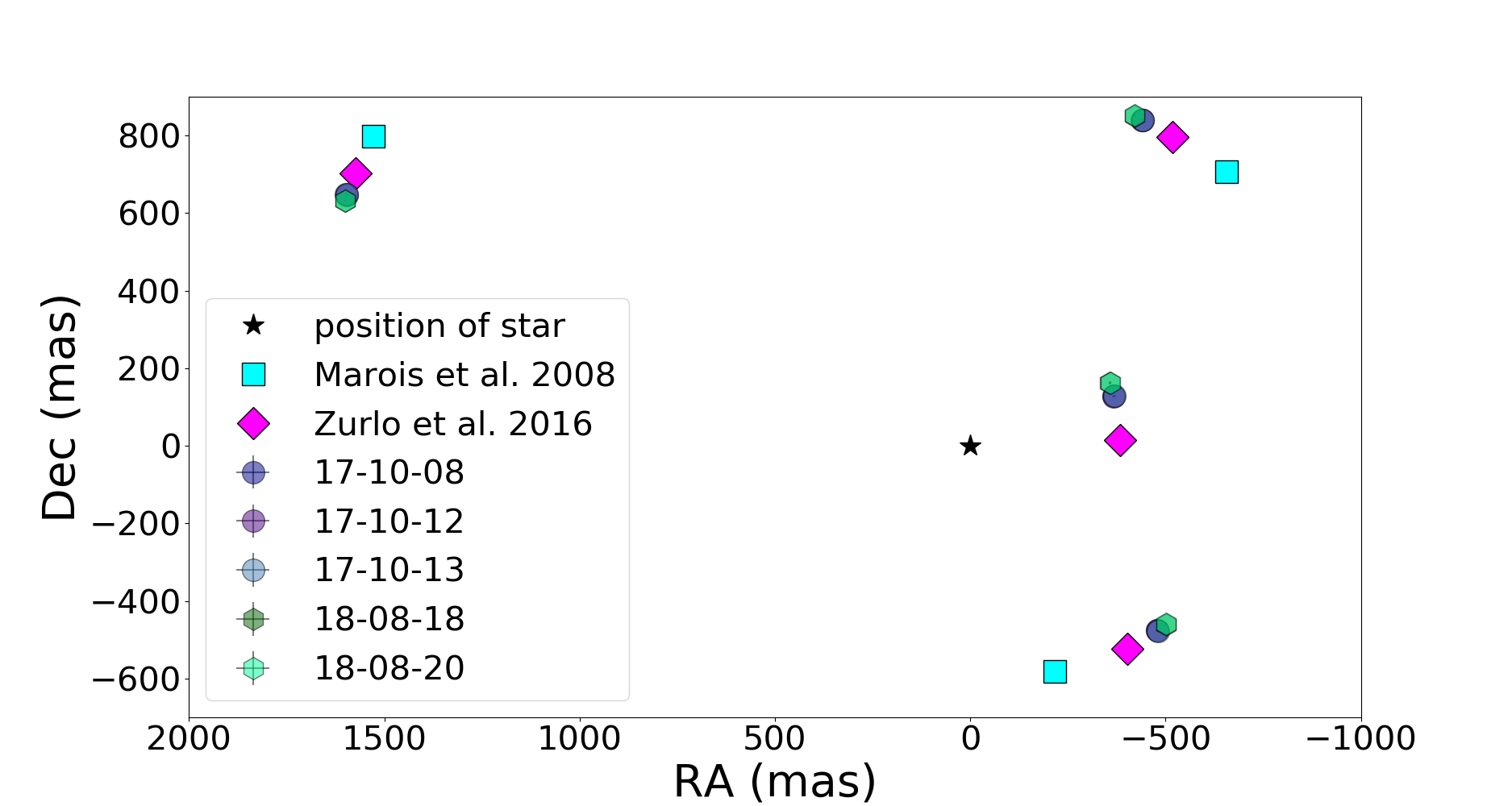}
\includegraphics[scale=0.16]{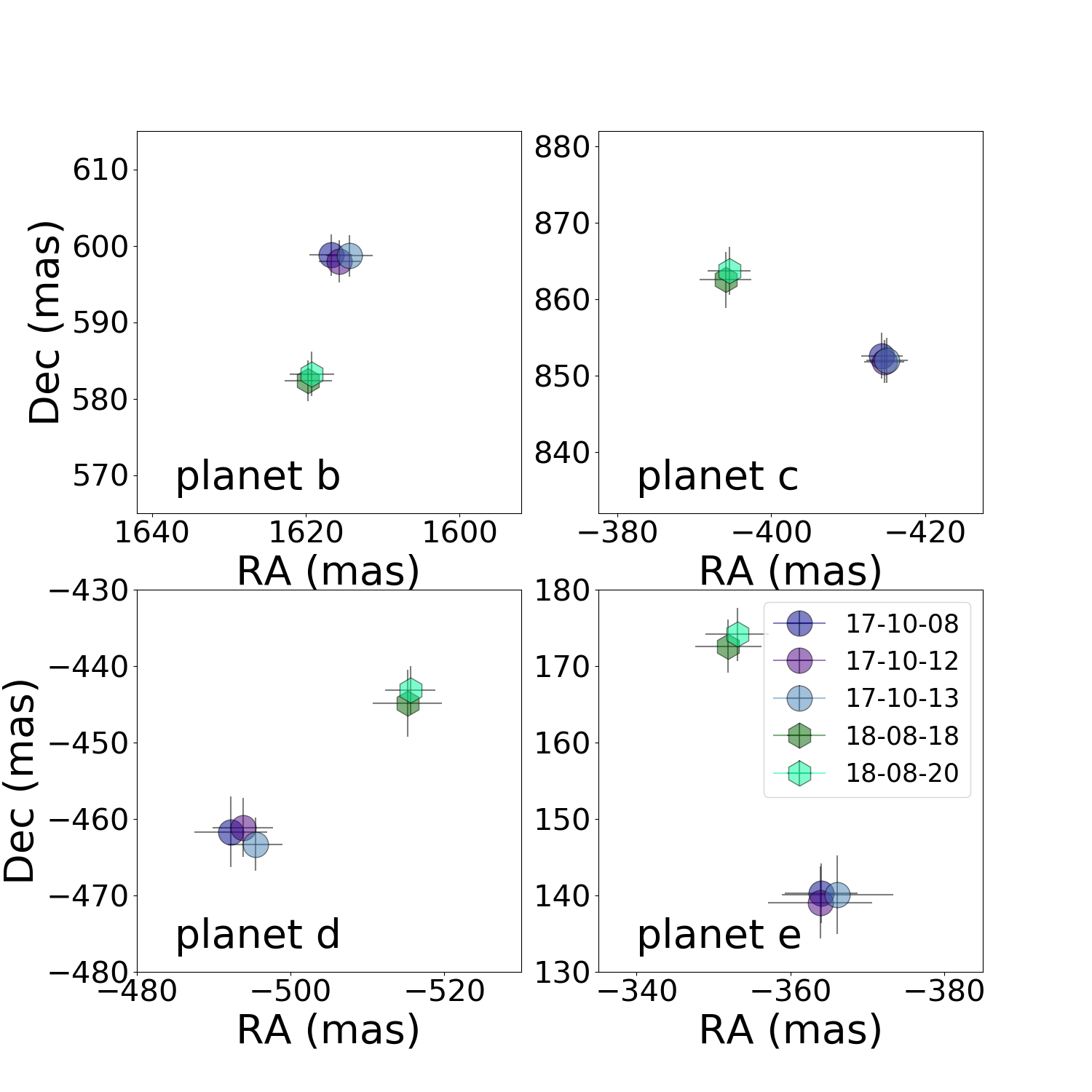}
\caption{{\it Left:} Astrometric comparison between our epochs, \citet{Mar08}, and \citet{Zur16}.  The error bars are smaller than the plot symbols; all four planets show significant on-sky motion even between our 2017 and 2018 epochs.  {\it Right}: Planet-by-planet view of astrometry from our October 2017 and August 2018 epochs.
}
\label{fig:astrom1}       
\end{figure*}

Astrometric results are plotted compared to literature astrometry in the left panel of Fig.~\ref{fig:astrom1}.  A zoomed-in plot on our epochs of astrometry is presented in right panel of Fig.~\ref{fig:astrom1} and the epoch-by-epoch astrometry for each planet is presented in Tab.~\ref{tab:astrometry}.  To convert the uncertainties derived from the negative simulated planet code from radius and position angle values to $\Delta$RA and $\Delta$DEC, we drew 100000 samples from Gaussians centered on the best radius and position angle values and with $\sigma$ given from the respective errors on from the negative simulated planet subtraction, converted these values to $\Delta$RA and $\Delta$DEC, then adopted the mean and standard deviations of these distributions as the corresponding $\Delta$RA, $\Delta$DEC values and 1-$\sigma$ uncertainties.  In this case, we conservatively adopt the 2-$\sigma$ values for uncertainties, which gives us the 
uncertainty on the relative planet position $\sigma_\mathrm{PP}$ at this epoch.  However, we must consider further astrometric uncertainties in our 
full error budget.  Following \citet{Zur16}, we adopt a plate scale uncertainty $\sigma_\mathrm{PS}$ of 2 mas and star center position $\sigma_\mathrm{SC}$ uncertainty of 1.2 mas; the full uncertainty is then given as the sum in quadrature of $\sigma_\mathrm{PP}$, $\sigma_\mathrm{PS}$, and $\sigma_\mathrm{SC}$. Our astrometric points are consistent with the already-observed orbital motion of these planets, however, we defer orbital fitting of these points and further epochs of SPHERE orbital monitoring to Zurlo et al. in prep. 

\begin{figure*}
\includegraphics[scale=0.3]{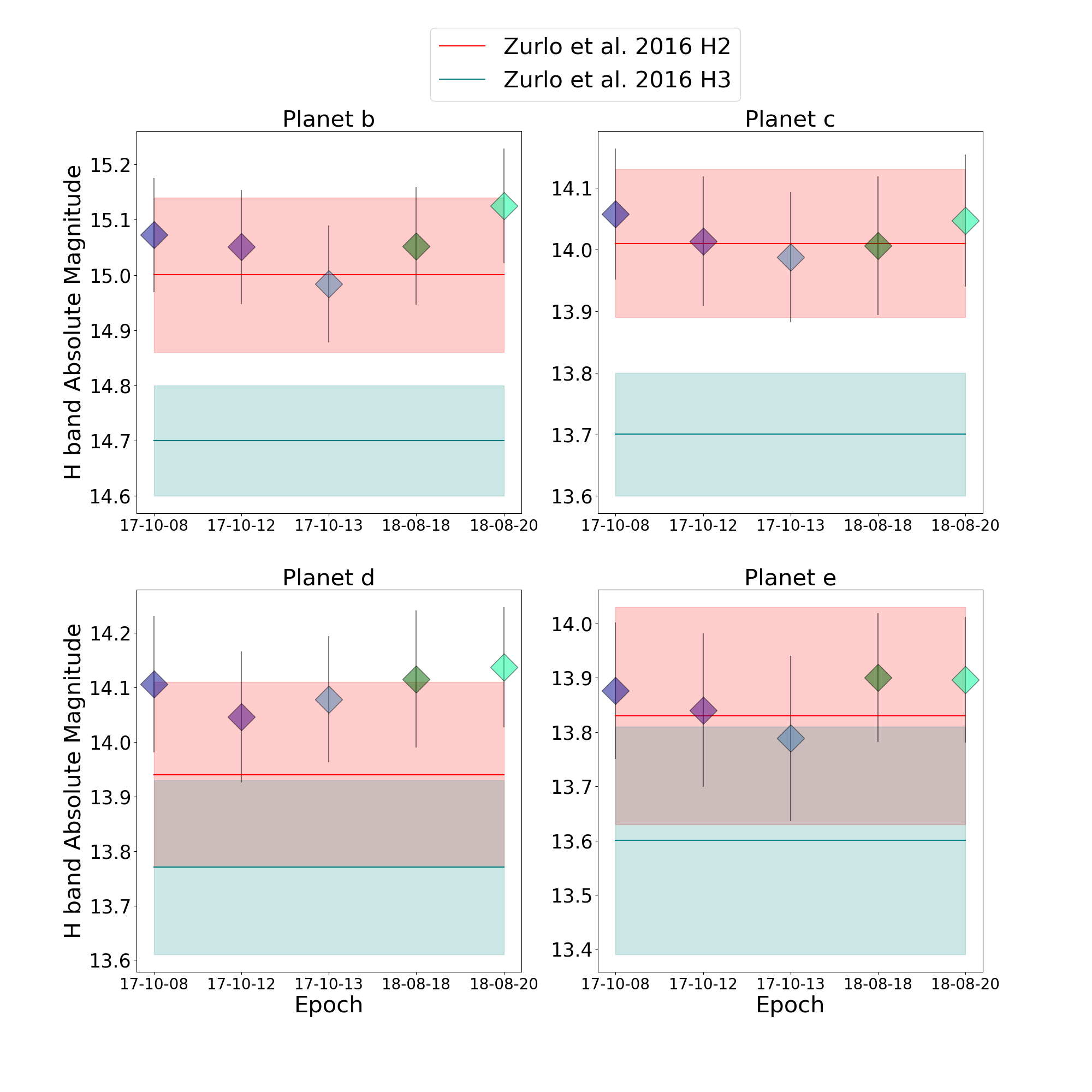}
\caption{Our broadband-$H$ epoch-by-epoch photometry compared to SPHERE $H2$ and $H3$ photometry from \citet{Zur16}.
}
\label{fig:fullphotSPHEREcomp1}       
\end{figure*}

\begin{figure*}
\includegraphics[scale=0.3]{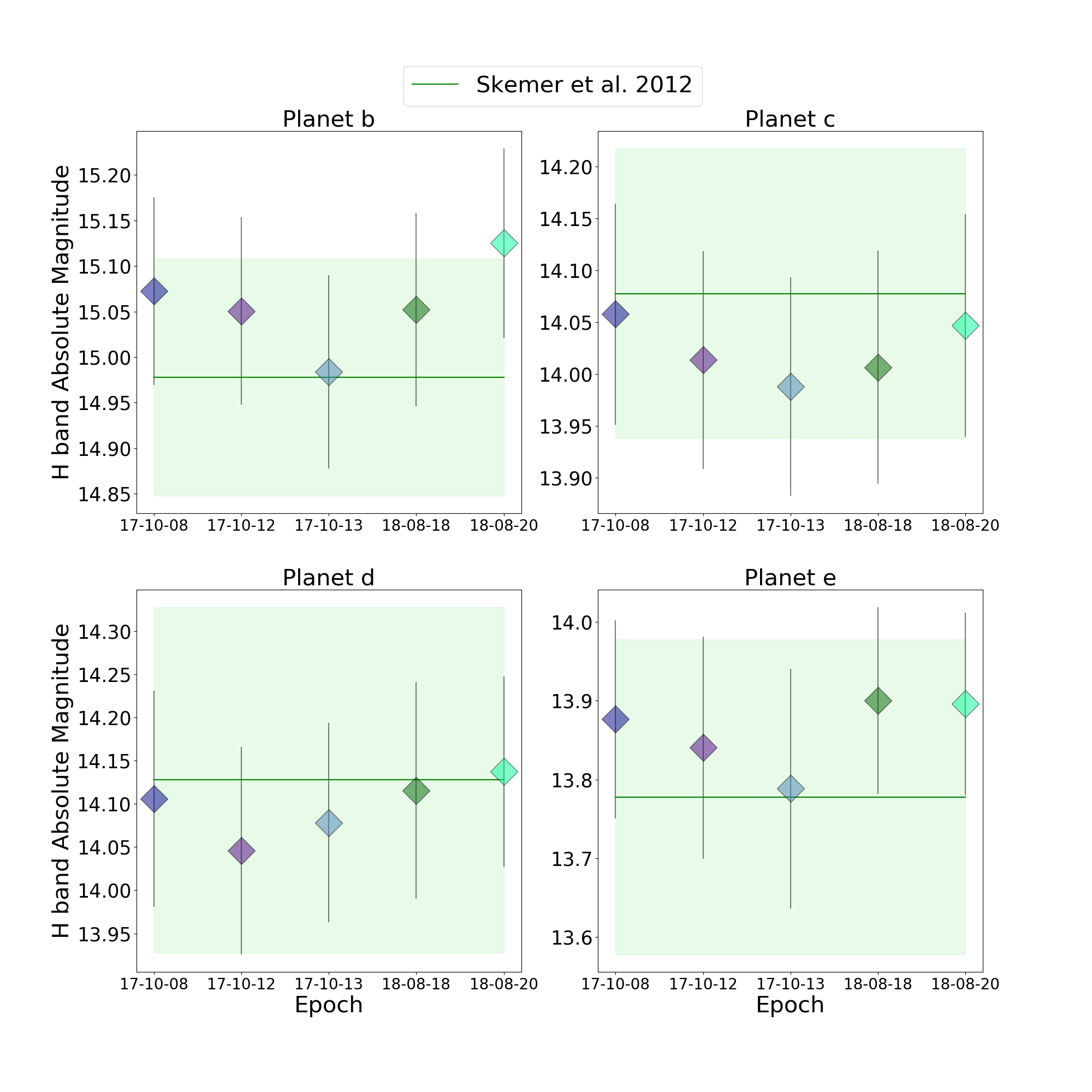}
\caption{Our broadband-$H$ epoch-by-epoch photometry compared to LBT $H$-band photometry from \citet{Ske12}.
}
\label{fig:fullphotSPHEREcomp2}       
\end{figure*}

\begin{figure*}
\includegraphics[scale=0.3]{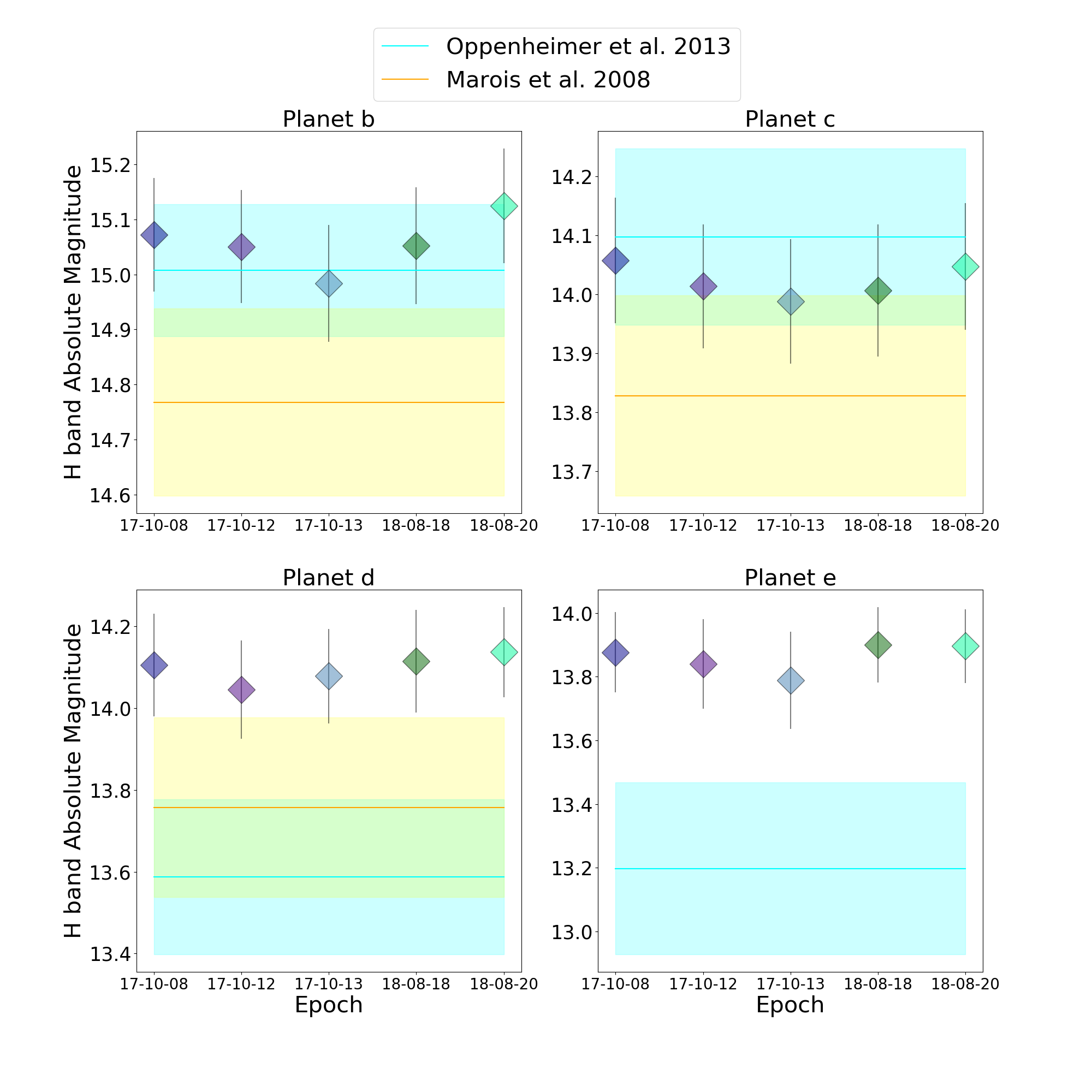}
\caption{Our broadband-$H$ epoch-by-epoch photometry compared to Palomar $H$-band photometry from \citet{Oppo13} and Keck $H$-band photometry from \citet{Mar08}. 
}
\label{fig:fullphotSPHEREcomp3}       
\end{figure*}

\begin{figure*}
\includegraphics[scale=0.3]{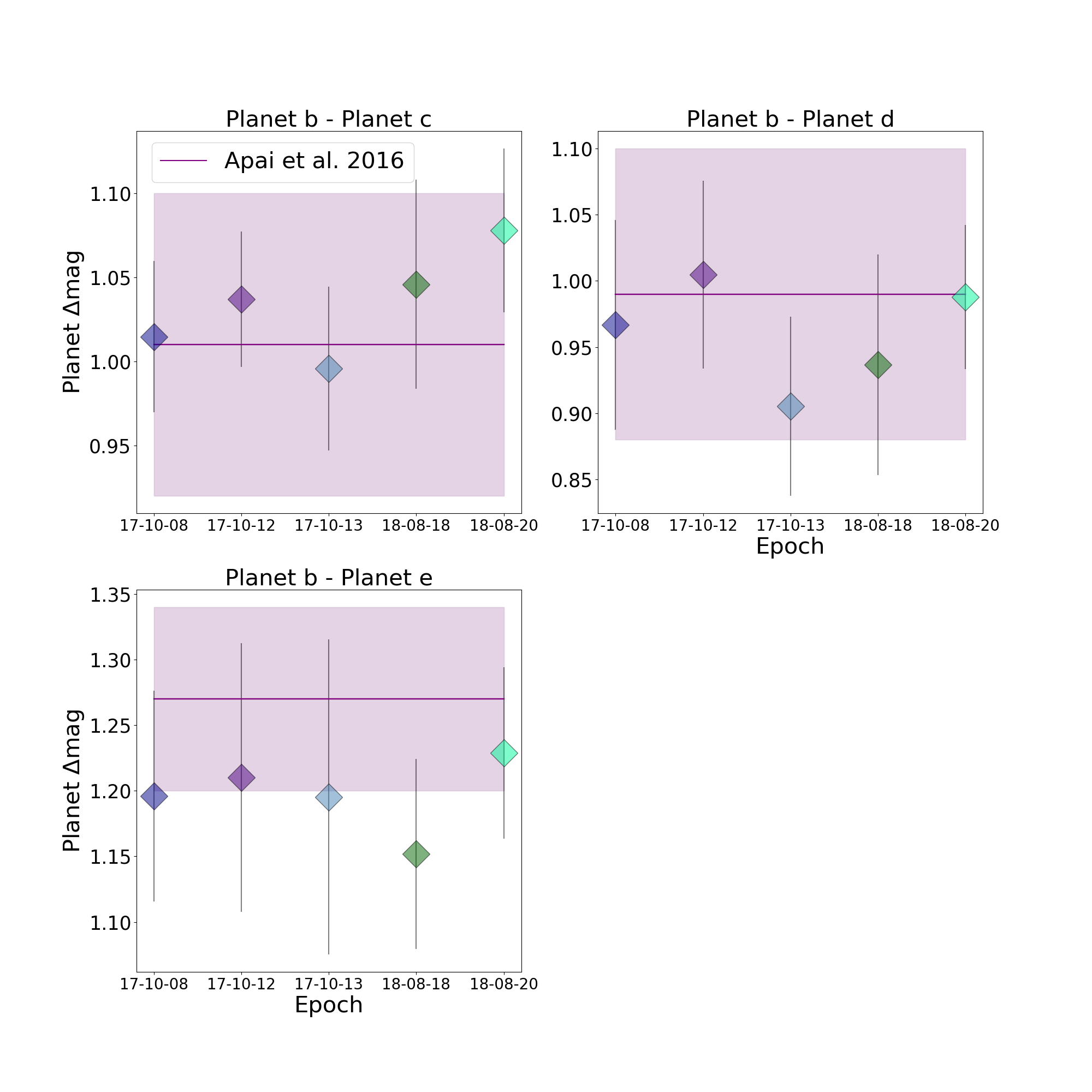}
\caption{Our broadband-H epoch-by-epoch photometry compared to \citet{Apa16}.
}
\label{fig:fullphotSPHEREcompsubtraction}       
\end{figure*}

Our epoch-by-epoch values of the absolute $H$ magnitude of each planet are compared to literature photometry in the $H$ band in Figs.~\ref{fig:fullphotSPHEREcomp1} to~\ref{fig:fullphotSPHEREcomp3} and presented in tabular form in Table~\ref{tab:photometryabsmag}.  To convert from $\Delta$mag to absolute magnitude in $H$, we adopt a parallax of 24.2175$\pm$0.0881 mas from Gaia DR2 \citep{GaiaDR2} and an $H$-band magnitude of 5.280$\pm$0.018 mag from the 2MASS all-sky survey \citep{Skrutskie2006}. Since the star is a known variable with an amplitude of up to 0.1 mag in the optical, we have added a conservative error of 0.1 mag on the magnitude of the star, in quadrature with the error we derive from our negative simulated planet subtraction photometry.  
The large errors endemic to converting from $\Delta$(mag) to absolute mag prevent a meaningful epoch-to-epoch comparison in absolute H, however, our photometry is constant to within error bars both between our own epochs and with $H2$ values from \citet{Zur16} (a somewhat different band), H-band values for all planets from \citet{Ske12}, and HR 8799b and c values from \citet{Oppo13}.  We see significant divergence from \citet{Mar08} and HR 8799d and e values from \citet{Oppo13}.  
Because of the inherently large uncertainties in converting from $\Delta$(mag) to absolute mag, \citet{Apa16} instead opt to consider $\Delta$(mag)$_{b}$ - $\Delta$(mag)$_{c}$, etc., in other words, the relative difference in contrast between different planets in the system.  We plot our measurements of this value in comparison to those from \citet{Apa16} in Fig.~\ref{fig:fullphotSPHEREcompsubtraction} and present our measurements in tabular form in Table~\ref{tab:photometrysub}; our values all agree within 1-$\sigma$ with those from \citet{Apa16}, with no compelling evidence of epoch-to-epoch variability. 

\section{Discussion}

\subsection{Use of satellite spots as photometric references}
As noted in Section~\ref{sec:system}, HR 8799 is a $\gamma$ Doradus pulsating variable, with a dominant periodicity of $\sim$0.5 day, and a peak-to-valley amplitude of up to $\sim$0.1 mag in the visible \citep{Sodor2014}.  SPHERE crucially enables monitoring of all planets as well as the satellite spots simultaneously, thus allowing us to disentangle any intrinsic variability of the star from variability of the planets.  As we do not expect the planets to have the same rotational period, phase or variability amplitude, any variability with the same period shared by multiple planets would then probably be intrinsic to the star.  Thus, our expectation was that the satellite spot light curves should follow the trend produced by the changing atmospheric conditions as well as any astrophysical variability from the star, as would the lightcurve of a non-variable exoplanet.  This did not prove to be universally true.  The lightcurves for HR 8799bc closely follow each other, suggesting that the variations observed are driven by the observing conditions or intrinsic variability of the star, and do not indicate variability in the planets.  However, while the lightcurves for HR8799bc closely follow that of the mean satellite spot lightcurve in the August 2018 data, there are significant deviations between the satellite spot and planet lightcurves in the October 2017 data.  This means that the satellite spots do not always provide appropriate photometric references for detrending the broadband $H$ exoplanet lightcurves; this seems to be correlated to the observing conditions and may be due to the fact that the planet spectra are similar to each other, but much redder than the spectrum of the star (which is shared by the satellite spots).  
To test whether this is the case, it would be interesting to observe a binary companion with a more similar (e.g. bluer) spectrum to its host, along with the satellite spots. If the binary companion displays the same trends as the satellite spots in a range of different observing conditions, the divergence between planet and satellite spot lightcurves we measure here would likely be due to the red colours of the planets relative to the star.  In this case, a narrow band or integral field spectroscopic search for variability would not have to contend with this particular issue, but would rather be limited by the brightness of the planet in a narrow bandpass.

However, the source of the divergence may instead be that the satellite spots are artificially introduced and do not fully share properties of imaged astrophysical objects.  The satellite spots for SPHERE are generated by introducing a waffle pattern onto the SPHERE deformable mirror, producing 4 artificial sources in a cross-pattern \citep{Lang13}.  In contrast, GPI uses a square grid in the pupil plane of the telescope to produce a similarly shaped cross-pattern of 4 first-order diffraction spots \citep{Sivaramakrishnan2006, Wang2014}.  It would be interesting to investigate whether these different methods to produce artificial references cause differences in the behaviours of these references; note however, that Figure 7 of \citet{Wang2014} seems to show similar correlations / anticorrelations between different satellite spots as we found in Section~\ref{sec:correlations} and Appendix~\ref{app:sscor}.  Unlike a real companion or stellar image, satellite spots in SPHERE and GPI are not incoherent with stellar residuals. The coherent satellite spots suffer from pinning effects (interference between the satellite spots and PSF residuals) and provide only reduced accuracy for photometry and astrometry. Recently, \citet{Sahoo2020} has experimented with introducing fiducial incoherent faint copies of the host star in the image plane of Subaru ScEXAO and then alternating the pattern of these copies between exposures.  This produces an incoherent photometric reference and may solve some of the issues of using satellite spots as photometric references encountered in this study.

\subsection{Sensitivity to variability / comparison with known variable objects}

HR 8799bcde share very similar spectra and colours to highly variable mid-to-late-L variable planetary mass objects such as VHS 1256-1257ABb \citep{Bowler2020, Zhou2020} and PSO J318.5-22 \citep{Liu13,Biller2015, Biller2018}.  Although variability amplitude can vary between epochs for these objects, these objects are always highly variable in the near-IR, and particularly in $J$-band -- VHS 1256-1257ABb has been observed to be variable by 24.7$\%$ over $\sim$8 hours at 1.27 $\mu$m \citep{Bowler2020} and PSO J318.5-22 has been observed to be 7-10$\%$ variable over $\sim$4 hours in the $J_S$ band \citep{Biller2015}.  These two objects are observed at high inclinations; \citet{Zhou2020} find that the $vsini$ and period of VHS 1256-1257ABb is most consistent with an equator-on viewing inclination and \citet{Biller2018} measure an inclination for PSO J318.5-22 of 56.2$\pm$8.1$^{\circ}$.  Field brown dwarfs \citep{Rad14a,Met15} display the highest amplitude variability in $J$ band ($\sim$1.1-1.3 $\mu$m), decreasing somewhat through the rest of the near-IR, and with considerably lower mid-IR (3-5 $\mu$m) amplitudes; VHS 1256-1257ABb shows a similar trend to the field brown dwarfs at all wavelengths, with the mid-IR amplitude less than half the near-IR amplitude \citep{Zhou2020}, while PSO J318.5-22 is a rare exception to this rule, with nearly equal near-IR vs. mid-IR amplitudes.\footnote{PSO J318.5-22 is also notably one of few brown dwarfs or exoplanet analogues with simultaneously measured near-IR and mid-IR amplitudes.}  As the HR 8799bcde planets share comparable atmospheric properties to these objects, they may also have high intrinsic variability amplitudes in the near-IR.

\begin{figure*}
\includegraphics[scale=1.3]{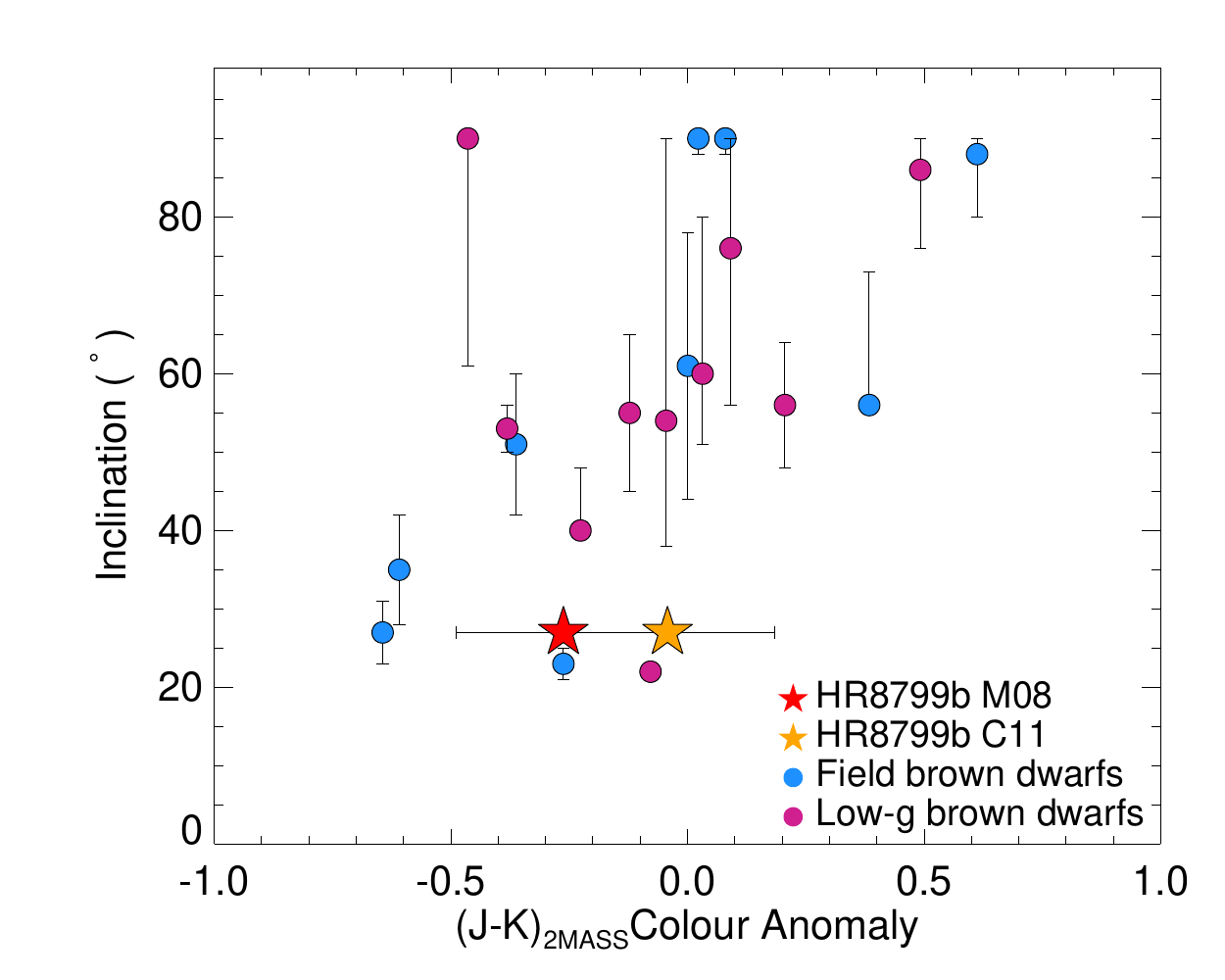}
\caption{Inclination vs. colour anomaly for HR 8799b as well as field and low surface gravity brown dwarfs from \citet{Vos2020}.  HR 8799b has been plotted assuming it shares the same 27$^{\circ}$ inclination as the HR 8799 disk\citep[][]{Matthews2014,Ruffio2019AJ, Wang2018b, Konopacky2016, Pueyo2015, Maire2015}, however, the inclination of this planet has not been directly measured.  Adopting a spectral type of L7, we calculate the colour anomaly (object $J-K$ - average $J-K$ for other objects with the same spectral type and gravity class) for HR 8799b using photometry from \citet{Mar08,Currie2011} converted into the 2MASS filter set using the conversion from \citet{Stephens2004}.  The colour anomaly for HR 8799b is consistent with a wide range of possible inclinations for this planet, including a nearly pole-on 27$^{\circ}$ inclination.  
}
\label{fig:colouranomaly}       
\end{figure*}

In our best epoch on 18 August 2018, for periods less than 10 hours, we were sensitive to 5-10$\%$ variability in HR 8799b and to $\geq$25$\%$ variability in HR 8799c in the broadband H filter.  However, the measured variability amplitude will be less than the intrinsic variability amplitude due to geometric dilution effects and attenuation of flux within the line-of-sight through the planet's own atmosphere.  \citet{Vos2017} estimate the contributions of geometric effects and atmospheric attenuation as:

\begin{equation}
A = A_O sini - \kappa \frac{dx}{sin i}
\label{eqn:amplitude}
\end{equation}

where $A$ is the measured variability amplitude, $A_O$ is the intrinsic variability amplitude, $i$ is the inclination of the planet, $\frac{dx}{sin i}$ is the path length of the light through the atmosphere of the planet, and $\kappa$ is the degree of attenuation through the atmosphere per unit length \citep[see~also~Fig.~4~in][]{Vos2017}. The first term gives the geometric dilution due to projection. The HR 8799 disk is inclined by $\sim$27$^{\circ}$ \citep[based~both~on~disk~studies~and~orbital~fits~for~the~individual~planets,~see~e.g.~][]{Matthews2014,Ruffio2019AJ, Wang2018b, Konopacky2016, Pueyo2015, Maire2015}.  For brown dwarfs with measured inclination angles, \citet{Vos2017,Vos2020} find a relationship between inclination angle and  $(J-K)_\textrm{2MASS}$ colour anomaly (object $J-K$ - average $J-K$ for other objects with the same spectral type and gravity class).  They find that objects viewed at high inclination (equator-on) are redder than the average object of the same spectral type and objects viewed at low inclination (pole-on) are comparatively bluer.  Assuming that HR 8799b shares the inclination of the disk (but noting that the inclination of the planet has not been measured), adopting an L7 spectral type, and calculating colour anomaly using photometry from \citet{Mar08,Currie2011} converted from MKO into 2MASS filters using the conversion from \citet{Stephens2004}, 
we plot HR 8799b alongside field and low-gravity brown dwarfs on the same inclination vs. colour anomaly plot in Fig.~\ref{fig:colouranomaly}.  The colour anomaly is consistent with a wide range of possible inclinations for this planet.

However, assuming the planets do share the inclination of the disk and ignoring additional attenuation from the atmospheres of the planets themselves implies that the intrinsic variability amplitude for the HR 8799 planets will be at least a factor of 2.2 greater than any measured variability amplitude.  Thus, if the planets are observed nearly pole-on, we are sensitive to {\bf intrinsic} variability amplitudes of 11-22$\%$ for HR 8799b, and $>$55$\%$ for HR 8799c.  Assuming that VHS 1256-1257ABb and PSO J318.5-22 are observed nearly equator-on (hence measured variability amplitude is nearly the same as the intrinsic variability amplitude), we could potentially have detected variability if HR 8799b was a VHS 1256-1257ABb analogue, but a PSO J318.5-22 analogue would have remained undetected (as we found from our simulated planet tests).  Taking atmospheric attenuation into account as well (second term in Equation~\ref{eqn:amplitude}) further reduces the measured variability amplitude as a function of decreasing inclination.  This term is sensitive to the wavelength being monitored; we expect light at near-IR wavelengths to be produced deeper in a given object than light at mid-IR wavelengths, thus, near-IR light will traverse a longer path length and will be more attenuated as a function of inclination.

While the observed photometric variability for brown dwarfs and planetary mass objects to date is consistent with inhomogeneous top-of-atmosphere features (e.g. quasiperiodic variability that evolves on timescales of a few rotational periods), a transiting companion to an exoplanet would provide another potential source of variability.
The geometry of the HR 8799 system is not ideal for this scenario -- assuming the planet and the companion to the planet are coplanar and have inclinations similar to that of the system, observable transits would be unlikely for this face-on system.  However, a companion to one of the HR 8799 planets with an orbit inclined perpendicularly to the inclination of the system could potentially produce transits in this system.  Lazzoni et al. in prep simulate transits due to companions to known planets for a range of planetary systems, focusing primarily on the edge-on $\beta$ Pic system.  For HR 8799b, simulating an ensemble of satellites around HR 8799 b with masses in the range [0.001,0.39]$\times$M$_{Jup}$ and semi-major axis in the range [0.001,$R_h$], where $R_h$ is the Hill radius of the planet yields numerous transit events with depths greater than 10$\%$.  However, the transit durations are often quite long -- up to tens of hours, with only a minority of simulated transits shorter than 10 hours. Thus, it is likely that ground-based observations with durations $<$10 hours, such as those reported here, would simply be too short to realistically detect such a transit.

\section{Summary and Conclusions}

Here we summarize the main results of this study:

\begin{enumerate}
    \item The satellite spots were well-illuminated ($\sim$2000 counts at peak) in all epochs except 2017-10-08 ($\sim$200 counts at peak).  For these well-illuminated epochs, background subtraction did not noticeably affect the mean satellite spot lightcurve, as the background level was so much lower than the satellite spot peak.  For 2017-10-08, subtracting the background noticeably detrended the mean satellite spot lightcurve.
    \item The degree of excursion (maximum vs. minimum satellite spot lightcurve value) varies strongly for different epochs of observations -- the August 2018 epochs are considerably more stable and have noticeably smaller ranges of values in the satellite spot lightcurves than the October 2017 epochs.  This leads to overall better sensitivity to variability and improved ability to detrend planet lightcurves.
    \item In contrast to \citet{Apa16}, we do not find a night-by-night stable pattern of correlations / anti-correlations between satellite spots.  In particular, for our best datasets (August 2018), we find low levels of correlation / anti-correlation between satellite spots, because the satellite spot photometry is very stable on those nights and do not cover a wide range of values (in other words, the satellite spot lightcurves are quite flat in these epochs, so do not appear very correlated at all).
    \item Ensuring 5$\times$FWHM sky rotation at the exoplanet radius to optimize speckle suppression but minimize self-subtraction, our $\sim$4.5 hour observations allow 6 equal parallactic angle temporal bins for HR 8799b, 5 such bins for HR 8799c, but only 3 for HR 8799de.  
    \item Planet photometry / astrometry for HR 8799bc was retrieved using negative simulated planet subtraction, inserting scaled unsaturated images of the star at the $\Delta$mag, position of each planet in each raw data frame, for 6 temporal bins for b and 5 temporal bins for c.  As HR 8799b is sufficiently far from the core of the speckle pattern, we were able to also retrieve direct aperture photometry for this planet.  In all epochs, HR 8799b aperture photometry and HR 8799bc negative simulated planet photometry share similar trends within uncertainties, except for the final temporal bin on 2018-08-18, where HR 8799c appears bright compared to HR 8799b.  This is likely due to uncorrected speckle noise at the radius of HR 8799c in this temporal bin, and is recreated in simulated planet injection tests using the satellite spots as the noise model.  The fact that HR8799bc generally show the same trends means that they are appropriate to use to detrend each other, to search for non-shared variations.
    \item Satellite spot lightcurves share the same trends as the planet lightcurves in the August 2018 epochs, however, the planets diverge significantly from the lightcurve trends in the October 2017 epochs.  The divergence may be due to the much redder colors of the planets relative to the satellite spots (which should share the spectrum of the star).  This means that the satellite spot lightcurves are only appropriate to detrend planet lightcurves for the August 2018 epochs, not the October 2017 epochs. 
    \item We found that the number of PCA modes removed had no real effect on our ability to retrieve photometry for HR 8799b, but HR 8799c is significantly affected by the number of modes subtracted, especially in the first and last temporal bin (also when conditions were worst / airmasses were highest).  For HR 8799c, removing 16 modes oversubtracts light from the exoplanet, but with 4 modes removed, obvious speckle noise remains in the residual images, thus we chose to subtract 10 modes as the best compromise.
    \item Considering results for HR 8799de using 3 temporal bins, these planets appear to follow satellite spot trends to within error bars, but too much speckle noise remains at these radii to meaningfully constrain variability to high precision.  
    \item Since the satellite spot lightcurves are not appropriate to detrend planet lightcurves at all epochs, we consider $\Delta(mag)_{b} - \Delta(mag)_{c}$ as a function of time during each observation -- essentially detrending the b lightcurve using that of c, or vice versa.  This traces non-shared variations between the two planets.  We used five temporal bins, to assure 20 pixels of movement on the sky for HR 8799b and c, with difference lightcurves presented in Fig.~\ref{fig:bcdeltadeltamag}.  Within errors, our results are in agreement with those from \citet{Apa16}.  The uncertainty on the measurement is dominated by the uncertainty in our measurement of $\Delta(mag)_{c}$. Depending on epoch, we rule out non-shared variability in $\Delta(mag)_{b} - \Delta(mag)_{c}$ to the 10-20$\%$ level over 4-5 hours.  
    \item To fully quantify our sensitivity to variability in the 2018-08-18 epoch (the best quality data of our search), we simulate variable and non-variable lightcurves by inserting and retrieving a suite of simulated planets at a similar radius from the star as HR 8799bc, but offset in position angle from the true planet position angle.  Simulated planets are assumed to have sinusoidal lightcurves, with noise estimated using the mean satellite spot lightcurve. We ran a complete grid of simulated planet lightcurves with periods from 2 to 20 hours, amplitudes from 3\% to 30\%, and phases from 0$^{\circ}$ to 330$^{\circ}$.  For HR 8799b, for periods $<$10 hours, we are generally quite sensitive to variability with amplitude $>5\%$.  For HR 8799c, our sensitivity is much lower, limited to variability $>25\%$ for similar periods.  Taking into account that these planets are likely observed nearly pole-on (inclination $<$27$^{\circ}$) and correcting for geometric dilution, these measured limits on variability amplitudes correspond to sensitivity to {\bf intrinsic} variability amplitudes of 11-22$\%$ for HR 8799b, and $>$55$\%$ for HR 8799c for periods of $<$10 hours, in the broadband H filter.
    \item  We derive epoch-by-epoch astrometry and photometry, and compare to literature values, using the data frames from the hour approaching meridian crossing and the hour after meridian crossing, thus covering the time period during each observation with the most field rotation and best airmass.  We again ran our full negative simulated planet code to retrieve the best $\Delta$mag, radius, and position angle value for each planet.  Our epoch-by-epoch photometry agrees within uncertainties to literature values which employ similar negative simulated planet subtraction techniques.  Unfortunately, the uncertainties inherent in converting from relative photometry within an epoch to $\Delta$(mag) for an entire epoch (i.e. variability of the star, error on the brightness of the star) prevent an epoch-by-epoch search for variability.
\end{enumerate}
While 8-m telescopes with extreme AO systems have the sensitivity to detect HR 8799bc with high S/N in a reasonably small amount of observing time, for the inner planets, the varying speckle noise floor effectively precluded variability studies at the required accuracy for a plausible detection.  This changing speckle background is driven by the fundamental instability of working from the ground; we must contend with light that has already traversed the atmosphere of the Earth and is only imperfectly corrected by our adaptive optics system. The limits of ground-based studies are also apparent in variability studies of field brown dwarfs and isolated planetary mass objects; ground-based variability monitoring is only sensitive to variations of $>$2-3$\%$ \citep{Rad14a, Vos2019}, whereas space-based monitoring with telescopes such as HST and Spitzer enable sensitivity down to variations of $>$0.1-0.5$\%$ \citep{Met15, Apa13}. However, existing space platforms such as HST do not have coronagraphs that can attain the contrasts or small inner working angles necessary for similar studies of close-companions such as the HR 8799 planets.  The upcoming James Webb Space Telescope (JWST) combines both the stability of a space platform with high-contrast imaging capability from a suite of coronagraphs and will enable much higher sensitivity to variability in these planets than we were able to achieve even in excellent conditions with VLT-SPHERE; thus, in the next few years, JWST will likely transform our understanding of the variability of young, directly-imaged giant exoplanet atmospheres. 

\section*{Data Availability Statement}
The raw data underlying this article were accessed from the ESO archive facility at http://archive.eso.org/cms.html and processed at the SPHERE Data Centre, jointly operated by OSUG/IPAG (Grenoble), PYTHEAS/LAM/CESAM (Marseille), OCA/Lagrange (Nice), Observatoire de Paris/LESIA (Paris), and Observatoire de Lyon. The derived data generated in this research will be shared on reasonable request to the corresponding author.

\section*{Acknowledgements}
We thank the anonymous referee for useful suggestions which improved this work.  B.B acknowledges funding by the UK Science and Technology Facilities Council (STFC) grant no.
ST/M001229/1.  J.M.V. acknowledges support by NSF Award Number 1614527 and Spitzer Cycle 14 JPL Research Support Agreement 1627378.
Support for this work was provided by NASA through the NASA Hubble Fellowship grant HST-HF2-51472.001-A awarded by the Space Telescope Science Institute, which is operated by the Association of Universities for Research in Astronomy, Incorporated, under NASA contract NAS5-26555.  A.Z. acknowledges support from the FONDECYT Iniciaci\'on en investigaci\'on project number 11190837 
Based on observations collected at the European Organisation for Astronomical Research in the Southern Hemisphere under ESO programmes 095.C-0689(A), 099.C-0588(A), and 0101.C-0315(A).
This work has made use of the SPHERE Data Centre, jointly operated by OSUG/IPAG (Grenoble),
PYTHEAS/LAM/CESAM (Marseille), OCA/Lagrange (Nice), Observatoire de Paris/LESIA (Paris), and Observatoire de Lyon.  
This research made use of Astropy,\footnote{http://www.astropy.org} a community-developed core Python package for Astronomy \citep{astropy:2013, astropy:2018}.  This research made use of Photutils, an Astropy package for
detection and photometry of astronomical sources (Bradley et al. 2019).

\bibliographystyle{mnras}
\bibliography{main}

\begin{center}
\begin{table*}
\begin{minipage}{160mm}
\begin{tabular}{ c | c | c | c | c | c | c | c | c | c }
Date & MJD at start  & filter set & obs. length & field rotation & seeing & airmass & median Strehl & DIT & NEXP \\ 
(UTC) & of observation &  & (hours) & (degrees) & (arcsec) & & & (s) &  \\ \hline
2015-07-30 & 57233.2178 & J2/J3 & 4.99 & 83.4 & 0.96 &  1.43 - 2.04 & 0.83 & 16 & 960 \\
2015-07-31 & 57234.1999 & K1/K2 & 5.96 & 90.9 & 1.30 & 1.43 - 2.54 & 0.69 & 16 & 1121 \\\hline
2017-10-08 & 58033.9927 & BB-H & 4.70 & 76.3 & 0.64 & 1.43 - 2.12 & 0.83 & 16 & 476 \\
2017-10-12 & 58038.0299 & BB-H & 4.50 & 75.8 & 0.60 & 1.43 - 2.03 & 0.87 & 16 & 882 \\ 
2017-10-13 & 58038.9936 & BB-H & 4.91 & 80.0 & 0.67 & 1.43 - 1.90 & 0.85 & 8 & 1856 \\\hline
2018-08-18 & 58348.1795 & BB-H & 4.50 & 73.9 & 0.65 & 1.43 - 2.02 & 0.89 & 16 & 896 \\
2018-08-20 & 58350.1367 & BB-H & 4.39 & 72.6 & 0.35 & 1.43 - 1.99 & 0.90 & 4 & 896 \\\hline
\hline
\end{tabular}
\caption{Log of IRDIS Observations \label{tab:observations}}
\end{minipage}
\end{table*}
\end{center}

\begin{center}
\begin{table*}
\begin{minipage}{160mm}
\begin{tabular}{ |c| c | c | c |}
Epoch & Planet & $\Delta$RA (mas) & $\Delta$Dec (mas)  \\ \hline
17-10-08 & HR8799b & 1616.72$\pm$2.81 & 598.78$\pm$2.74\\
17-10-08 & HR8799c & -414.39$\pm$2.72 & 852.59$\pm$3.01\\
17-10-08 & HR8799d & -492.29$\pm$4.72 & -461.70$\pm$4.61\\
17-10-08 & HR8799e & -363.98$\pm$4.74 & 140.31$\pm$3.91\\ \hline
17-10-12 & HR8799b & 1615.61$\pm$2.73 & 597.93$\pm$2.77\\
17-10-12 & HR8799c & -414.72$\pm$2.60 & 851.83$\pm$2.83\\
17-10-12 & HR8799d & -493.82$\pm$3.92 & -461.12$\pm$3.85\\
17-10-12 & HR8799e & -363.87$\pm$6.79 & 139.06$\pm$4.73\\ \hline
17-10-13 & HR8799b & 1614.33$\pm$3.01 & 598.67$\pm$2.74\\
17-10-13 & HR8799c & -415.03$\pm$2.69 & 852.03$\pm$2.94\\
17-10-13 & HR8799d & -495.43$\pm$3.51 & -463.32$\pm$3.45\\
17-10-13 & HR8799e & -366.11$\pm$7.27 & 140.13$\pm$5.12\\ \hline
18-08-18 & HR8799b & 1619.67$\pm$3.05 & 582.29$\pm$2.67\\
18-08-18 & HR8799c & -394.05$\pm$3.31 & 862.52$\pm$3.72\\
18-08-18 & HR8799d & -515.20$\pm$4.53 & -444.86$\pm$4.35\\ 
18-08-18 & HR8799e & -351.91$\pm$4.29 & 172.60$\pm$3.51\\ \hline
18-08-20 & HR8799b & 1619.21$\pm$2.87 & 583.23$\pm$2.90\\
18-08-20 & HR8799c & -394.53$\pm$2.79 & 863.73$\pm$3.15\\
18-08-20 & HR8799d & -515.60$\pm$3.27 & -443.16$\pm$3.15\\
18-08-20 & HR8799e & -353.08$\pm$4.13 & 174.17$\pm$3.47\\ \hline
\hline
\end{tabular}
\caption{Epoch-by-Epoch Astrometry\label{tab:astrometry}}
\end{minipage}
\end{table*}
\end{center}

\begin{center}
\begin{table*}
\begin{minipage}{160mm}
\begin{tabular}{ |c| c | c | c |}
Epoch & Planet & H$_{abs} (mag)$  \\ \hline
17-10-08 & HR8799b & 15.07$\pm$0.10 \\
17-10-08 & HR8799c & 14.06$\pm$0.11 \\
17-10-08 & HR8799d & 14.11$\pm$0.13 \\
17-10-08 & HR8799e & 13.88$\pm$0.13 \\ \hline
17-10-12 & HR8799b & 15.05$\pm$0.10 \\
17-10-12 & HR8799c & 14.01$\pm$0.11 \\
17-10-12 & HR8799d & 14.05$\pm$0.12 \\
17-10-12 & HR8799e & 13.84$\pm$0.14 \\ \hline
17-10-13 & HR8799b & 14.98$\pm$0.11 \\
17-10-13 & HR8799c & 13.99$\pm$0.11 \\
17-10-13 & HR8799d & 14.08$\pm$0.12 \\
17-10-13 & HR8799e & 13.79$\pm$0.15 \\ \hline
18-08-18 & HR8799b & 15.05$\pm$0.11 \\
18-08-18 & HR8799c & 14.01$\pm$0.11 \\
18-08-18 & HR8799d & 14.12$\pm$0.13 \\
18-08-18 & HR8799e & 13.90$\pm$0.12 \\ \hline
18-08-20 & HR8799b & 15.12$\pm$0.10 \\
18-08-20 & HR8799c & 14.05$\pm$0.11 \\
18-08-20 & HR8799d & 14.14$\pm$0.11 \\
18-08-20 & HR8799e & 13.90$\pm$0.12 \\ \hline
\hline
\end{tabular}
\caption{Epoch-by-Epoch Photometry\label{tab:photometryabsmag}}
\end{minipage}
\end{table*}
\end{center}

\begin{center}
\begin{table*}
\begin{minipage}{160mm}
\begin{tabular}{ |c| c | c | c |}
Epoch & b -c (mag) & b - d (mag) & b - e (mag) \\ \hline
17-10-08 & 1.01$\pm$0.16 & 0.97$\pm$ 0.18 & 1.20$\pm$0.18 \\
17-10-12 & 1.04$\pm$0.16 & 1.00$\pm$ 0.17 & 1.21$\pm$0.19 \\
17-10-13 & 1.00$\pm$0.19 & 0.91$\pm$ 0.20 & 1.20$\pm$0.22 \\
18-08-18 & 1.05$\pm$0.19 & 0.94$\pm$ 0.20 & 1.15$\pm$0.20 \\
18-08-20 & 1.08$\pm$0.17 & 0.99$\pm$ 0.18 & 1.23$\pm$0.18 \\
\hline
\end{tabular}
\caption{Epoch-by-Epoch subtracted Photometry\label{tab:photometrysub}}
\end{minipage}
\end{table*}
\end{center}

\appendix

\section{Satellite spot photometry}
\label{app:ssphot}

\begin{figure*}
\includegraphics[scale=0.4]{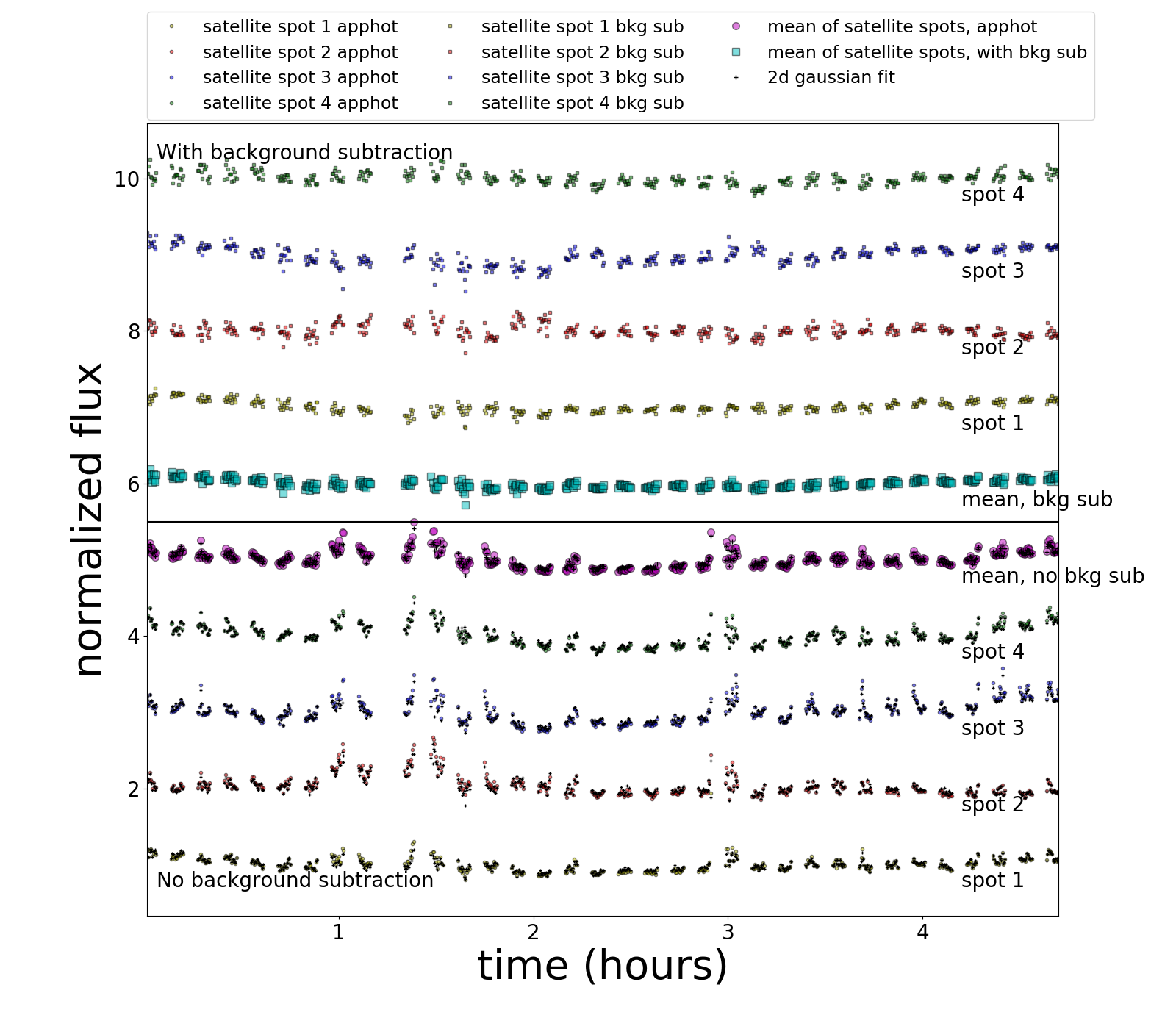}
\caption{Satellite spot photometry for 2017-10-07 (normalized
by the median value across the whole observation for each
satellite spot) as a function of time for each satellite spot, with
an offset in flux included between each satellite spot lightcurve for clarity.
Results with background subtraction are shown on the top half of the plot, while results without 
background subtraction are plotted on the bottom half of the plot.
The amplitudes from a 2-d gaussian fit to each satellite spot are plotted as 
black crosses, elliptical aperture photometry is plotted
as filled circles, and elliptical aperture photometry with an 
additional background subtraction is plotted as filled squares.  
The mean of the elliptical aperture photometry for 
all 4 satellite spots are plotted in the middle of the 
figure, with magenta circles for the elliptical aperture photometry 
without background subtraction, and cyan square for elliptical 
aperture photometry with background subtraction.
}
\label{fig:SS_Oct7}       
\end{figure*}

\begin{figure*}
\includegraphics[scale=0.4]{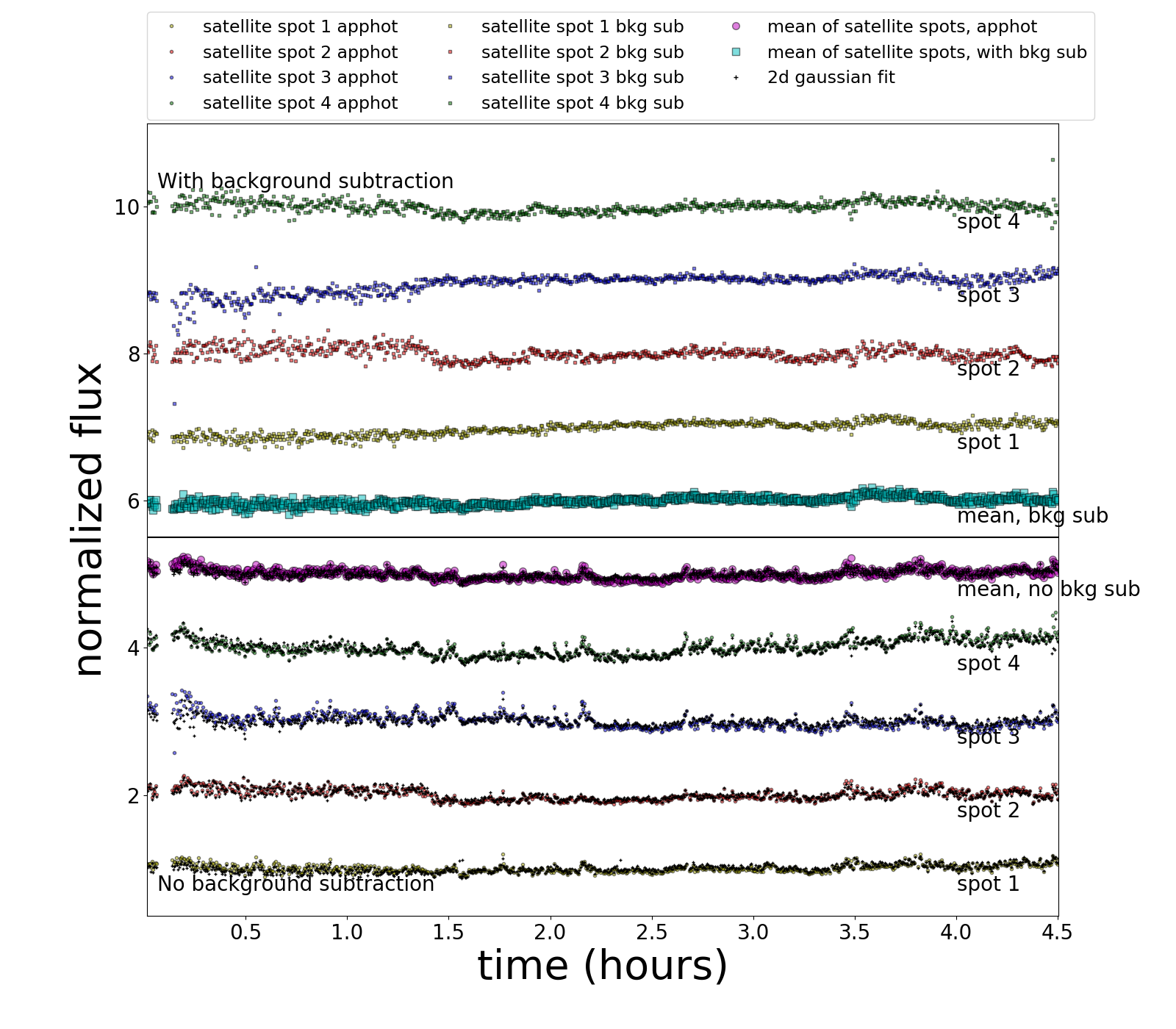}
\caption{Satellite spot photometry for 2017-10-12, plot axes and labels 
are the same as in Fig.~\ref{fig:SS_Oct7}.
}
\label{fig:SS_Oct10}       
\end{figure*}

\begin{figure*}
\includegraphics[scale=0.4]{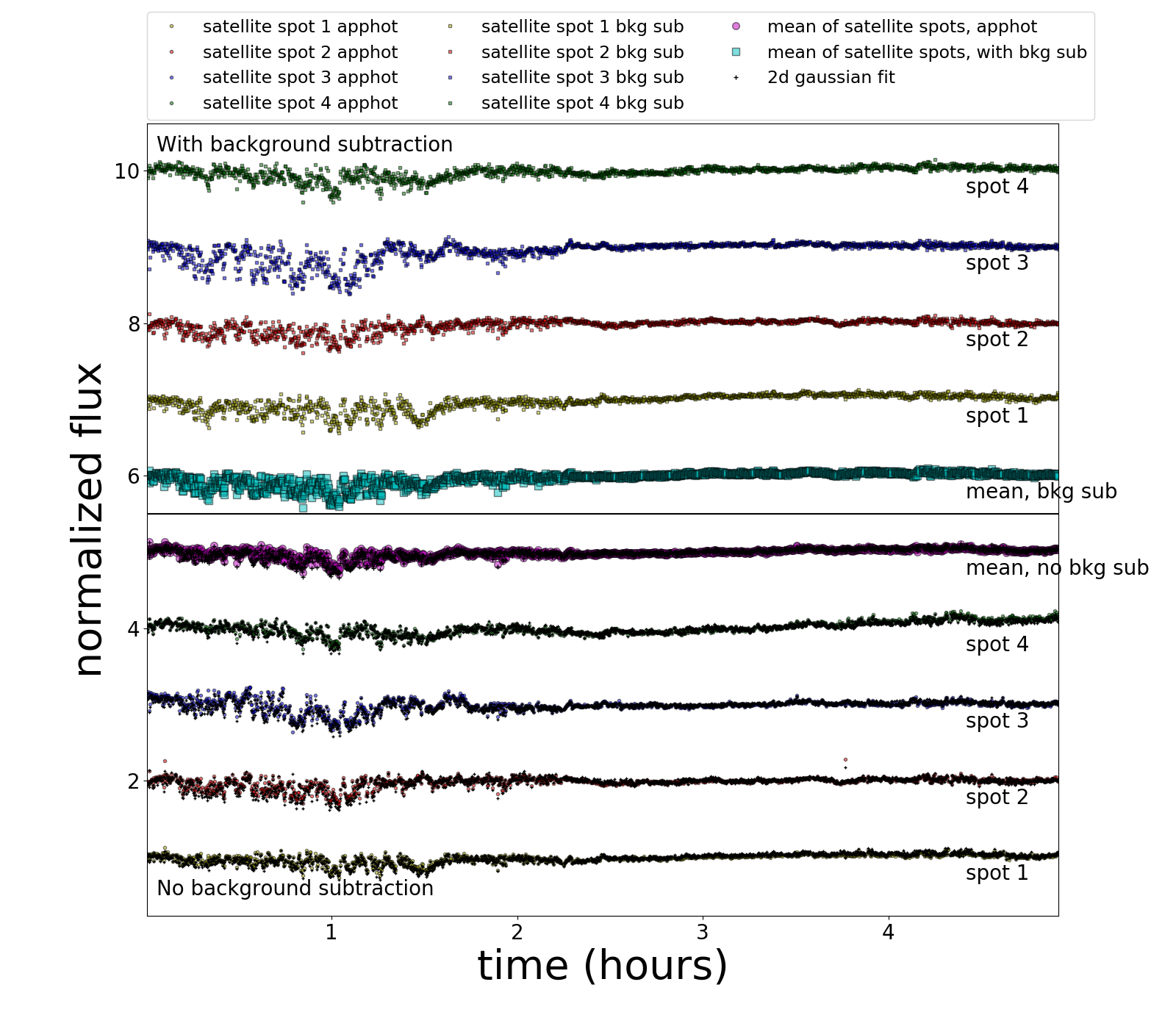}
\caption{Satellite spot photometry for 2017-10-13, plot axes and labels 
are the same as in Fig.~\ref{fig:SS_Oct7}.
}
\label{fig:SS_Oct13}       
\end{figure*}

\begin{figure*}
\includegraphics[scale=0.4]{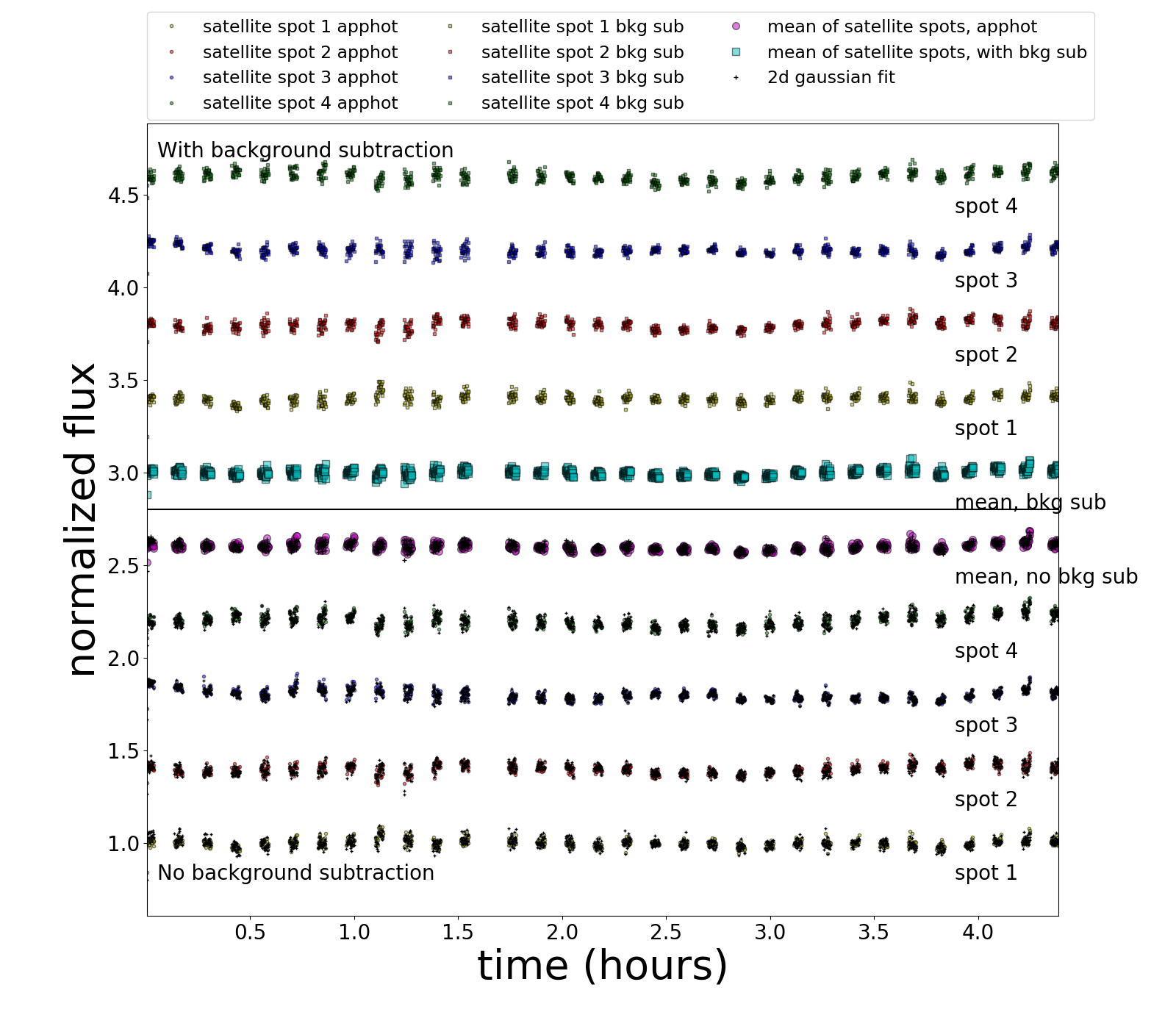}
\caption{Satellite spot photometry for 2018-08-20, plot axes and labels 
are the same as in Fig.~\ref{fig:SS_Oct7}.
}
\label{fig:SS_Aug20}       
\end{figure*}

We present here satellite spot lightcurves for the epochs of 2017-10-07, 2017-10-12, 2017-10-13, and 2018-08-20
(Figs.~\ref{fig:SS_Oct7} through~\ref{fig:SS_Aug20}).
The satellite spot photometry for the August 2018 epochs is notably more stable and less variable than that for the October 2017 epochs; note the smaller plot range used for the August 2018 epochs.

\section{Effect of conditions on satellite spot photometry}
\label{app:ssphotconditions}

\begin{figure*}
\includegraphics[scale=0.45]{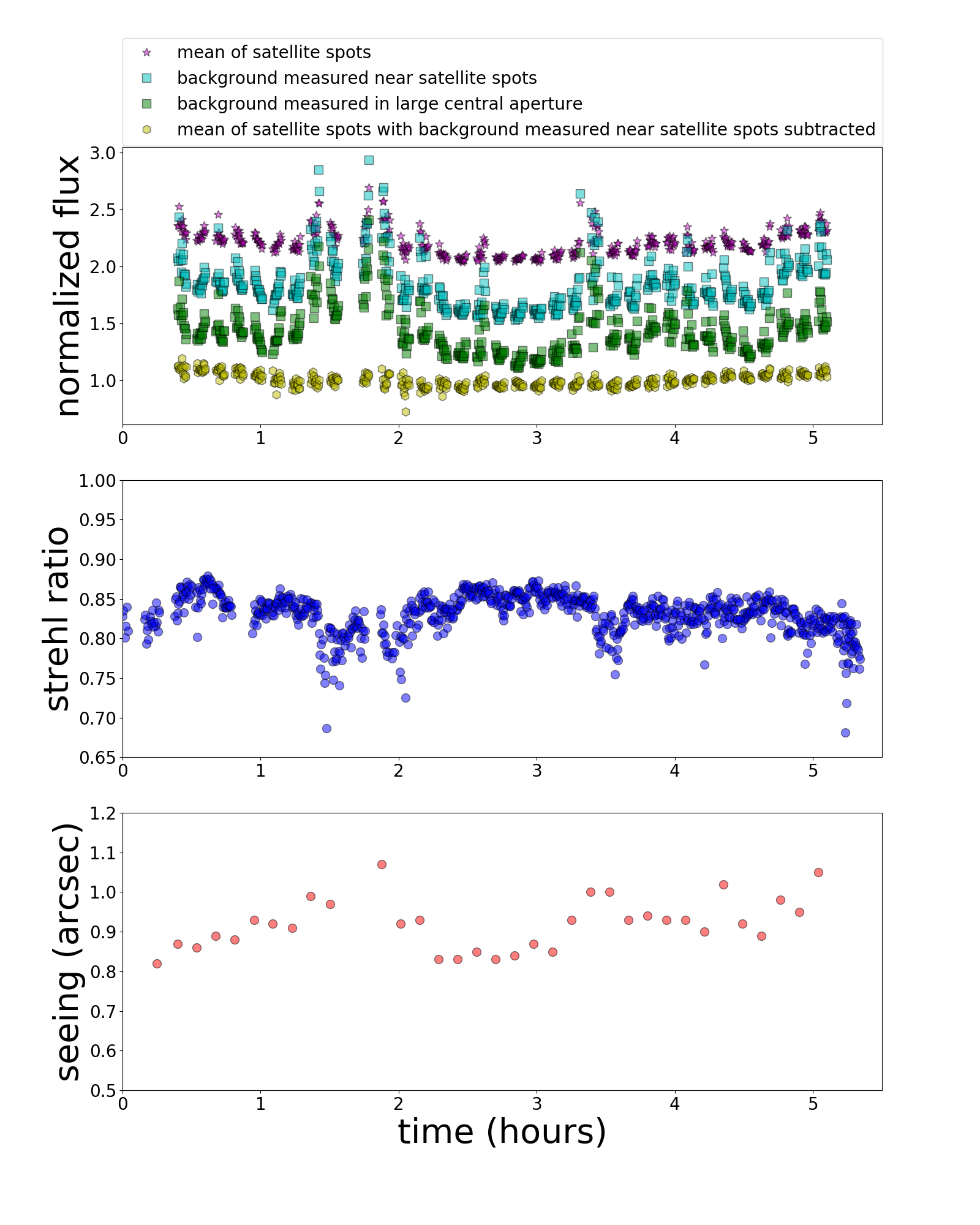}
\caption{Mean satellite spot photometry, background photometry, strehl ratio, and seeing vs. time for 2017-10-08.
}
\label{fig:Strehl_Oct7}       
\end{figure*}

\begin{figure*}
\includegraphics[scale=0.45]{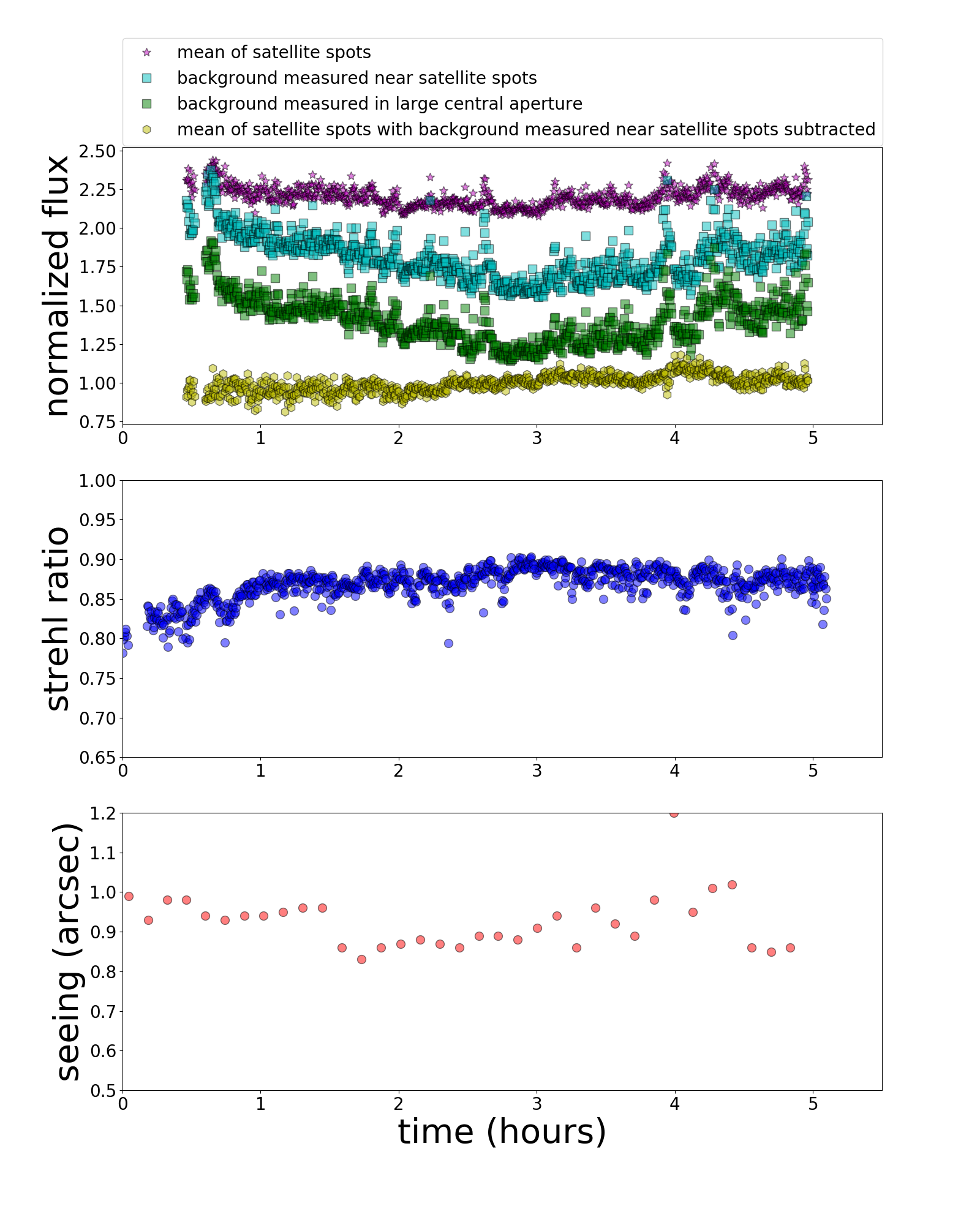}
\caption{Mean satellite spot photometry, background photometry, strehl ratio, and seeing vs. time for 2017-10-12.
}
\label{fig:Strehl_Oct12}       
\end{figure*}

\begin{figure*}
\includegraphics[scale=0.45]{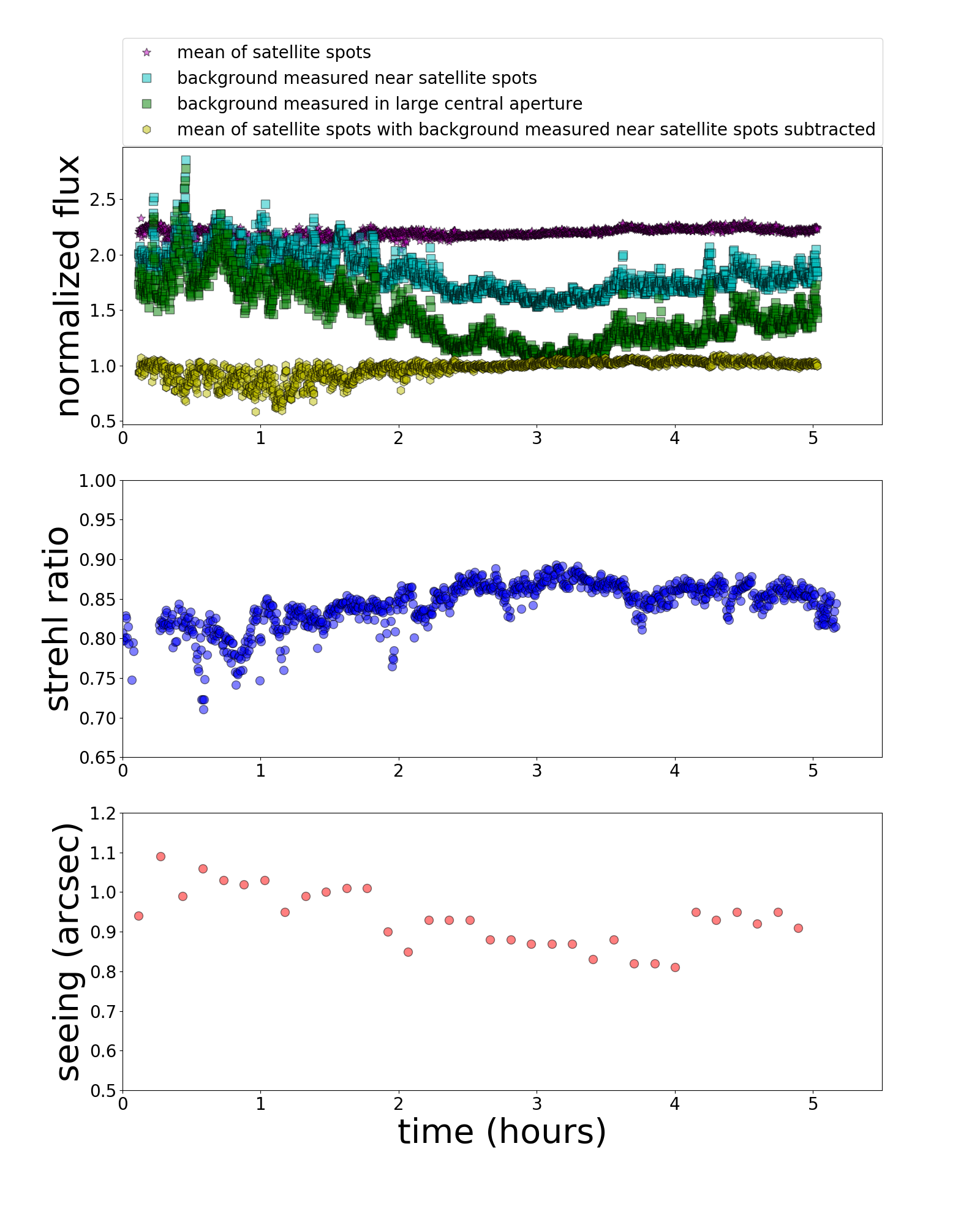}
\caption{Mean satellite spot photometry, background photometry, strehl ratio, and seeing vs. time for 2017-10-13.
}
\label{fig:Strehl_Oct13}       
\end{figure*}

\begin{figure*}
\includegraphics[scale=0.45]{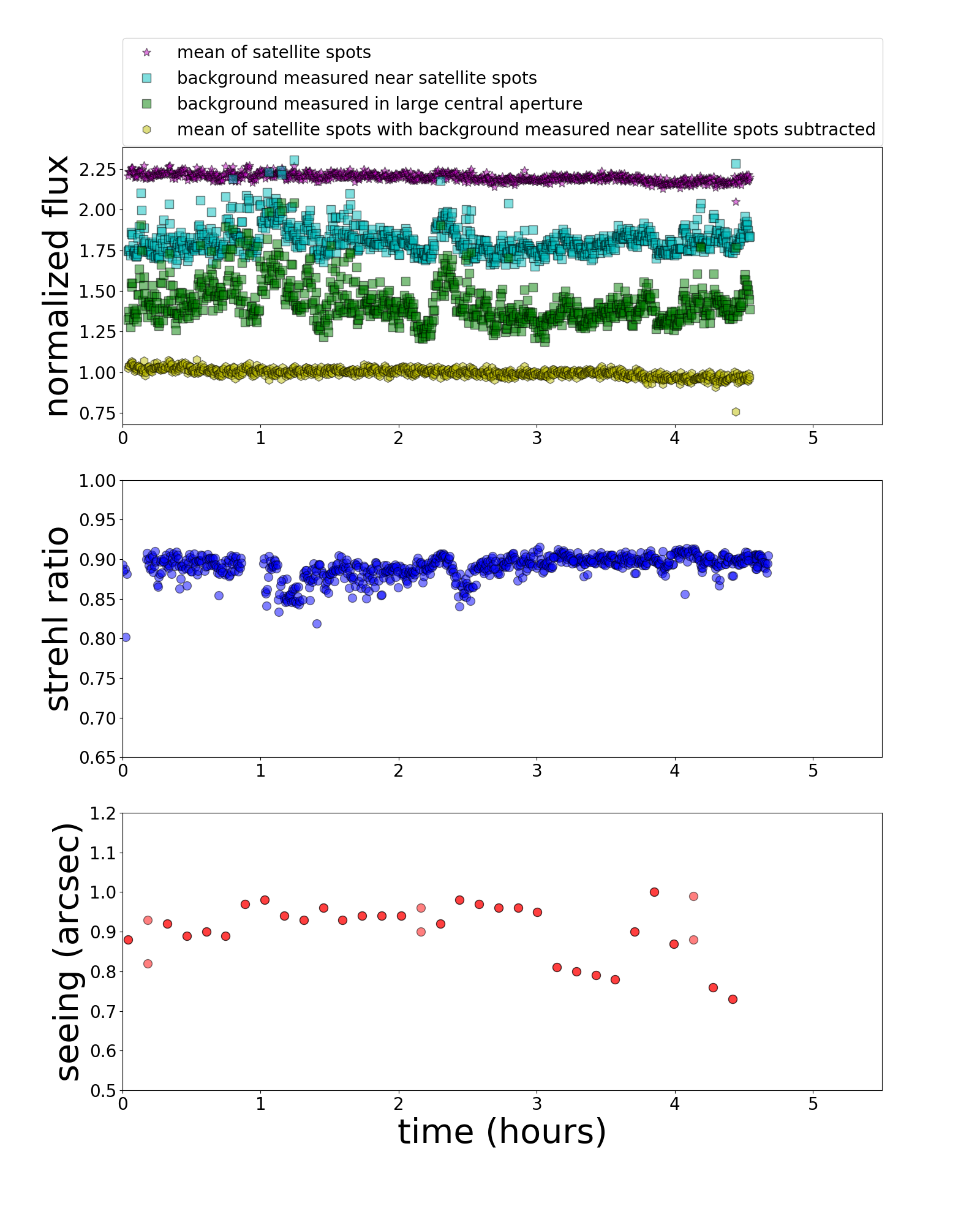}
\caption{Mean satellite spot photometry, background photometry, strehl ratio, and seeing vs. time for 2018-08-18.
}
\label{fig:Strehl_Aug18}       
\end{figure*}

\begin{figure*}
\includegraphics[scale=0.45]{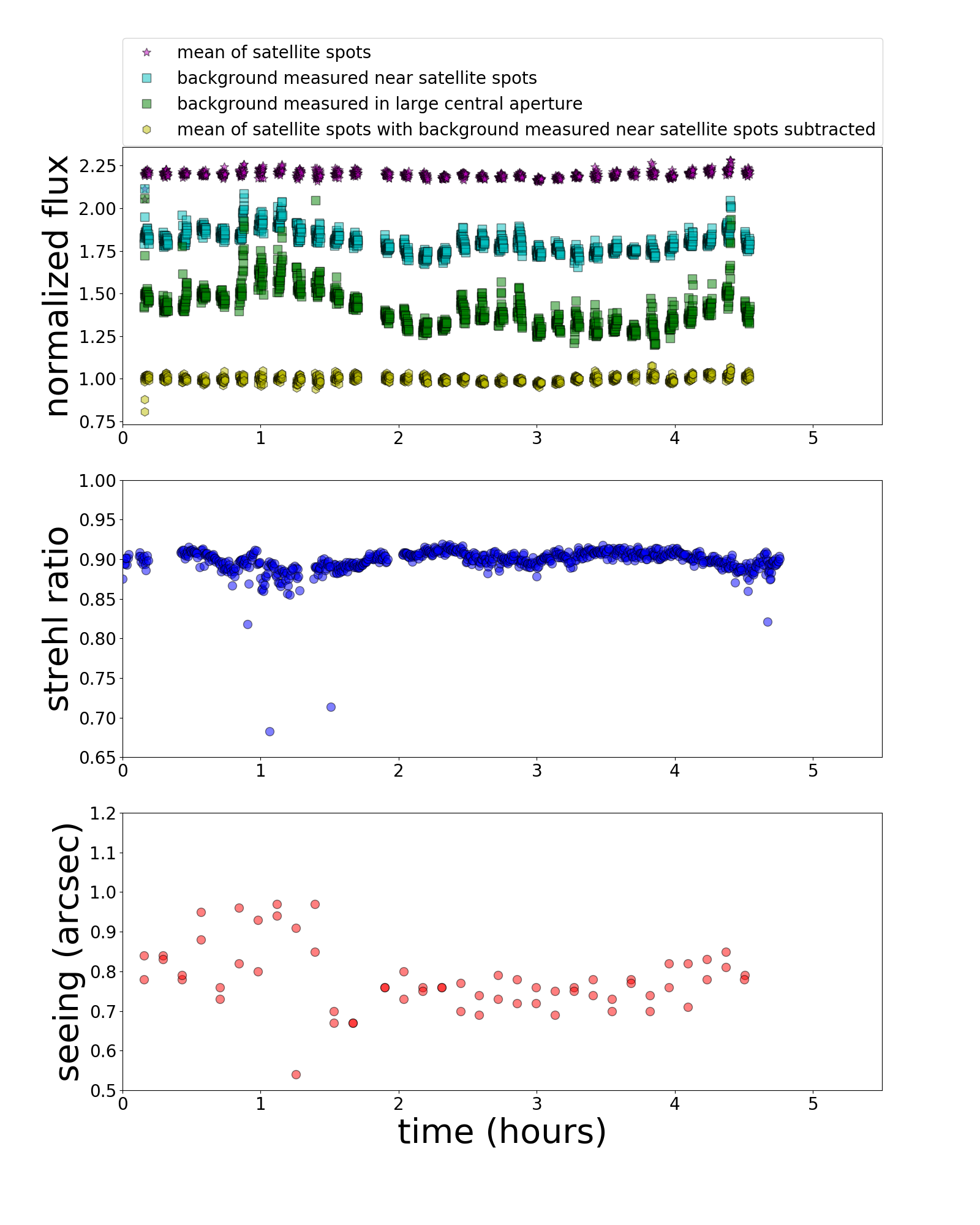}
\caption{Mean satellite spot photometry, background photometry, strehl ratio, and seeing vs. time for 2018-08-20.
}
\label{fig:Strehl_Aug20}       
\end{figure*}

To search for correlations in the satellite spot lightcurves as a function of ambient conditions, we overplot the median flux in the satellite spots, measured background flux, obtained Strehl ratio and seeing for the 2017-10-12 (average conditions) and 2018-08-18 (excellent conditions) epochs in Figs.~\ref{fig:Strehl_Oct7} 
through~\ref{fig:Strehl_Aug20}.  All satellite spot and background lightcurves have been normalized to themselves; the background levels are generally quite a bit lower than the flux in the satellite spots, but here we consider variations relative to the normalization of the component in question.  As expected, background level is anti-correlated with the Strehl ratio; as Strehl ratio decreases, more starlight is lost from the core of the PSF and instead appears as a higher background level.  Seeing is also roughly anti-correlated with Strehl ratio, commensurate with the expectation of poorer adaptive optics correction during poorer seeing conditions.   Other than the 2017-10-08 epoch (when, as noted previously, the satellite spots were under-illuminated), the satellite spot light curves do not appear to be obviously anti-correlated with the background level.  However, this may be because the AO residual background scales as (1 - Strehl ratio), whereas the satellite spots should scale directly with the Strehl ratio.  Thus, for instance, a drop in Strehl ratio from 90\% to 80\% would have a factor of 2 effect on the AO residual background, but only a 10$\%$ effect on the satellite spot brightness. 

\section{Satellite Spot Correlation plots} 
\label{app:sscor}

\begin{figure*}
\includegraphics[scale=0.15]{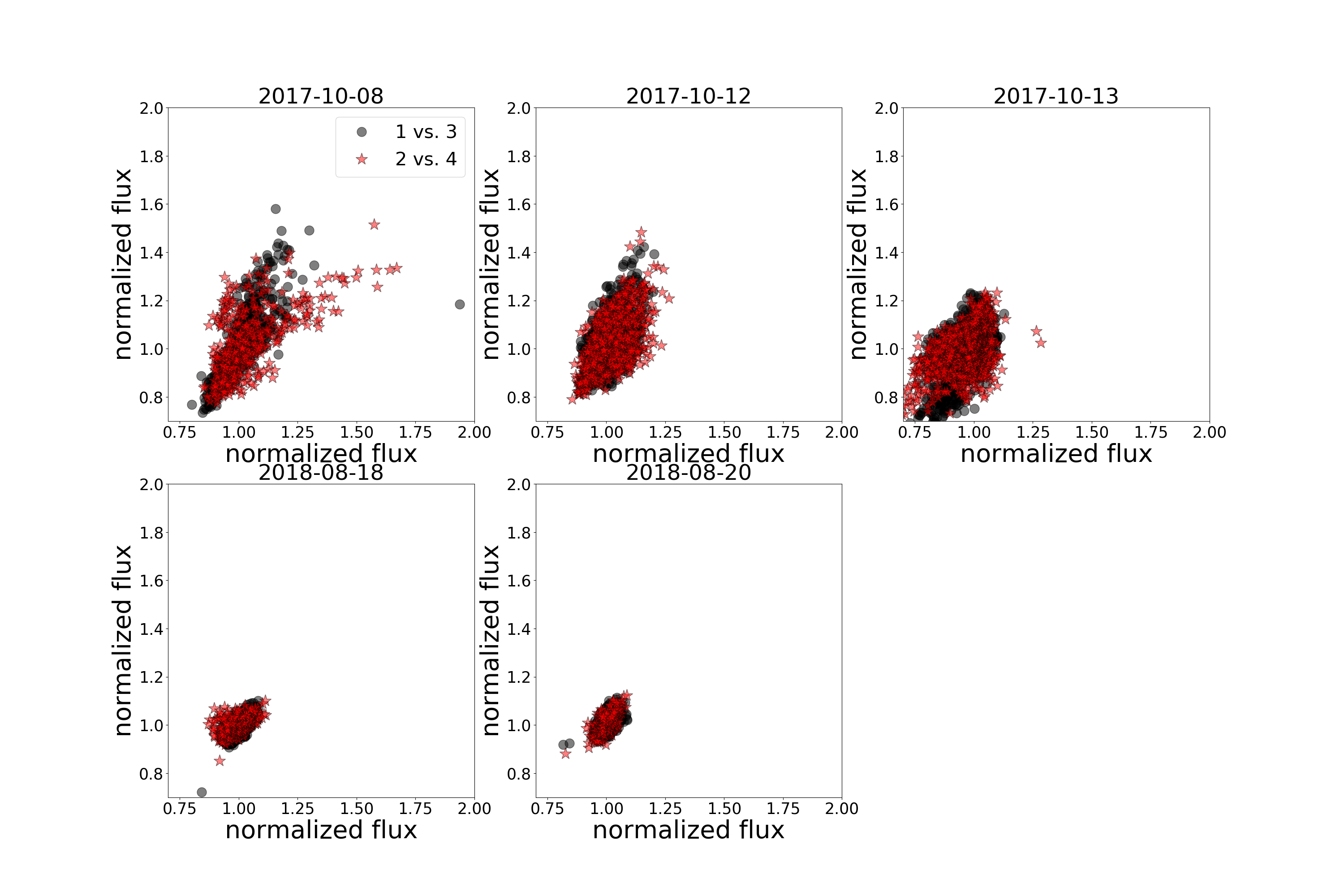}
\caption{Correlation plots for satellite spots across from each other.
}
\label{fig:corr_cross_nomed}       
\end{figure*}

\begin{figure*}
\includegraphics[scale=0.15]{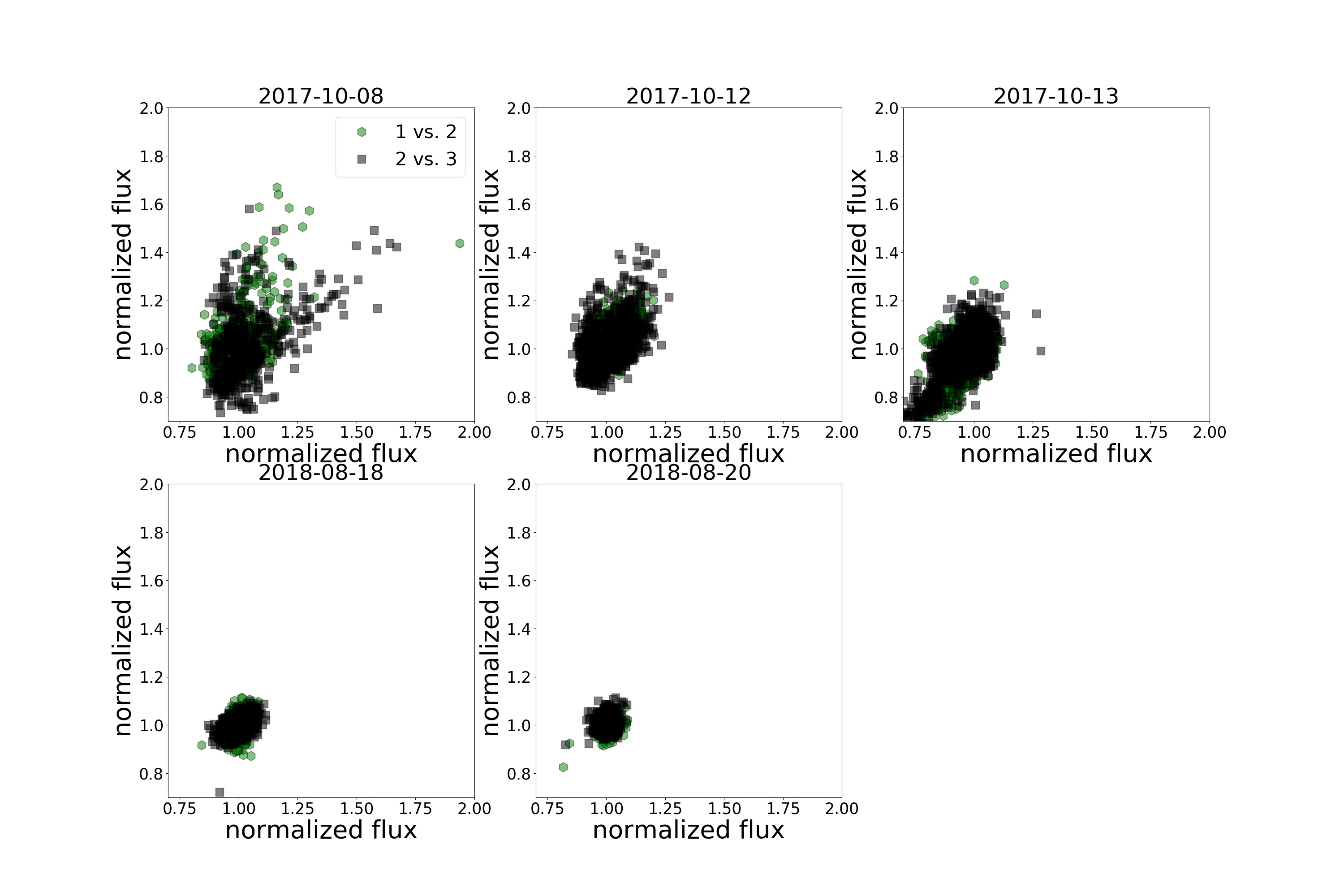}
\caption{Correlation plots for satellite spots on the sides.
}
\label{fig:corr_sides_nomed1}       
\end{figure*}

\begin{figure*}
\includegraphics[scale=0.15]{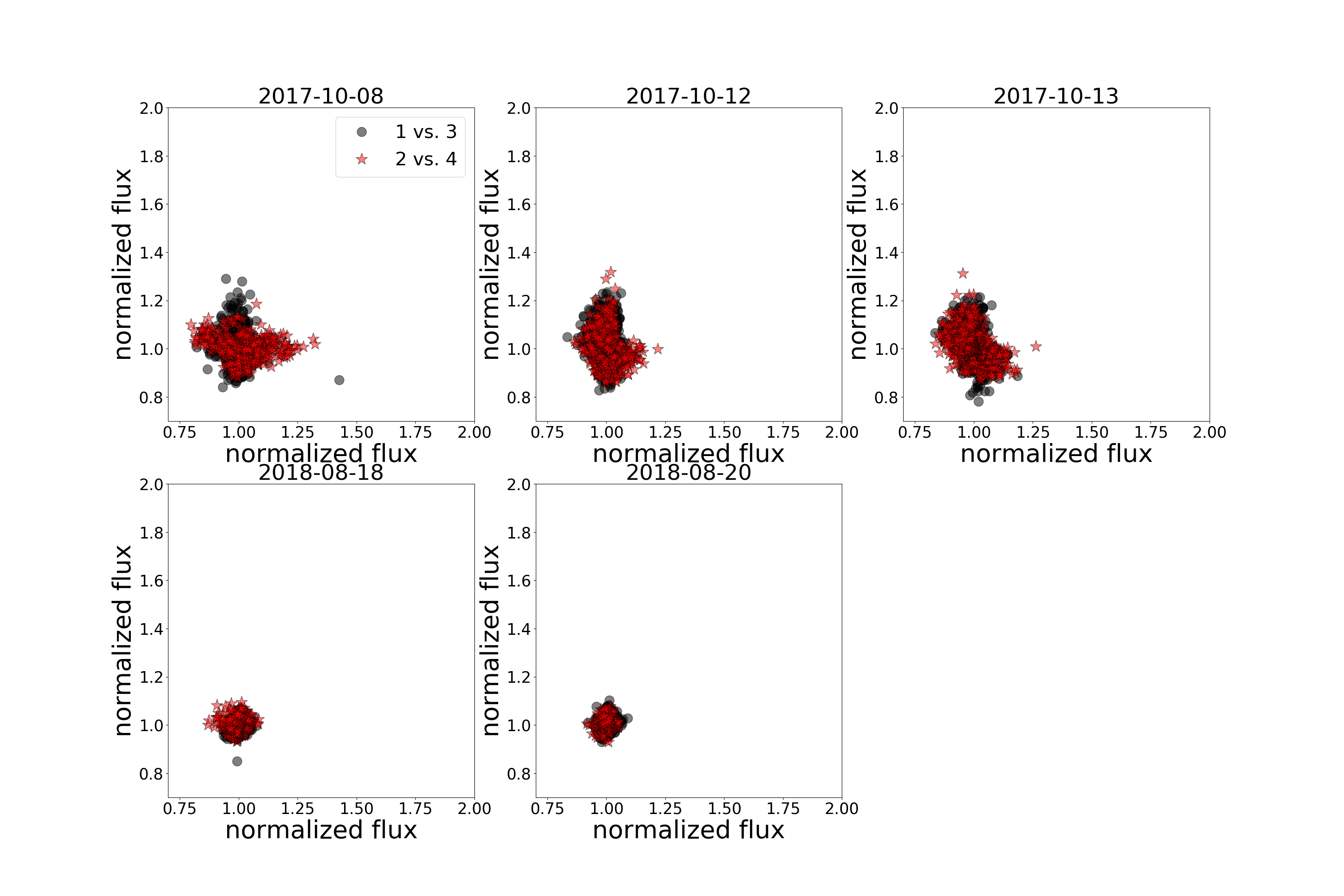}
\caption{Correlation plots for satellite spots across from each other, mean
satellite spot lightcurve removed.
}
\label{fig:corr_cross_med}       
\end{figure*}

\begin{figure*}
\includegraphics[scale=0.15]{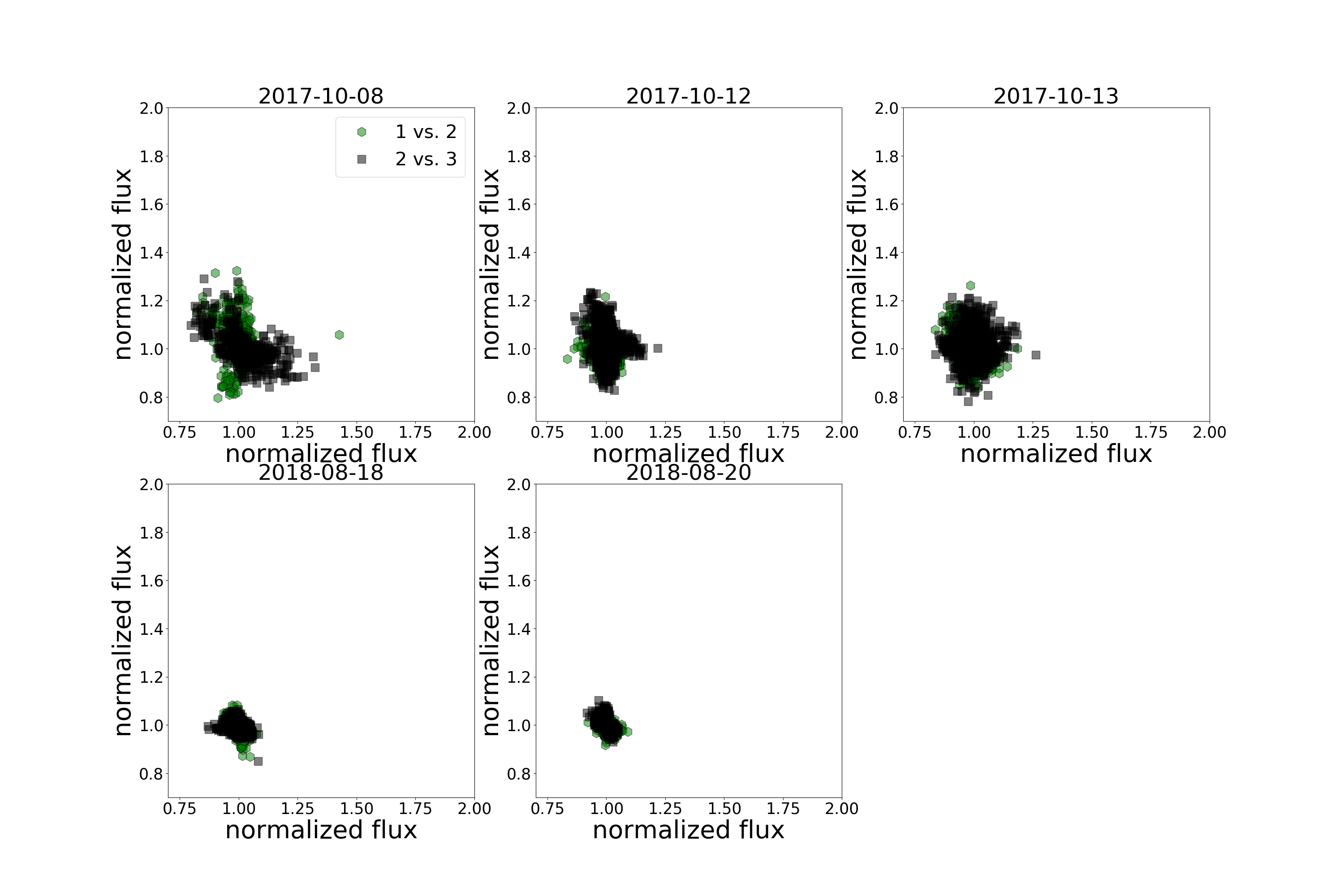}
\caption{Correlation plots for satellite spots on the sides, mean
satellite spot lightcurve removed.
}
\label{fig:corr_sides_med1}       
\end{figure*}

\begin{figure*}
\includegraphics[scale=0.15]{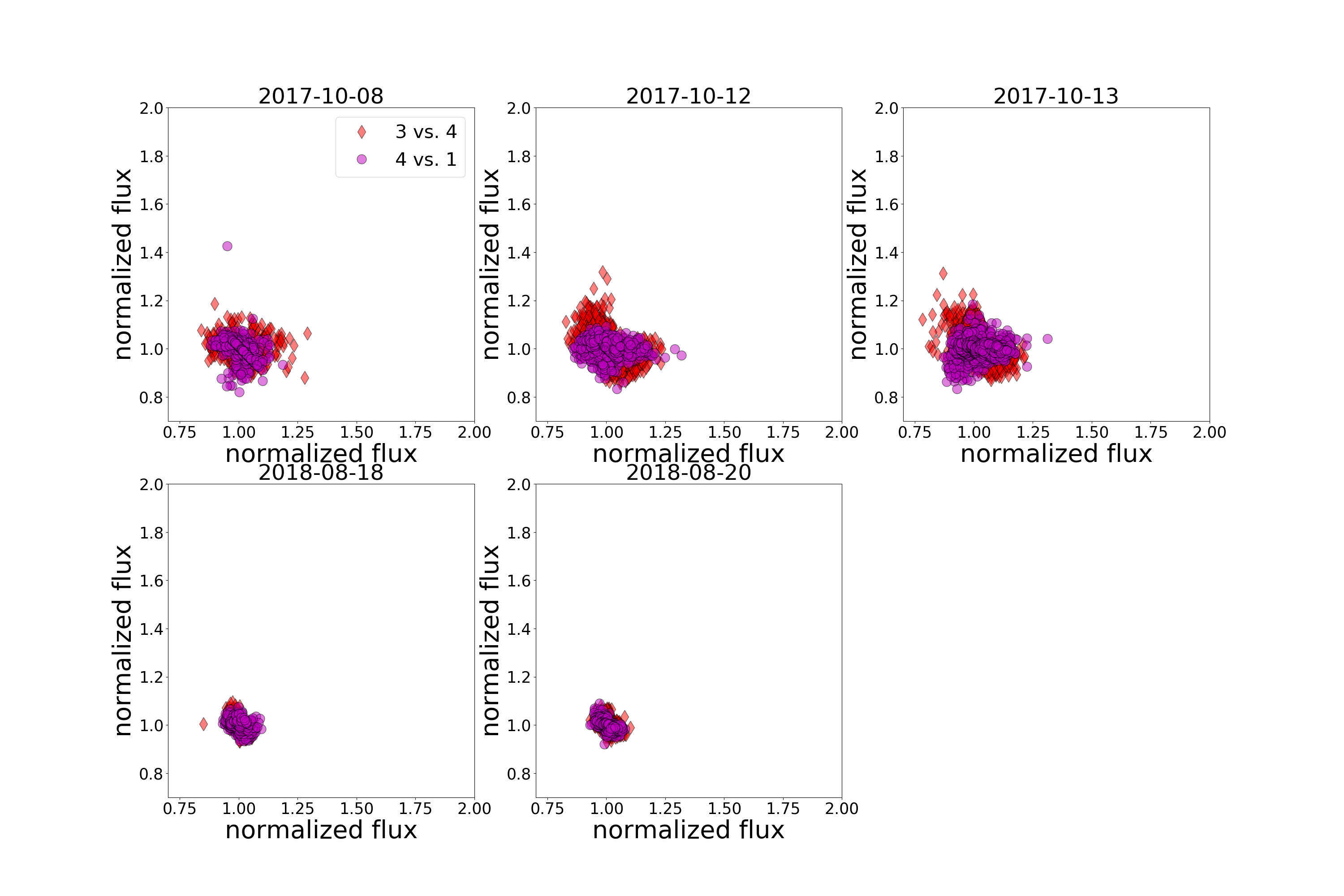}
\caption{Correlation plots for satellite spots on the sides, mean
satellite spot lightcurve removed.
}
\label{fig:corr_sides_med2}       
\end{figure*}

We present additional satellite spot correlation plots.
Correlation plots without the mean lightcurve removed are shown in Fig.~\ref{fig:corr_cross_nomed} and~\ref{fig:corr_sides_nomed1}.  Plots with the mean lightcurve removed are shown in Figures~\ref{fig:corr_cross_med}
through~\ref{fig:corr_sides_med2}.
The goal of measuring satellite spot photometry is to eventually remove trends due to non-astrophysical variability (i.e. changes in AO correction, seeing, flexure) as well as astrophysical trends from the star.  Thus, we measure two different types of trends in our satellite spot data: how much the satellite spots follow a monotonic function between each other (i.e. smooth trends in time due to changes in observing conditions) -- and how much inherent spread is left in the satellite spots after removing common variations. Larger trends with time in the satellite spot data shared between spots follow major changes in Strehl, seeing, coherence time, etc -- better data, such as those from August 2018 show smaller trends over the observation compared to data from October 2017, as conditions were inherently more stable.  Plotting satellite spots against each other then mostly highlights the inherent scatter in the data, as for the best datasets, we covered only a small range of different star spot values.  This produces a smaller Spearman $\rho$ value, even though the data is of better quality.  

On 2018-10-18, and 2018-08-20, we find the strongest anti-correlations between adjacent satellite spots 1-4 and 2-3, weaker anti-correlations between satellite spots 1-2 and 3-4, and no or slight positive correlation between the two opposite sets of satellite spots. On 2017-10-08, we find that adjacent satellite spots 1-4 and 2-3 show relatively strong anti-correlation (but not much correlation between other adjacent satellite spots).  On 2017-10-12, we find anti-correlation between adjacent spots 1-2 and 3-4.  On 2017-10-13, adjacent spots 1-2 and 3-4 are somewhat anti-correlated, as well as opposite spots 1-3 and 2-4.  These anti-correlations are directly visible in the satellite spot lightcurves in most cases, especially in the August 2018 data.
These results differ from \citet{Apa16}, who found a fairly constant pattern of no or weak anti-correlation between most of the satellite spots, with relatively strong anti-correlations between one set of directly opposite spots, and two sets of adjacent spots.  However, given the small range of flux values covered in our observations, we do not consider these Spearman $\rho$ values to be robust. Additionally, in this study we have considered observations taken only in a fairly narrow range of observing conditions -- thus, we have not considered Spearman $\rho$ for a sufficient range of different conditions to justify strong conclusions regarding satellite spot behavior.  It would be interesting to consider correlation plots for all archival SPHERE-IRDIS sequences using satellite spots throughout the sequence, to search for overarching trends as a function of observing conditions.

\bsp	
\label{lastpage}
\end{document}